\documentclass{aa}  

\usepackage{graphicx}
\usepackage{txfonts}
\usepackage{lineno}

\usepackage[symbol]{footmisc}

\usepackage{xcolor}
\definecolor{darkorange}{RGB}{153, 76, 0}

\usepackage{stackengine}

\usepackage{hyperref}
\hypersetup{colorlinks,allcolors=[rgb]{0,0,0.8}}

\defcitealias{gatti21}{Gatti \& Sheldon et al. 2021}
\defcitealias{myles21}{Myles \& Alarcon et al. 2021}
\defcitealias{cordero22}{Cordero et al. 2022}
\defcitealias{maccrann22}{MacCrann  et al. 2022}
\defcitealias{McClintock19}{McClintock \& Varga et al. 2019}
\defcitealias{bellagamba19}{Bellagamba et al. 2019}
\begin{document} 

   \title{The SRG/eROSITA All-Sky Survey}

    \subtitle{Dark Energy Survey Year 3 Weak Gravitational Lensing by eRASS1 selected Galaxy Clusters}
\titlerunning{DES Y3 WL of eRASS1 clusters}
   \author{S.~Grandis
          \inst{1}\and  V.~Ghirardini\inst{2} \and  S.~Bocquet\inst{3}\and C.~Garrel\inst{2}\and J.~J.~Mohr\inst{3,2}\and A.~Liu\inst{2}\and M.~Kluge\inst{2}\and L.~Kimmig\inst{3}\and T.~H.~Reiprich\inst{4}\and \newline
          A.~Alarcon\inst{9}
% ...
\\
           (The Dark Energy Survey Collaboration and the eROSITA-DE Consortium)
          }

   \institute{Affiliations at the end of the paper, \email{sebastian.grandis@uibk.ac.at}}

   \date{Received---; accepted ---}

 
  \abstract
   {Number counts of galaxy clusters across redshift are a powerful cosmological probe, if a precise and accurate reconstruction of the underlying mass distribution is performed -- a challenge called \emph{mass calibration}. With the advent of wide and deep photometric surveys, weak gravitational lensing by clusters has become the method of choice to perform this measurement.}
   {We measure and validate the weak gravitational lensing (WL) signature in the shape of galaxies observed in the first 3 years of the Dark Energy Survey (DES Y3) caused by galaxy clusters and groups selected in the first all-sky survey performed by SRG (Spectrum Roentgen Gamma)/eROSITA (eRASS1). These data are then used to determine the scaling between X-ray photon count rate of the clusters and their halo mass and redshift.}
   {We empirically determine the degree of cluster member contamination in our background source sample. The individual cluster shear profiles are then analysed with a Bayesian population model that self-consistently accounts
   for the lens sample selection and  contamination, and includes marginalization
   over a host of instrumental and astrophysical systematics. To quantify the accuracy of the mass extraction of that model, we perform mass measurements on
   mock cluster catalogs with realistic synthetic shear profiles. This allows us to establish that hydro-dynamical modelling uncertainties at low lens redshifts ($z<0.6$) are the dominant systematic limitation.  At high lens redshift the uncertainties of the sources' photometric redshift calibration dominate.} 
   {With regard to the X-ray count rate to halo mass relation, we determine its amplitude, its mass trend, the redshift evolution of the mass trend, the deviation from self-similar redshift evolution and the intrinsic scatter around this relation.  }
   {
   The mass calibration analysis performed here sets the stage for a joint analysis with the number counts of eRASS1 clusters to constrain a host of cosmological parameters. We demonstrate that WL mass calibration of galaxy clusters can be performed successfully with source galaxies whose calibration was performed primarily for cosmic shear experiments, opening the way for the cluster cosmological exploitation of future optical and NIR surveys like \textit{Euclid} and LSST.
   }


   \maketitle

%

\section{Introduction}

General Relativity describes the gravitational force as curvature of space-time, induced by mass and energy densities. This curvature determines the so-called null geodesics on which photons travel through space time.  As a consequence, the path of the light from a distant source to the observer is deflected by intervening gravitational potentials, with a deflection angle that is proportional to the path integral of the transverse component of the negative gradient of the potential. Indeed, \citet{dyson20} measured the predicted displacement in the positions of
distant stars when observed close to the Sun during a solar eclipse. In the cosmological context, the original position of the sources is not known. In general, also the gravitational potential of the intervening lenses is unknown, 
because the majority of cosmic matter is believed to be invisible and its density is among the targets of our experiments.

On cosmic scales, gravitational lensing of distant galaxies thus relies on the fact that the differential deflection of neighboring light paths locally induces a magnification and a distortion of the source galaxy image \citep[for a pedagogical introduction see][]{schneider05}.
In this regime, large-scale gravitational potentials create spatially coherent, small distortions in the observed shapes of background galaxies, 
a process called \emph{weak gravitational lensing} (hereafter WL). A special case is the WL caused by massive gravitationally collapsed, virialised structures, 
termed halos, which dominate the line-of-sight-integrated gravitational potential \citep[for a review see][]{umetsu20}. These objects induce a tangential distortion in the shapes of the background galaxies which traces the density contrast of the 2-d projected density profiles. The 3-d density profiles of halos are well understood in simulations, and are found to tightly follow a parametric class of functions with two free parameters: the total halo mass and the typical halo scale \citep{NFW}. Fitting predictions derived from this family of models to the tangential reduced shear profiles thus enables the direct measurement of the halo mass. 

The most massive halos host galaxy clusters, whose observed properties display tight correlations among themselves 
\citep{mohr97}, and with the host halos mass \citep{angulo12}, 
enabling a clean halo selection. For such samples of galaxy clusters, the WL mass information is then used to reconstruct the differential number density of halos as a function of mass and redshift, the halo mass function. It traces the growth of structure in the Universe, and is therefore a sensitive probe to the total matter density, the accelerated expansion, and gravity itself \citep{haiman01, majumdar04, allen11}. Examples of such analyses are \citet{mantz15}, who used X-ray selected  galaxy clusters with pointed WL observations, and \citet{bocquet19}, who used galaxy clusters selected at millimeter wavelengths.  Both these analyses result in competitive cosmological constraints, which are largely complementary and independent of other cosmological probes \citep[for a recent review, see][]{huterer2022}.

Our ability to detect larger numbers of massive halos has recently been transformed by eROSITA, which performed its first All-Sky Survey from December 2019 to June 2020 \citep{Predehl2021, Sunyaev2021}, detecting more than 1 million X-ray sources in the Western Galactic Hemisphere \citep{merloni2023}. Among these X-ray sources, $\sim 12$k are confirmed as clusters of galaxies through significant extended X-ray emission and associated red galaxy members in the 12791~deg$^{2}$ footprint covered by the DESI Legacy Survey DR9 and DR10 data \citep{bulbul23, kluge23}. In this work, we will restrict ourselves to the 5263 clusters selected for the \emph{cosmology} sample, via a higher extent likelihood cut. They have percent accurate photometric redshifts, in the range of 0.1 to 0.8, and a well-calibrated selection function \citep{clerc23}. 

We complement the eROSITA data with the wide photometric Dark Energy Survey year 3 (DES-Y3) data \citep{sevilla-noarbe21}. From 
these data we use more than 10$^8$ galaxy shape measurements measured in the $riz$ bands by \citetalias{gatti21} in an effective survey area of 
4,143~deg$^2$ of the southern sky. The strength of the WL signal around galaxy clusters depends not only on their mass, but also on the geometrical configuration of lens and source. To estimate statistically the 
distance of the latter we leverage the exquisite accuracy of the photometric redshift calibration of the DES-Y3 sources by \citetalias{myles21}. 

Besides the uncertainty on the shape and photometric redshift measurements for the source galaxies, percent accurate cluster WL measurements need to account for the cluster mis-centring errors and contamination of the source sample by cluster galaxies \citepalias[for a complete analysis on DES year 1 data, see][]{McClintock19}. Even if all observational systematics are well-calibrated and accounted for, the modelling 
uncertainties of baryonic feedback processes alter the mass distribution of galaxy clusters. This poses a theoretical limit to our ability to calibrate cluster masses via WL \citep{grandis+21}. As argued in that work, setting up many realisations of realistic, synthetic cluster WL mock datasets, and analysing them with the same mass extraction method as the real data, allows 
one to calibrate both the absolute values, and -- crucially -- also the uncertainty of the \emph{WL mass bias}. This approach has already been successfully demonstrated in work by \citet{chiu22a} in the context of Hyper Supreme Camera (HSC) observations of the eROSITA Final Equatorial-Depth Survey (eFEDS) used for science verification.

This paper is organized as follows: in Section~\ref{sec:data} we describe the lens and source catalogs, as well as the calibration products used in this analysis. Section~\ref{sec:3} then describes the measurement process for the tangential reduced shear, and the suite of associated validation and calibration steps. 
In Section~\ref{sec:4} we then describe the shear profile model used for the mass measurement, as well as its calibration on realistic synthetic data. These two sections follow in significant part the analysis performed by \citet{bocquet+ipa} of the DES Y3 WL signal around South Pole Telescope (SPT) selected clusters. The actual mass calibration is then performed and validated in Section~\ref{sec:mass_calibration},
and we discuss our results in Section~\ref{sec:discussion}. Halo masses in this work are reported as spherical over density masses $M_{\Delta \text{c}}$, with $\Delta=500, 200$. This means that they are defined via the radius $R_{\Delta \text{c}}$ enclosing a sphere of average density $\Delta$ times the critical density of the Universe at that redshift, i.e. $M_{\Delta \text{c}} = \frac{4 \pi}{3} \Delta \rho_\text{crit}(z)R_{\Delta \text{c}}^3$. We use a flat $\Lambda$CDM cosmology with parameters $\Omega_\text{M}=0.3$, and $h=0.7$ as reference cosmology, where not specified otherwise. Throughout this work, we will use lens redshift $z_\text{l}$, and cluster redshift $z_\text{cl}$ interchangeably, depending on which is more applicable to the context of the discussion.

\section{Data}\label{sec:data}

\begin{figure}
  \includegraphics[width=\columnwidth]{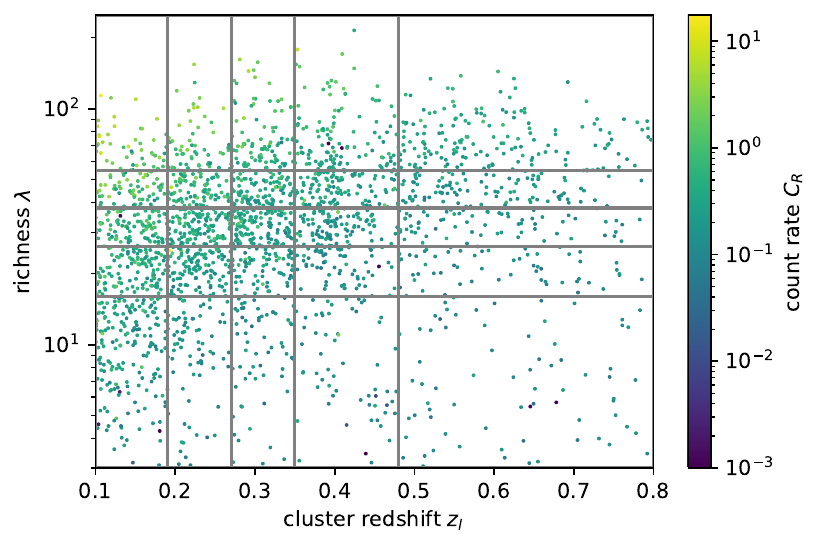}
  \caption{Lens sample composed of the eRASS1 cosmology cluster and group sample with DES WL data, plotted in redshift against richness, with color coded the X-ray count rate. 
  The gray lines mark the edges of the richness--redshift bins used for the calibration and validation of the WL measurement. Cosmology ready WL 
  data
  products are however produced for each cluster individually.}
  \label{fig:lens_sample}
\end{figure}

\subsection{eRASS1 Cluster and Group sample}

As a lens catalog, we use the cluster catalog acquired from the first eROSITA All-Sky Survey on the Western Galactic Hemisphere,
completed on June 11, 2020. The detection properties of the X-ray sources detected in this survey, including extended sources, are discussed in detail in \citet{merloni2023, bulbul23}. We use the cosmology catalog described in detail in \citet{bulbul23}, comprised of X-ray sources detected as significantly extended, with reliable photometric confirmation and redshift estimation using the DESI Legacy Survey DR10 \citep{dey19}.
Specifically, \citet{kluge23} run an adapted version of the \texttt{redMaPPer} algorithm \citep{rykoff14, rozo15, rykoff16} on the X-ray cluster candidate position, thus measuring their richness and photometric redshift. Of the 5,263 clusters and groups in that catalog, 2,201 have DES Y3 shape and photo-$z$ information  (described below), and are therefore used for the WL analysis in this work \citep[see][Fig.~1 for a comparison of the two survey footprints]{ghirardini23}. Their distribution in redshift, richness and X-ray photon count rate is shown in Fig.~\ref{fig:lens_sample}. The bulk of the lens sample 
lies at low redshift, ideally placed for WL 
studies. Some objects at low richness have redshifts much larger then expected for an X-ray selected sample. These are contaminants, 
which we account for in our mass calibration (see Section~\ref{sec:mixture_model}).

\subsection{DES WL data}

The Dark Energy Survey is an approximately 5,000~deg$^2$ photometric survey in the optical bands $grizY$, carried out at the 4m Blanco telescope at the Cerro Tololo Inter-American Observatory (CTIO), Chile, with the Dark Energy Camera \citep[DECam][]{flaugher15}.  In this analysis we utilize data from the first three years of observations (DES~Y3), covering the full survey footprint.

\begin{figure}
  \includegraphics[width=\columnwidth]{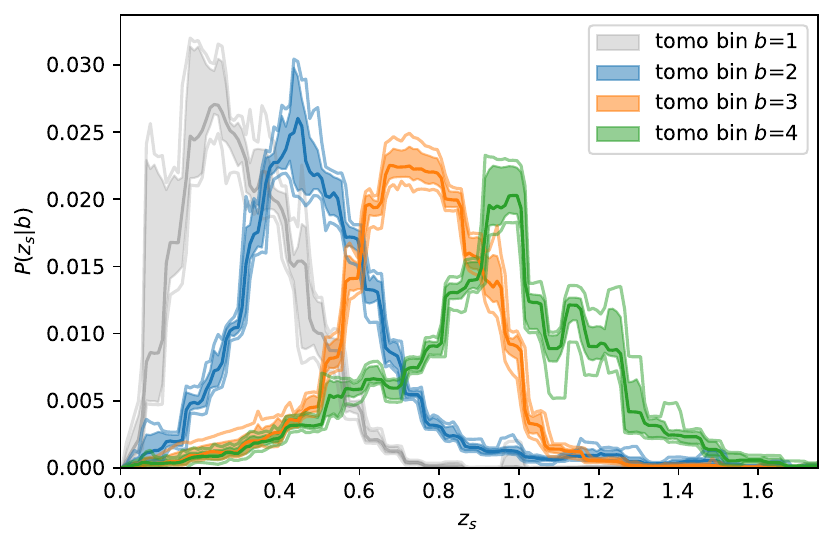}
  \caption{Source redshift distributions of the tomographic redshift bin of the DES Y3 data. We plot the mean of the 1000 realisations (full line), together with the 16-84 percentile (filled region), and the 2.5-97.5 percentile (faded lines). The first tomographic bin is not used, while the others are weighted 
  so as to ensure a lens dependent background selection.}
  \label{fig:source_distr}
\end{figure}

\subsubsection{The shape catalog}

The DES~Y3 shape catalog \citepalias{gatti21} is built from the $r,i,z$-bands using the \textsc{Metacalibration}
pipeline \citep{huff&mandelbaum17, sheldon&huff17}.
Other DES~Y3 publications contain more detailed information about the photometric 
dataset \citep{sevilla-noarbe21}, the Point-Spread Function modeling \citep{jarvis21}, and image simulations \citep{maccrann22}. We refer the reader to these works for more information.
After application of all source selection cuts, the DES~Y3 shear catalog contains about 100~million galaxies over an area of 4,143~square-degrees. Its effective source density is 5--6~arcmin$^{-2}$, 
depending on the exact definition. 
In Sections~\ref{sec:3}, we will derive the exact values of effective source density.

\subsubsection{Source redshift distributions and shear calibration}
\label{sec:DES_Pz}

Our analysis uses the same selection of lensing source galaxies in tomographic bins as the DES 3x2pt analysis \citep{DES_Y3_3x2pt}. This selection is defined in \citetalias{myles21}.
In that work, source redshifts are estimated with Self-Organizing Maps, 
and the method is thus referred to as SOMPz.
The final calibration accounts for the (potentially correlated) systematic uncertainties in source redshifts and shear measurements, as determined in the image simulations by \citet{maccrann22}.
For each tomographic source bin, the estimated redshift distribution is provided, and the systematic uncertainties on this estimate are captured through 1,000 realizations (see Fig.~\ref{fig:source_distr}). These realisations are indexed via the hyper parameter $\mathcal{H}$, which takes integer values between 0 and 999 \citep{cordero22}.
For convenience, these redshift distributions are normalized to varying values $(1+m)$, where the varying $m$ spans the range of the multiplicative shear bias values. Note that the DES Y3 survey only has minor depth variations over the survey area. Survey averaged quantities, like the source redshift distributions, are thus applicable to the joint eROSITA-DE -- DES Y3 footprint.

We will also use \textsc{dnf} \citep{devicente16} and \textsc{bpz} \citep{benitez00} source redshift measurements to constrain the amount of cluster member contamination in our source sample (see Section~\ref{sec:cluster_mbr_cnt}). While \textsc{bpz} is a Bayesian template fitting code, \textsc{dnf} is based on a nearest neighbor interpolation on the color magnitude space of spectroscopic reference samples. As such, \textsc{bpz} is more robust against the incompleteness of the spectroscopic reference samples, while \textsc{dnf} is more data driven, and avoids the biases that occur from the template choice \citep[see][Section 6.3, for more details]{sevilla-noarbe21}.  

\section{Measurement}\label{sec:3}

In this section, we shall outline 1) the methods used to measure the WL signal around eROSITA selected clusters, 2) the steps undertaken to correctly calibrate this measurement, and 3) the use of this measurement to determine the mass scale of the eROSITA selected clusters and groups.  Note that the methods presented in this work draw significantly upon the methods presented in \citet{ bocquet+ipa}.

\subsection{WL by massive halos}

Assuming that background galaxies are isotropically oriented intrinsically, the coherent tangential distortion induced by the gravitational potential on background galaxy images at a radius $R$ from the halo center can be obtained, to linear order, by averaging the tangential component of their ellipticities, 
\begin{equation}
  g_\text{t}(R) = \mathcal{R}^{-1}\langle e_\text{t} \rangle_R, 
\end{equation}
 where $g_\text{t}(R)$ is the reduced tangential reduced shear, and $\mathcal{R}$ is the average response of the measured ellipticity to a shear (shear response). In practical applications, instrumental effects and noise make it different from 1. It is determined in the process of galaxy shape measurement and validated through image simulations. We discuss below how we include this effect in our measurements, and how we account for non-linear responses. While the ellipticity field itself is not Gaussian, our reduced shear estimator is constructed by averaging a large number of sources. 
 The central limit theorem thus ensures that the statistical error of the shear estimate can be directly derived from the effective dispersion $\sigma_\text{eff}$ in the ellipticity of the source galaxies, and their effective number $N_\text{eff}$. We will discuss below how we estimate these two quantities.

Given a source galaxy at redshift $z_\text{s}$, the lens redshift $z_\text{l}$, and the surface mass density $\Sigma(\textbf{x})$ in the lens plane, the tangential reduced shear in any position on the sky $\boldsymbol{\theta} = \textbf{x}/D_\text{l}$ is given by
\begin{equation}
    g_\text{t}(\boldsymbol{\theta}) = \frac{ \gamma_\text{t}(\textbf{x}) }{ 1 - \kappa(\textbf{x})} \text{ with }\kappa(\textbf{x}) = \Sigma_\text{crit}^{-1}\Sigma(\textbf{x}),
\end{equation}
 where $\gamma_\text{t}(\textbf{x})$ is the tangential
 component of the
 shear, $\kappa(\textbf{x})$ the convergence, $D_\text{l}$ the angular diameter distance to the lens, and $\Sigma_\text{crit}^{-1}$ the inverse of the critical lensing surface density. The latter expresses the geometrical configuration of the source and lens, and is given by
 \begin{equation}\label{eq:sigmacritinv_ls}
     \Sigma_\text{crit,ls}^{-1} = \frac{4 \pi G}{c^2} \frac{D_\text{l}}{D_\text{s}} \max \left[0, D_\text{ls} \right],
 \end{equation}
where $G$ is the gravitational constant, $c$ the speed of light. Furthermore, we use the angular diameter distance between observer and source ($D_\text{s}$), and lens and source ($D_\text{ls}$). Note that for sources in front of the lens, $D_\text{ls}$ becomes negative. These sources are not lensed, which we account for by setting their inverse critical lensing surface density to zero, hence the term $\max \left[0, D_\text{ls} \right]$.

In General Relativity, the azimuthally averaged tangential  shear 
at projected radius $R$ is proportional to the density contrast, 
\begin{equation}
    \gamma_\text{t}(R) = \Sigma_\text{crit}^{-1} \Big[ \langle \Sigma(<R) \rangle - \Sigma(R) \Big],
\end{equation}
while the orthogonal, B-mode like component $\gamma_\text{x}$, called cross-shear  (see below for the exact definition of the decomposition), averages to zero when integrated over closed paths such as radial bins, 
\begin{eqnarray}
    \gamma_\text{x}(R) = 0.
\end{eqnarray}
We will use the latter as a validation test for our measurement.

\subsection{Measurement}

\begin{figure*}
  \includegraphics[width=\textwidth]{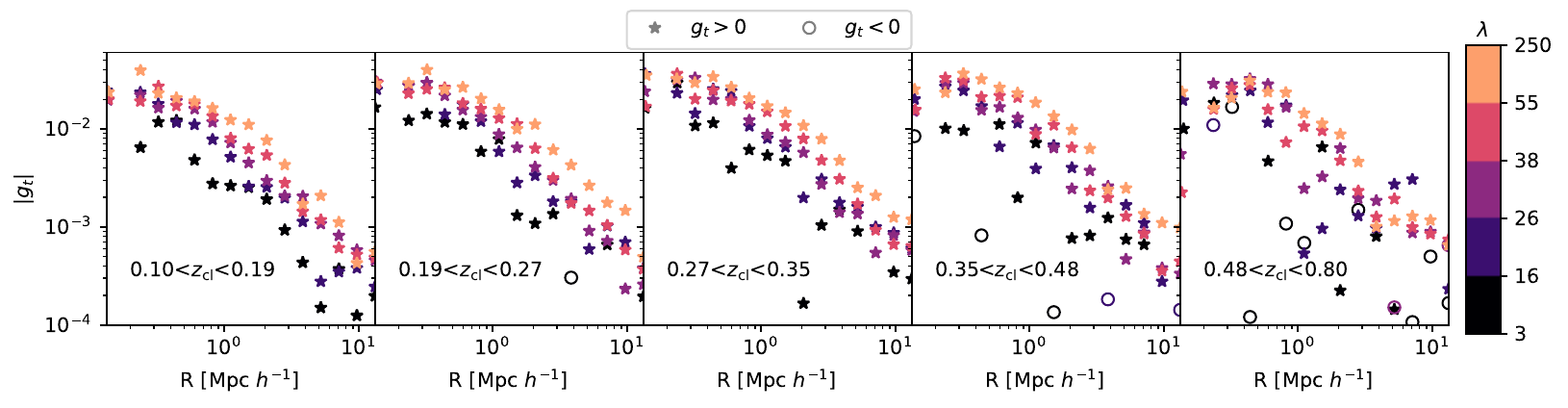}
  \caption{Absolute value of the raw tangential reduced shear profile, Eq.~\ref{eq:raw_shear}, in bins of cluster redshift (panels: lowest redshift left, higher redshift right) and of richness (color coding), for our background selection. Stars denote positive values, while empty circles denote negative values. The profiles show a clear increase towards the cluster centers, as well as a trend with richness, in line with our qualitative expectations.
  }
  \label{fig:raw_gt}
\end{figure*}

To perform the measurement, for each lens $l$ we query the shape catalog for all sources at a 
projected
physical distance of $<15$ Mpc $h^{-1}$ in the reference cosmology ($h=0.7$, $\Omega_\text{M}=0.3$). For each 
source lens pair $(i, l)$, we compute the position angle $\varphi$ at which the source is seen from the lens' position. 

We then project the ellipticity components $e_{1,2}$ onto the tangential and cross components as
\begin{equation}
    \begin{split}
    e_{\text{t}} &= - e_1 \cos 2 \varphi + e_2 \sin 2 \varphi \\
    e_{\text{x}} &= e_1 \sin 2 \varphi + e_2 \cos 2 \varphi, \\
    \end{split}
\end{equation}
We use here the tangential/cross decomposition for the case where the $e_2$-component is defined w.r.t. right ascension, which is the case in the DES Y3 shape catalog.\footnote{The prevalent notation \citep[e.g.][Eq. 83]{umetsu20} is valid for $e_2$ defined for a right-handed, local coordinate system ($x \propto - \text{RA}$). It can be recovered by the transformation $e_2 \mapsto - e_2$.}
We also record the source weight $w^s_i$, the response $\mathcal{R}_{i} $, as presented in \citetalias[][Section~4.3]{gatti21}, and the photometric redshift estimates in form of the SOMPz cell $c_{\text{SOM},i}$ of the source, and the photometric redshift estimates $\hat z_{\textsc{dnf},i}$ and $ \hat z_{\textsc{bpz},i}$.

To select only sources in the background of each lens, we discard the first tomographic redshift bin and apply a weight to each source depending on the tomographic redshift bin $b$ it resides in, 
\begin{equation}\label{eq:tomo_weights}
    w^b = \begin{cases}
			\left\langle \Sigma_\text{crit,ls}^{-1} \right\rangle_{\text{s}\in b} &\text{for } z_\text{l} < z_{\text{med},b} \text{ and } b>1\\
            0. & \text{otherwise},
		 \end{cases}
\end{equation}
where the average is taken using the mean source redshift distribution of the tomographic bin, as reported by \citetalias{myles21}. Our background selection is thus a weighted sum of the tomographic bin selections. This is convenient, as the shear and photo-$z$ calibration \citepalias{myles21, cordero22, maccrann22} is only valid for the specific selection criteria of the tomographic redshift bins. It can be extended to our analysis by using the same weights as above.

To validate and calibrate the cluster WL measurement, we define lens richness bins $(3, 16, 26, 38, 55, 250)$, and redshift bins $(0.10, 0.19, 0.27, 0.35, 0.48, 0.8)$, shown as grey lines in Fig.~\ref{fig:lens_sample}, and tuned to result in approximately 
even occupations of the lenses.
In these bins we measure the following quantities for radial bins whose upper edges are log-equally spaced between 0.2--15~$h^{-1}$Mpc at the lens redshift in the reference cosmology:
\begin{itemize}
\item the raw tangential and cross-shear, with the tangential components shown in Fig.~\ref{fig:raw_gt},  
\begin{equation}\label{eq:raw_shear}
    g_\mathrm{\alpha,\,raw} = \frac{\sum_{b=2,3,4}  w^b \sum_{i\in b}  w^\mathrm{s}_i e_{\mathrm \alpha,i}  }{\sum_{b=2,3,4} w^b \sum_{i\in b} w^\mathrm{s}_i \mathcal{R}_{i} } \text{ for } \alpha\in(t,x),
\end{equation}
 where $i \in b$ stands for the source $i$ in the tomographic bin $b$ and in the respective lens richness, redshift and distance bins.
\item the effective number of sources
\begin{equation}\label{eq:N_eff}
    N_\mathrm{eff} = \frac{ \left( \sum_{b=2,3,4} \sum_{i\in b}   w^b  w^\mathrm{s}_i  \mathcal{R}_{i}  \right)^2}{ \sum_{b=2,3,4} \sum_{i\in b} \left(  w^b  w^\mathrm{s}_i \mathcal{R}_{i} \right)^2 },
\end{equation}

\item the effective dispersion of source ellipticities
\begin{equation}\label{eq:sigma_eff}
    \sigma^2_\mathrm{\alpha,\,eff} = \frac{ \sum_{b=2,3,4} \sum_{i\in b} \left(  w^b  w^\mathrm{s}_i \right)^2 e_{\mathrm \alpha,\,b,i}^2 }{ \sum_{b=2,3,4} \sum_{i\in b} \left(  w^b  w^\mathrm{s}_i \mathcal{R}_{i} \right)^2 } \text{ for } \alpha\in(t,x), \text{ and}
\end{equation}

 \item the source redshift distribution for fine source redshift bins with edges $(z_{\text{s}-}, z_{\text{s}+})$, reading
 \begin{equation}
     \hat P_\beta(z_{s}) = \frac{\sum_{b=2,3,4} w^b  \sum_{i\in b}   w^\mathrm{s}_i \mathcal{R}_{i} \, \text{I}_s(\hat z_{\beta,i})}{\sum_{b=2,3,4} w^b \sum_{i\in b} w^\mathrm{s}_i \mathcal{R}_{i} } \text{ for } \beta\in(\textsc{bpz},\textsc{dnf}),
 \end{equation}
 where $\text{I}_s( \hat z_{\beta,i})=1$ when $z_{\text{s}-}<\hat z_{\beta,i}<z_{\text{s}+}$. Practically speaking, this is a weighted histogram of the photo-$z$ estimates $\hat z_{\beta,i}$.
\end{itemize}

It is important to note that 
also 
the estimates of the effective source density, the shape dispersion and the redshift distribution need to take 
the shear response $\mathcal{R}$ 
into account. 
For a derivation for Eq.~\ref{eq:N_eff},
and Eq.~\ref{eq:sigma_eff}, we refer the reader to App.~\ref{app:Nsigma_eff}, where we present the derivation of the effective number of sources and their effective shape dispersion in the presence of a shear response and un-normalized weights.

\subsection{Calibration and validation}

We shall now discuss several calibration and validation steps we take to ensure the quality of our WL measurements.

\subsubsection{Cluster member contamination}\label{sec:cluster_mbr_cnt}

\begin{figure*}
  \includegraphics[width=\textwidth]{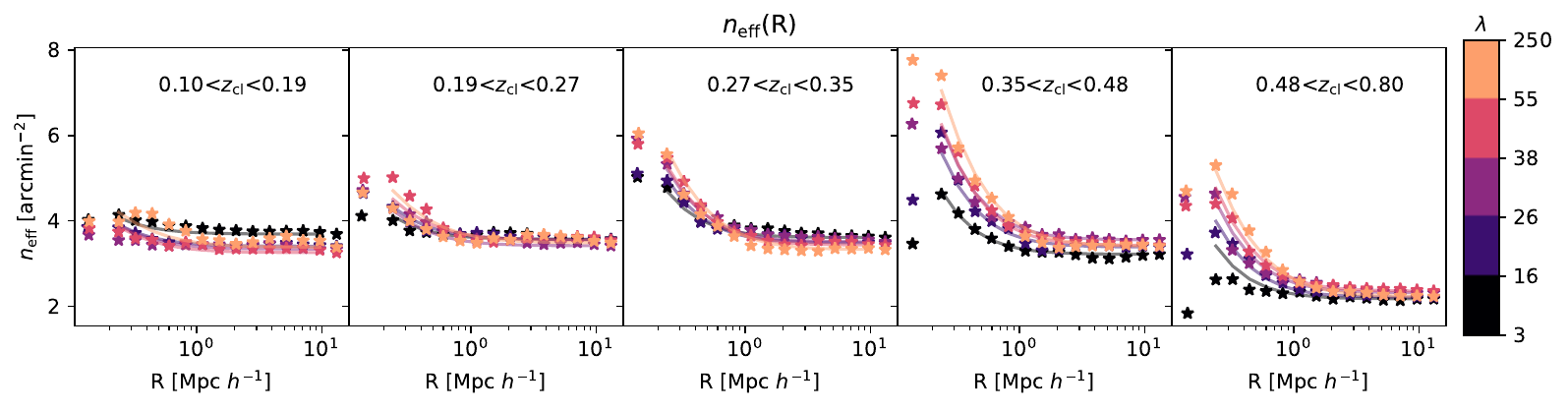}
  \caption{Effective source density estimated from the effective number of source, Eq.~\ref{eq:N_eff}, and the geometric area of the radial bins stacked in bins of cluster redshift (panels, lowest redshift left, higher redshift right) and of richness (color coding) as stars, for our background selection. Cluster members contaminating our background source sample lead to an increase of the effective source density towards the cluster center which depends on both cluster richness and redshift. This signal needs to be determined, as cluster member contaminants are not sheared and dilute the shear estimator. Our model (lines) captures this trend well.
  }
  \label{fig:clmcont_neff}
\end{figure*}

\begin{figure}
  \includegraphics[width=\columnwidth]{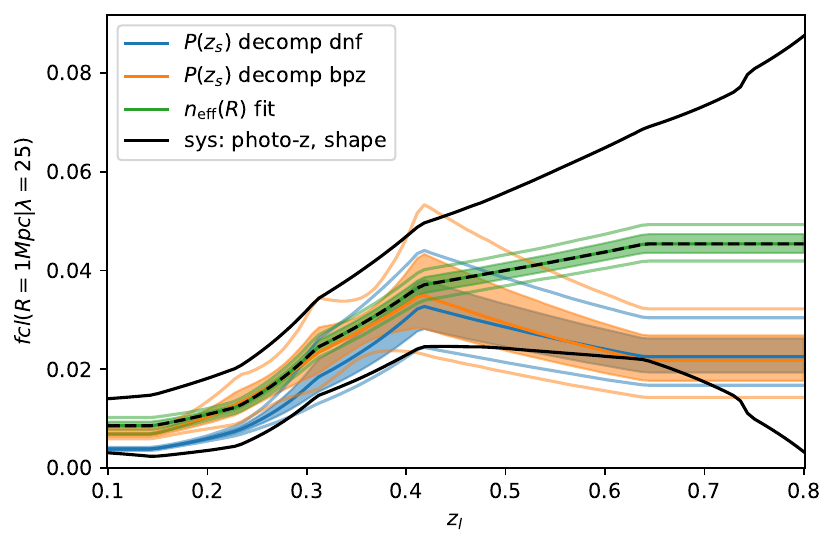}
  \caption{Cluster redshift trend of the cluster member contamination of a richness $\lambda=25$ object at 1 Mpc from the cluster center. In green the fit to the effective source number density, while in blue and orange the constraints from the \textsc{dnf} and \textsc{bpz} source redshift distribution decomposition are shown. The latter method is independent of masking. While we find marginal agreement between the contamination fractions determined by the two methods, the difference is smaller than the relative error induced by shape and photo-$z$ systematics -- plotted in black extending from the source density fit for comparison.  
}
  \label{fig:measurement_sys}
\end{figure}

As suggested by their name, galaxy clusters are substantial over-densities not only in matter, but also in the galaxy field.  As a result, an a priori unknown fraction of the cluster members contaminates any source background selection.  
These cluster member contaminants carry no shear distortion signal from the cluster potential, and therefore dilute the shear signal \citep{hoekstra12, gruen14, dietrich19, varga19}. This effect is equivalent to source-lens clustering in other WL applications \citep[compare, for instance, with][Section III.B]{prat22}.
Current simulations are not accurate enough to reconstruct the colors and magnitudes of cluster member galaxies with sufficient fidelity to calibrate this effect directly in simulations.
We therefore resort to empirical calibration methods. We shall outline in the following paragraphs the model for the cluster member contamination, and the two methods we use to fit for it, as well as a comparison of the fit results.

\paragraph{Cluster member contamination model} In modeling the cluster member contamination we adopt an approach that was developed previously in \citet{paulus21}.  We parametrize the radial fraction of cluster contaminants as  
\begin{equation}\label{eq:fcl_profile}
\begin{split}
    &f_\text{cl}(R | \lambda, z) = \frac{A(\lambda, z, R)}{1+A(\lambda, z, R)} \text{ with }\\ 
    &A(\lambda, z_j, R) = e^{A_{j}} \left( \frac{\lambda}{25} \right)^{B_\lambda} \Sigma^\text{norm}_\text{NFW}\left(R \Big| r_S = c^{-1} \left( \frac{\lambda}{20}  \right)^{1/3}\right),
\end{split}
\end{equation}
where $A_{j}$ is a free amplitude parameter for each cluster redshift bin $j$ we consider, $B_\lambda$ describes the richness trend of the cluster member contamination, and $\Sigma^\text{norm}_\text{NFW}(R | r_S)$ is a 2d 
projected NFW profile, normalized to 1 at its scale radius $r_S$, which we parameterize via a concentration $c$ and a scaling with richness. 

As shown in App.~\ref{app:fcl_model}, $A(\lambda, z, R)$ is proportional to the radial number density profile of the cluster member contaminants, which we therefore model as a 
2d projected NFW profile. Given that at this stage of the analysis, the cluster mass is still unknown, we resort to using the richness as a mass proxy. For ICM-selected cluster samples, the richness is known to scale approximately linearly with mass \citep{saro15, bleem20, grandis20, grandis21b}. Furthermore, the richness provides a convenient proxy also for the total number density of cluster member galaxies. It also has contributions from correlated structures along the line of sight, that would also contribute almost unsheared contaminants to the background sample.

\paragraph{Regularisation} Given the complex interactions of cluster redshift, cluster member colors, photo-$z$ estimation and background selection, we opt for a non-parametric fit for the redshift evolution by fitting independent amplitudes $A_{j}$ for each cluster redshift bin. To impose a data-driven smoothness on the redshift evolution, we place a Gaussian prior with unknown variance $\sigma_\text{reg}^2$ on the difference between the amplitudes of neighboring redshift bins, implemented via the Gaussian likelihood  
\begin{equation}
    \ln \mathcal{L}_\text{reg} = \frac{n_\text{z-bins}-1}{2} \ln \sigma_\text{reg}^2 - \frac{1}{2 \sigma_\text{reg}^2} \sum_j \left( A_{j} - A_{j+1}\right)^2,
\end{equation}
where $n_\text{z-bins}$ is the number of redshift bins used. This likelihood adds an extra fit parameters, $\sigma_\text{reg}$, when added to the primary likelihood of the cluster member contamination.

 \paragraph{Source density fit} The cluster member contamination can be fitted from the effective source density as a function of projected distance from the clusters (which are our lenses) stacked in bins of cluster redshift and richness, which we present in Fig.~\ref{fig:clmcont_neff}. This density is computed by dividing the effective number of sources, Eq.~\ref{eq:N_eff}, by the geometric area of the radial 
bin.
Given the presence of cluster member galaxies obstructing the detection of background galaxies, and of detection masks, this results in a underestimation of the actual source density, as explored in detail by \citet{kleinebreil23} in the context of Kilo Degree Surveys (KiDS) WL around SPT selected clusters. This can be seen most clearly in the reported number density in the most central radial bin $R<0.2$ $h^{-1}$~Mpc. On these radial scales, cluster lines of sight are typically dominated by the massive central galaxy and its stellar envelope. We exclude this radial bin from the following analyses.

\begin{figure*}
  \includegraphics[width=\textwidth]{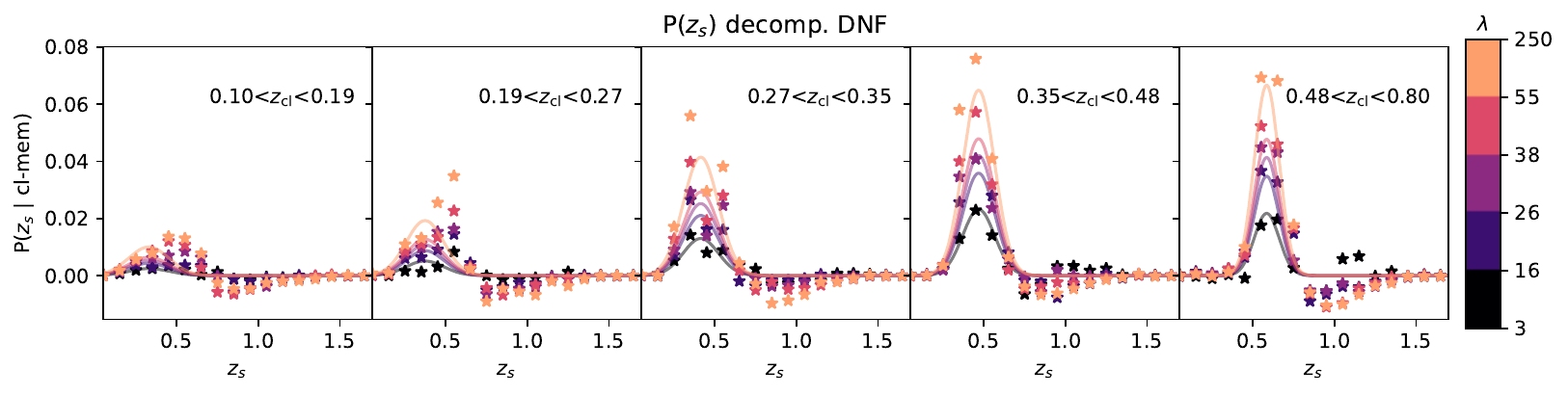}
  \includegraphics[width=\textwidth]{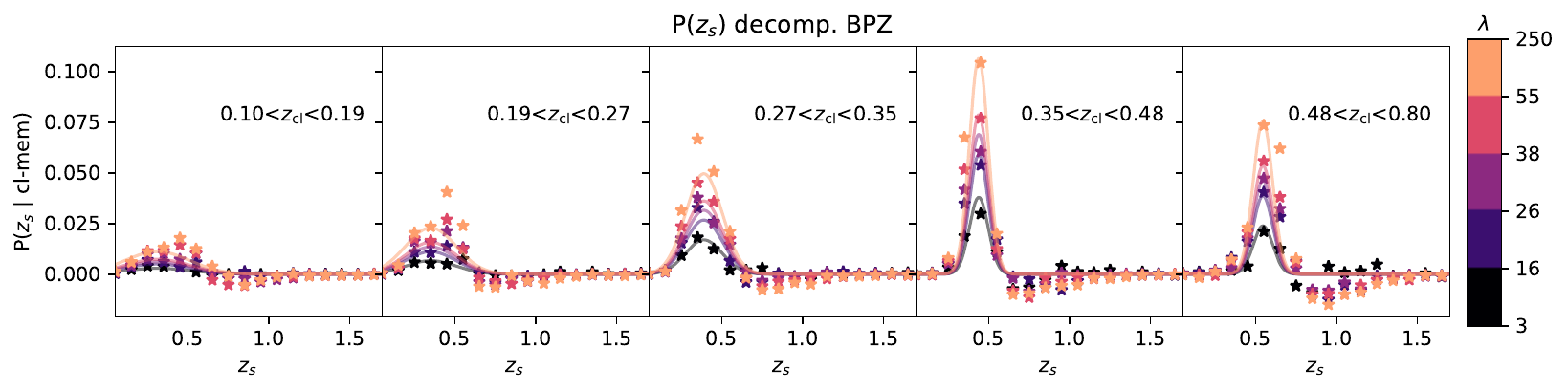}
  \caption{Validation of the source redshift distribution decomposition for \textsc{dnf} (top) and \textsc{bpz} (bottom) source redshift estimates. In different richness (color coded) and redshift bins (different panels), we show the difference between the source redshift distribution measured along the cluster 
  line of sights
  and the field distribution extracted at a safe distance from the cluster center, shown as stars. Both are shown here for the 
  projected
  cluster-centric distance of 1~Mpc in our reference cosmology. The resulting increment is well modelled by a Gaussian component (lines) caused by the cluster members contaminating the source sample. The amplitude of this component scales with cluster richness.
  }
  \label{fig:Pz-valid}
\end{figure*}

We fit the effective number density of cluster members as a function of projected radius $R$ for a cluster of richness $\lambda$ and redshift $z$ as
\begin{equation}\label{eq:neff_clmc}
    n_\text{eff}(R | \lambda, z)= \frac{n_\text{field}(z)}{1-f_\text{cl}(R | \lambda, z)},
\end{equation}
where the effective number density in the field $n_\text{field}(z)$ is extracted around our clusters in the outer regions, $11~h^{-1}$Mpc $<R<15$~$h^{-1}$~Mpc.

For the fit of the parameters of the cluster member contamination, $\boldsymbol{p}_\text{cl}=$ $(A_{j}, B_\lambda, c)$, we set up a Gaussian likelihood for the effective number density measured in the above mentioned richness-redshift bins. Our data vector, the effective source density, results from the weighted sum of all tomographic redshift bins (see Eq.~\ref{eq:N_eff}). We therefore fit one global cluster member contamination for our overall background selection. We assume a covariance matrix which combines Poisson noise from the effective number of sources and the variance we find in the outer radial bin, as a proxy for the cosmic variance in the source density. Sampling this likelihood together with the regularization likelihood provides us with constraints on the parameters of the cluster member contamination.

\paragraph{Source redshift distribution decomposition} The estimate from the effective source density around clusters needs to be validated independently, as in the current implementation we did not consider source obstruction and masking. As demonstrated by \citet{varga19}, and applied by \citetalias{McClintock19}; \citet{paulus21, shin21, bocquet+ipa},
cluster member contamination can also be determined by decomposing the source redshift distribution as a function of radius into a field component and a Gaussian component of cluster members, i.e., 
\begin{equation}
    P(z_\text{s}|R, \lambda, z) = (1-f_\text{cl}(R | \lambda, z)) \hat P_\text{field}(z_\text{s}| z) + f_\text{cl}(R | \lambda, z) P_\text{cl}(z_\text{s}| z), 
\end{equation}
where the field source redshift distribution $\hat P_\text{field}(z_\text{s}| z)$ is measured in the outer radial bin. The cluster member component is modelled as a Gaussian $P_\text{cl}(z_\text{s}| z) =\mathcal{N}(z_\text{s}|z+\mu(z), \sigma^2(z))$ with mean offset $\mu(z) = \mu_0 + \mu_z (z-z_0)$, and standard deviation $\sigma(z) = \sigma_0 + \sigma_z (z-z_0)$, with pivot $z_0=0.3$, close to the cluster redshift median. This technique is an adaptation of the method proposed by \citet{gruen14} to wide photometric surveys with readily available photometric redshift estimates. Given that we fit for the shape of a normalized source redshift distribution, the amount of masking does not impact this inference. The likelihood is
\begin{equation}
    \ln \mathcal{L}_\beta = \sum N_\text{eff} \hat P_\beta(z_{s}) \ln P(z_\text{s}|R, \lambda, z) \text{ for } \beta\in(\textsc{bpz},\textsc{dnf}),
\end{equation}
where the sum runs over richness, cluster redshift, cluster-centric distance and source redshift bins.\footnote{This follows directly from the likelihood of a sample $x_i$ being drawn from a model distribution $P(x)$, $\ln \mathcal{L} = \sum_{i\in b} \ln P(x_i)$. In our case, the model distribution is $P(z_\text{s}|R, \lambda, z)$, and the number of samples in each richness, cluster-centric distance, source redshift bin is $ N_\text{eff} \hat P_\beta(z_{s})$.} Sampled together with the regularization likelihood, this provides us with constraints on the parameters of the cluster member contamination. We perform this fit for redshift distributions constructed both for \textsc{bpz} and \textsc{dnf} photo-$z$s. 

\paragraph{Results and Comparison} The best-fit results of the number density fit are over plotted as lines on the effective source density data in Fig.~\ref{fig:clmcont_neff}, demonstrating that our model is able to capture the trends in the data well. Furthermore, we find a concentration for the contaminants profile of $c=2.5\pm 0.1$. This matches the concentration values found for the galaxy populations of massive clusters \citep{hennig17}. Our results also indicate a slight richness slope of the contamination fraction of $B_\lambda = 0.47 \pm 0.1$. The redshift evolution is shown in Fig.~\ref{fig:measurement_sys} in green (with the filled band corresponding to the 1-sigma uncertainty region, and the green lines indicate the 2-sigma). For a cluster at richness pivot $\lambda=25$ at a cluster-centric distance of $R=1$ Mpc, we find that the contamination fraction increase from around $f_\text{cl}\approx 0.01$ at $z_\text{cl}=0.1$ to $f_\text{cl}\approx 0.05$ for $z_\text{cl}>0.5$, with the strongest increase happening between redshifts $0.25<z<0.4$. We will use this setting to gauge the impact of the different cluster member contamination fits w.r.t. other systematics.

The source redshift distribution decomposition fits are validated by plotting differences between the measured source redshift distribution and the re-scaled field distribution, as shown in Fig.~\ref{fig:Pz-valid} for $1$ Mpc from the center. This subtraction clearly highlights an approximately Gaussian component, that is well described by our model, plotted as full lines. The  
low amplitude negative dips at source redshifts slightly larger than the Gaussian component show that our field re-scaling, and thus our cluster member estimate, is not perfect. These residuals, however, are of the scale of the fluctuations in the data. We therefore accept the source redshift distribution decomposition as a valid fit.

Constraints on the concentration of the contaminants profile, and the richness trend of the contamination fraction, are in statistical agreement between the fit to the effective number density and to the source redshift distributions. The contamination fraction resulting from the source redshift distribution is plotted in Fig~\ref{fig:measurement_sys}. It shows qualitative agreement with the contamination fraction constrained from the effective source density at low redshift. We will argue in the following, why the high redshift differences will not bias the mass calibration.

\begin{figure*}
  \includegraphics[width=\textwidth]{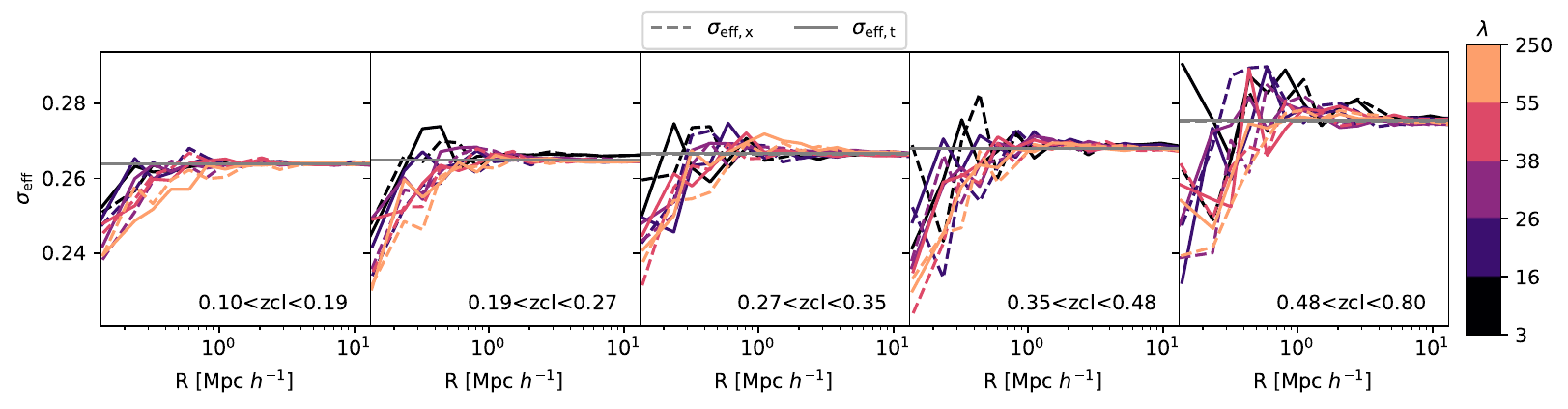}
  \caption{Effective shape noise as a function of cluster-centric distances stacked in bins of cluster redshift (panels: lowest redshift left, higher redshift right) and of richness (color coding). Despite significant scatter, increasing towards the cluster center, a trend of lower shape noise towards the center is visible. As grey horizontal lines, the effective shape noise in the field, used for the semi-analytical covariance of the shear profiles is shown.  }\label{fig:effectice-shape-noise}
\end{figure*}

\begin{figure}
  \includegraphics[width=\columnwidth]{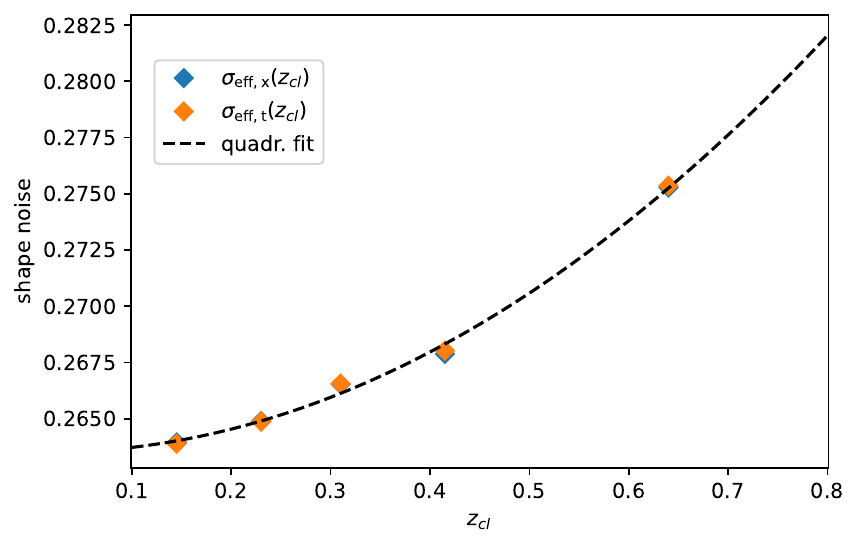}
  \caption{Cluster redshift trend of the effective shape noise in the field as blue and orange diamonds for the cross and tangential reduced shear respectively. Given our cluster-redshift-dependent background selection, the mix of source galaxies changes as a function of cluster redshift. The trend is well fitted by a quadratic expression, shown as a black dashed line.
}
  \label{fig:stats_fit}
\end{figure}

\subsubsection{Shape measurement and photo-$z$ uncertainty}\label{sec:des_photoz_shape_unc}

We compare the difference between the cluster member contamination estimates to the shape and photo-$z$ measurement uncertainties of the DES Y3 data as follows. The shape and photo-$z$ uncertainties of the tomographic redshift bin $b$ are summarized by realisations of the source redshift distributions, $P_\mathcal{H}(z_s | b)$, running over a hyper parameter $\mathcal{H}$.
The variation in the shape of these distributions expresses the uncertainty in the photo-$z$ calibration, while variation in the normalization expresses uncertainties in the multiplicative shear bias $1+m$, as described in more detail in \citet{cordero22}. Folding this together with our background selection, implemented via lens-redshift dependent weights $w_b$ for the different tomographic redshift bins $b$ 
(e.g., Eq.~\ref{eq:tomo_weights}), we find a fractional uncertainty of mean lensing efficiency 
\begin{equation}
    r_{g_\text{t}, \text{s}} = \frac{\sqrt{\text{Var}_\mathcal{H}\left[ \sum_b w_b \int \text{d}z_\text{s}\, P_\mathcal{H}(z_\text{s} | b) \Sigma_\text{crit,ls}^{-1}  \right] }}{\text{E}_\mathcal{H}\left[ \sum_b w_b \int \text{d}z_\text{s}\, P_\mathcal{H}(z_\text{s} | b) \Sigma_\text{crit,ls}^{-1}  \right]  },
\end{equation}
which we take to express the impact of shear and photo-$z$ measurement uncertainties on the measured shear profiles.
It increases from $0.55\%$ at $z_\text{cl}=0.1$ to $4.3\%$ at $z_\text{cl}=0.8$. We plot these systematic uncertainties around the best-fit values for our cluster member contamination from the effective source density in Fig.~\ref{fig:measurement_sys}. The difference between our different cluster member contamination estimates is smaller than this systematic uncertainty. We therefore conclude that we can determine the cluster member contamination to a higher accuracy than provided by the shape and photo-$z$ measurement uncertainties.

\subsubsection{Shape noise}\label{sec:shape_noise}

In Fig.~\ref{fig:effectice-shape-noise} we show the measured effective shape noise as a function of cluster redshift and richness, and as a function of cluster-centric 
projected
distance. The effective shape noise increases with cluster redshift and richness, and decreases towards the cluster center. In conjunction with the cluster member contamination estimate from the previous section, we can reconstruct the effective shape noise of cluster members from this decrement. To this end, we fit the measured effective shape noise with a squared sum of field shape noise determined in the outskirt of the clusters, and an unknown cluster-member shape noise. The latter is found to be between 18\% (for $0.1<z_\text{cl}<0.19$) and 6\% percent (for $0.48<z_\text{cl}<0.80$) lower than the field shape noise, with the trend steadily decreasing with redshift. Given that most of the signal-to-noise in cluster mass measurements comes from scales larger than $R>1$ $h^{-1}$Mpc, we ignore this effect. In future work, we aim to investigate if this is due to cluster member galaxies being inherently rounder, if the survey-averaged shear response is impacted by the stronger blending in the crowded cluster lines of sight, or if cluster members are just brighter, and therefore have a smaller shape measurement uncertainty contribution to the effective shape noise.

 For our shape noise modelling, we instead focus on these larger scales, where we measure that the effective shape noise increases with source redshift, as shown in Fig.~\ref{fig:stats_fit}, both for the tangential and the cross component. This trend matches the survey averaged effective shape dispersion found by \citet{friedrich21} for the tomographic redshift bins. For later use in our semi-analytical covariance matrix modelling, we fit this trend with the quadratic expression 
\begin{equation}\label{eq:fit_to_shapenoise}
    \sigma_\text{eff}(z_\text{cl}) = 0.2635 + 0.0300 \,z_\text{cl}^2 - 0.0008 \,z_\text{cl}.
\end{equation}
This fit is shown as a dashed line in Fig.~\ref{fig:stats_fit}, and excellently matches the trend we measure in the data. Extrapolated to $z_\text{cl}\rightarrow 0$ it matches closely the value reported by \citetalias{gatti21}. In absence of a reliable estimate for the uncertainty of the effective shape noise, we forego reporting error bars on this fit, using it as an effective smoothing to avoid noise on the error estimate of our target quantity, the reduced shear profile.

\begin{figure}
  \includegraphics[width=\columnwidth]{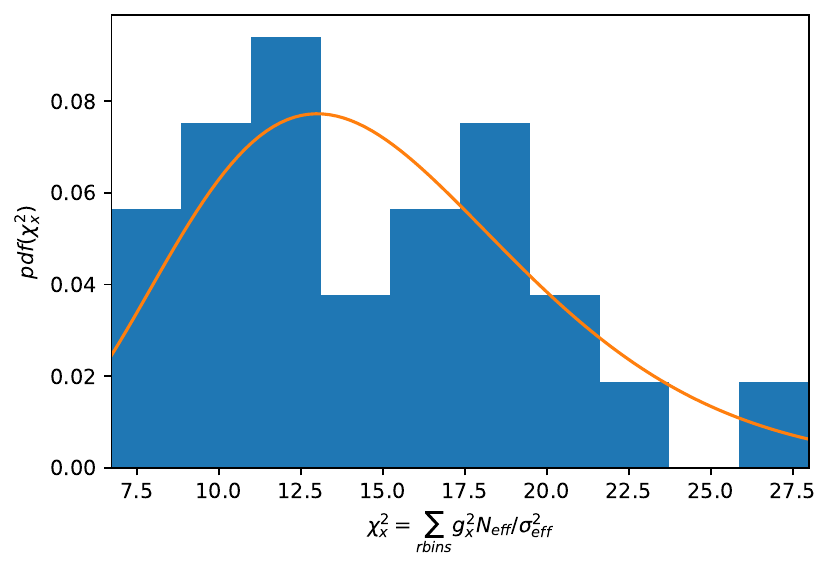}
  \caption{Distribution of the uncertainty scaled squared residuals of the cross-shear profile w.r.t. zero signal, as a blue histogram. In orange, the chi-squared distribution for 15 degrees of freedom, matching the number of radial bins. The squared losses distribution is well described by the chi-squared distribution, indicating that the cross-shear profiles are consistent with zero given our shape noise modelling.
}
  \label{fig:gx_chisquared}
\end{figure}

\subsubsection{Cross-shear}

\begin{figure*}
  \includegraphics[width=\textwidth]{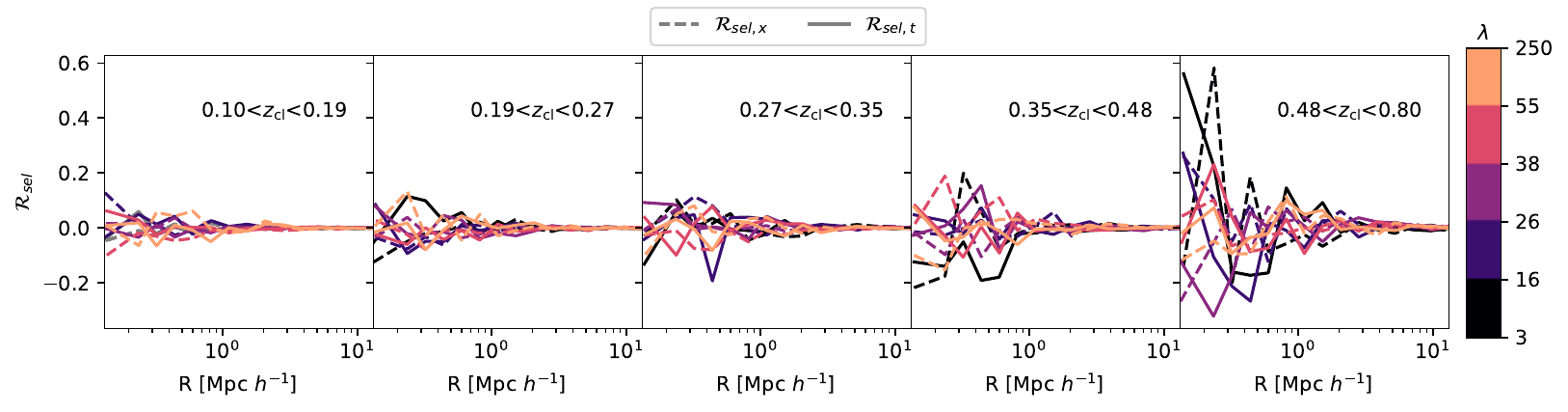}
  \caption{Selection response of our tangential (full lines) and cross (dashed lines) shear estimator, averaged over artificial shears in the two Cartesian coordinates, as a function of cluster-centric distance, stacked in bins of cluster richness (color coded), and cluster redshift  (panels: lowest redshift left, higher redshift right). The selection response is very noisy, but shows no significant trends with richness or cluster-centric distance. 
  }
  \label{fig:selection_response}
\end{figure*}

\begin{figure*}
  \includegraphics[width=\textwidth]{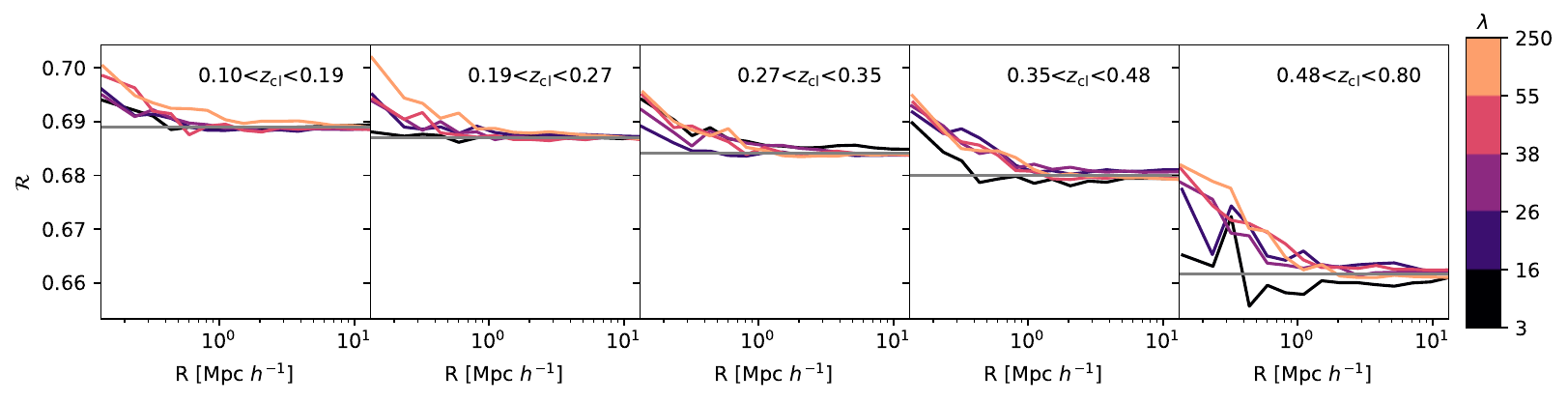}
  \caption{Average shear response, as a function of cluster-centric distance, stacked in bins of cluster richness (color coded), and cluster redshift  (panels: lowest redshift left, higher redshift right). Noticeably, the shear response increases towards the cluster center. Interestingly this signal seems to affect all lens redshift bins equally, independently on their cluster member contamination levels. We account for this effect in the cosmology ready data products.
  }
  \label{fig:shear_response}
\end{figure*}

Using our fitting formula for the effective shape noise, for each bin in richness and lens-redshift we can compute the consistency of the cross-shear signal with zero as
\begin{equation}
    \chi^2_\text{x} = \sum_\text{r-bins} \frac{g_\text{x}^2 N_\text{eff}}{\sigma^2_\text{eff}}.
\end{equation}
These squared losses follow a chi-squared distribution with $n_r$ degrees of freedom, where $n_r$ is the number of radial bins we consider, as shown in Fig.~\ref{fig:gx_chisquared}.  This demonstrates that our cross-shear component averages to zero, consistently with the expectation of well-calibrated shape measurements distorted by gravity. We also visually inspect the cross-shear profiles in the above defined richness, redshift bins, and find no significant deviations from zero. Computing instead the global degree of consistency of the raw tangential reduced shear profiles with zero, we find a signal-to-noise of 92.\footnote{Here we use the definition of signal-to-noise $\text{S/N}=\sqrt{\sum \chi^2_\text{t}}$, summed over all bins. This quantity does not enter our analysis, and is simply used to give a rough estimate of the constraining power of our dataset.}

\subsubsection{Selection response}\label{sec:sel_resp}

The quantities that we use for our source selection are impacted by the shear we try to measure. For instance, the flux of a sheared galaxy is different from the flux of the original galaxy if it was not sheared. 
This means that the shear we try to measure impacts our selection of source galaxies, and therefore biases our estimators. To correct for this bias, we have to measure how our estimator fares on catalogs selected from artificially sheared versions of the actual DES Y3 images. Four such catalogs have been created, each with an artificial shear of $\Delta \gamma=0.01$ positive/negative (+/-) on the first/second Cartesian component. The selection response is given by
\begin{equation}
\begin{split}
    \mathcal{R}_\mathrm{\alpha,\,sel} &= \frac{\mathcal{R}_\mathrm{\alpha,\,sel, 1} + \mathcal{R}_\mathrm{\alpha,\,sel, 2} }{2} = \\
    &=\frac{1}{2 \Delta \gamma} \left( \langle g_\mathrm{\alpha,\,raw} \rangle^{1+}-\langle g_\mathrm{\alpha,\,raw} \rangle^{1-} + \langle g_\mathrm{\alpha,\,raw} \rangle^{2+} -\langle g_\mathrm{\alpha,\,raw} \rangle^{2-} \right),
    \end{split}
\end{equation}
for $\alpha\in(t,x)$. Here $\langle g_\mathrm{\alpha,\,raw} \rangle^{1+}$ denotes the tangential reduced shear estimator on the source sample extracted from the images that have been artificially sheared in positive ($+$) direction along the first ($1$) Cartesian shear component, and similarly for the negatively sheared images (-), and the second Cartesian component (2), respectively. The radial, richness and redshift trends of the selection response are shown in Fig.~\ref{fig:selection_response}. Especially for high cluster redshift bins, the selection response is very noisy
such that no evident trends can be made out.
A reliable estimate can thus only be obtained after 
averaging over richness and radius. We find that the global selection response is positive and 
bounded 
by $\mathcal{R}_\mathrm{\alpha,\,sel}<0.001$ for $z_\text{cl}<0.48$, and $\mathcal{R}_\mathrm{\alpha,\,sel}\sim0.004$ for $0.48<z_\text{cl}<0.8$. 

We directly compare the selection response values to the mean shear responses given by
\begin{equation}
    \mathcal{R} = \frac{\sum_{b=2,3,4}  w^b \sum_{i\in b}  w^\mathrm{s}_i \mathcal{R}_{i}  }{\sum_{b=2,3,4} w^b \sum_{i\in b} w^\mathrm{s}_i  } \text{ for } \alpha\in(t,x),
\end{equation}
shown in Fig~\ref{fig:shear_response}. 

 Notably, the shear response increases towards the cluster center, while showing also a gentle trend with richness. This signal seems to affect all lens redshift bins equally, independently 
 of their cluster member contamination levels. Another contributing effect could be that cluster member contaminants have different sizes and signal-to-noise ratios compared to the field galaxies. We will account for changes in shear response in the cosmology-ready data products, as shown below. The shear response used here is a smooth function of the source relative size w.r.t. the PSF and its signal-to-noise, and might therefore be impacted by the magnification effect. We plan to investigate how these properties are impacted by cluster lines of sight in future work.

Direct comparison between the mean shear responses and the selection response shows that the latter is of order $0.1 \%$ of the former for $z_\text{l}<0.4$, increasing to $0.5 \%$ for the redshift bin $0.48<z_\text{l}<0.8$. In both cases, this is a factor of 
a few less than the uncertainty induced by shape and photo-$z$ measurements, and as such can be ignored. This assessment is in contrast 
to the impact of the selection response on galaxy-galaxy lensing studies \citep{prat22}. The main difference is that we do not consider each tomographic bin independently, but 
rather make a selection based on the lens redshift. For low lens redshifts, $z_\text{l}<0.48$,  we thus include a large fraction of the total source sample, reducing the impact of selection effects compared to a tomographic redshift bin selection. At high lens redshift, $0.48<z_\text{l}<0.8$, 
the close proximity of the source sample to the lens increases dramatically the effects of photo-$z$ uncertainties on the lensing efficiency, dwarfing the effect of the selection response.

\begin{table*}
\caption{\label{tab:sims}
Overview of the simulation inputs used for the calibration of the WL bias and scatter.}
\begin{tabular}{lll}
name & use case & Section \\
\hline
 TNG300 & projected surface mass density of massive halos at different redshift snapshots  & \ref{sec:hydro_input}  \\
 \textsc{Magneticum Pathfinder} & offsets between X-ray surface brightness peak and projected halo center & \ref{sec:intr_miscentering} \\
 eROSITA all-sky survey twin & offsets between X-ray surface brightness peak and measured X-ray position & \ref{sec:obs_miscent} \\
 & mapping between halo mass and redshift, and detection likelihood and extent & \ref{sec:synth_prof} \\
 Monte Carlo realisations & propagation of systematic uncertainties to error budget on the WL bias  & \ref{sec:wl_biassactter} \\
 \hline
 \multicolumn{3}{p{\linewidth-12pt} }{\small The respective references, as well as detailed explanations, are given in the respective sections. }
\end{tabular}
\end{table*}

\subsection{Cosmology-ready data products}\label{sec:cosmo_rdy_prods}

Given the various distance dependencies of the WL signal, the mass calibration is cosmology dependent. To marginalise over this dependency in this work, and to enable the WL mass calibration of our sample self-consistently with the cosmological number counts experiment in \citet{ghirardini23}, we need to define cosmology independent data products. Furthermore, given that we pursue a Bayesian population analysis in which each individual cluster likelihood is evaluated, we construct the following data products for each cluster in the cosmology sample that has DES Y3 lensing information:
\begin{itemize}
    \item the measured reduced shear profile 
    \begin{equation} \label{eq:cosmo_rdy_gt}
    \hat g_\mathrm{t} = \frac{\sum_{b=2,3,4}  w^b \sum_{i\in b}  w^\mathrm{s}_i e_{\mathrm t,\,b,i}  }{\sum_{b=2,3,4} w^b \sum_{i\in b} w^\mathrm{s}_i \mathcal{R}_{i} },
\end{equation}
in radial bins with $0.5 h^{-1}$ Mpc $<R<3.2 (1+z_\text{cl})^{-1}$ $h^{-1}$Mpc, in our reference cosmology. The outer bin is chosen following \citet{grandis+21} 
to limit our extraction to the 1-halo term region. We use only the shear response $\mathcal{R}_i$, and ignore the selection response, as argued in Section~\ref{sec:sel_resp};
\item the statistical uncertainty on the reduced shear,
\begin{equation}\label{eq:cosmo_rdy_dgt}
    \delta g_\mathrm{t} = \frac{\sigma_\text{eff}(z_\text{cl})}{\sqrt{N_\text{eff}}} \text{ with } N_\text{eff}=\frac{ \left( \sum_{b=2,3,4} \sum_{i\in b}   w^b  w^\mathrm{s}_i  \mathcal{R}_{i}  \right)^2}{ \sum_{b=2,3,4} \sum_{i\in b} \left(  w^b  w^\mathrm{s}_i \mathcal{R}_{i} \right)^2 },
\end{equation}
in the same bins. Note here that we use the quadratic fit to the global shape noise given in Eq.~\ref{eq:fit_to_shapenoise} to suppress the noise in the individual shear variance estimators;
\item the average angular distance of the binned sources from the cluster position in the corresponding radial bins
\begin{equation}\label{eq:cosmo_rdy_theta}
  \theta =  \frac{\sum_{b=2,3,4} w^b  \sum_{i\in b}   w^\mathrm{s}_i \mathcal{R}_{i} \, \theta_i}{\sum_{b=2,3,4} w^b \sum_{i\in b} w^\mathrm{s}_i \mathcal{R}_{i} }.
\end{equation}
We report the angular scale, 
because in the cosmological inference, the distance--redshift relation needs to be self-consistently re-calculated, while the data binning is frozen; and
\item the field source redshift distribution extracted in the outer regions of the cluster ($11h^{-1}$ Mpc $<R<15 $$h^{-1}$Mpc) based on the SOM-Pz redshift estimation method,
\begin{equation}\label{eq:cosmo_rdy_pz}
P(z_\text{s}) =  \frac{\sum_{b=2,3,4} w^b  \sum_{i\in b}   w^\mathrm{s}_i \mathcal{R}_{i} \, P(z_\text{s}| \hat c_i)}{\sum_{b=2,3,4} w^b \sum_{i\in b} w^\mathrm{s}_i \mathcal{R}_{i} },
\end{equation}
where $\hat c_i$ is the SOM cell in which the respective source falls. This extraction ensures that our source redshift distribution is a representation of the local field sources.
\end{itemize}

Using these scale cuts, we reduce the signal-to-noise ratio in the tangential reduced shear to 65. These cuts are necessary to avoid the increased systematic uncertainty towards the cluster centers \citep[see][Section 3.2]{grandis+21}, and to limit our WL observable to the 1-halo regime. As such, we avoid the 2-halo regime with its more complicated cosmology dependence and possible effects of assembly bias.

\section{Calibration of mass extraction}\label{sec:4}

Proper cosmological exploitation of the data products derived above 
must accurately map the WL signal to the halo mass used in the number counts experiment. In this work, we follow the approach of \citet{grandis+21} by simulating synthetic shear profiles starting from hydro-dynamical simulations, WL survey specifications and characteristics of the lens sample. The actual mass measurement is performed on the synthetic shear profiles, resulting in the so-called \emph{WL mass} (hereafter WL mass), which displays a bias and scatter w.r.t. the true halo mass. Priors on the WL bias and scatter are derived from Monte-Carlo realisations of the synthetic shear profiles, which sample the range of systematic uncertainties on the input specifications. The different simulation inputs used for this calibration are summarized in Table~\ref{tab:sims}.

\subsection{Hydro-dynamical simulations input}\label{sec:hydro_input}

We base the creation of our synthetic shear profiles on the surface mass maps of halos with $M_\text{200c}> 3\times 10^{13}\,h^{-1} $ M$_\odot$, extracted with the technique outlined in \citet{grandis+21} from the cosmological hydro-dynamical TNG300 simulations \citep{Pillepich2018MNRAS.475..648P, Marinacci2018MNRAS.480.5113M, Springel2018MNRAS.475..676S, Nelson2018MNRAS.475..624N, Naiman2018MNRAS.477.1206N, Nelson2019ComAC...6....2N} at redshifts $z_\text{snap}=0.24,\,0.42,\,0.64,\,0.95$. To boost the statistical sample size, for each halo we create three maps by projecting along each Cartesian coordinate. As described in \citet{grandis+21}, these maps are processed into scaled tangential shear $\Gamma_t(R,\,\phi | R_\text{mis})$ and scaled convergence $\Sigma(R,\,\phi | R_\text{mis})$ maps, binned in polar coordinates $(R, \phi)$ around isotropically mis-centered positions for a range of mis-centering distances $R_\text{mis}$. $\Sigma(R,\,\phi | R_\text{mis})$ is the surface mass density in units of $h~M_\odot$ Mpc$^2$, which multiplied by the inverse critical surface density gives the convergence, $\kappa = \Sigma_\text{crit}^{-1}\Sigma$. The definition of $\Gamma_t$ has been introduced in \citet{grandis+21} to signify the result of applying the inverse of Kaiser-Squires algorithm to the surface mass density $\Sigma(\vec{x})$, and then determining the tangential component w.r.t. the chosen center. As both these operations are linear, the resulting quantity is directly related to the tangential shear map as $\gamma_t(R,\,\phi | R_\text{mis}) = \Sigma_\text{crit}^{-1}\Gamma_t(R,\,\phi | R_\text{mis})$. Being independent from the specific source redshift distribution, this is a convenient simulation output. 

This data format conserves the azimuthal anisotropy sourced by the halo triaxiality and by the mis-centering. Possible inaccuracies induced by the correlation between halo shape and mis-centring direction, recently highlighted by \citet{sommer23}, have been shown to not significantly impact the cosmological results derived from the mass calibration presented in this work \citep[see][Section 5.1]{ghirardini23}, but likely need to be accounted for in future analyses with larger statistical constraining power. 

Note also that we employ in this work the strategy to mitigate hydro-dynamical modelling uncertainties proposed by \citet{grandis+21}, and used by \citet{chiu22a, chiu22b, bocquet+ipa}. The hydro-dynamical simulation from which we source the surface density maps have gravity-only twin runs with the same initial conditions. It is well established that both runs form exactly the same halos \citep{castro+21, grandis+21}, albeit with different masses. We purposefully choose the fictitious mass in the gravity-only simulation, as the halo mass functions used in the number counts experiment are calibrated on gravity-only simulations. We can thus use these calibrations 
to absorb the effects of hydro-dynamical feedback on the mass distributions of massive halos in the mapping between 
the (gravity-only) halo mass and 
the WL mass, and account for 
the uncertainty in this mapping
in the systematic error budget (see Section~\ref{sec:wl_biassactter}). 

\begin{figure}
  \includegraphics[width=\columnwidth]{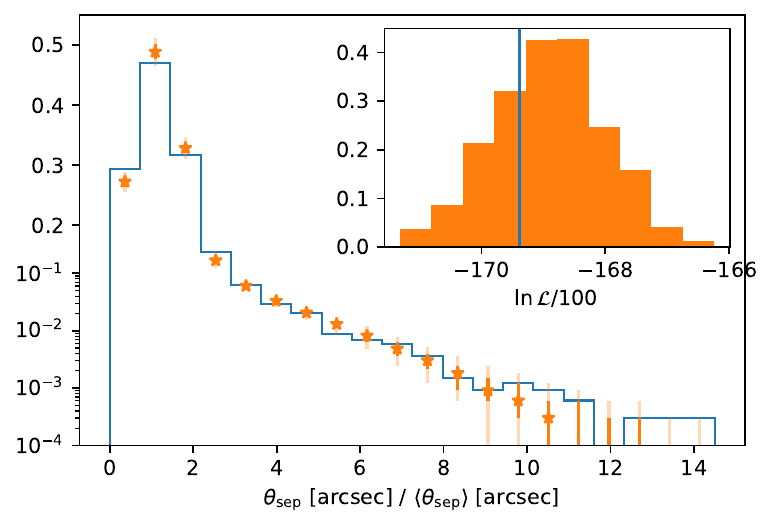}
  \caption{Observational mis-centering between the input X-ray surface brightness peaks and the output positions of clusters in the eROSITA digital twin, $\theta_\text{sep}$, scaled by the best-fit average offset, $\left \langle \theta_\text{sep}\right \rangle$, shown in blue as a histogram. Overlaid in orange are realizations of mock offsets drawn from the posterior of our model. The data is indistinguishable from the mock draws. Our model is also validated by comparing, in the insert, the likelihood of the data in blue with the likelihoods of the mocks, orange.}
  \label{fig:micentering-distr}
\end{figure}

\subsection{Mis-centering}\label{sec:miscentering}

A faithful simulation of synthetic shear profiles requires us to account for the fact that the observed position of the lens is mis-centered from the projected position of the true halo center, which is defined consistently through this work as the position of the most-bound particle. The surface mass maps are centered around this position \citep{grandis+21}.

We distinguish two sources of mis-centering: 1) the offset between the noise-free peak of the 2d X-ray surface brightness profile and the 2d position of the halo center (hereafter called \emph{intrinsic}), and 2) the offset between the noise-free peak of the 2d X-ray surface brightness profile and the measured X-ray position (hereafter called \emph{observational}). The latter displacement is induced by the photon shot noise and the PSF of the eROSITA cameras and is a special case of the astrometric uncertainty discussed in \citet{merloni2023}.

\subsubsection{Intrinsic mis-centering} \label{sec:intr_miscentering}
The intrinsic mis-centering is studied through the use of \textit{Box2b/hr} of the hydro-dynamical cosmological simulation suite \textsc{Magneticum Pathfinder}\footnote{www.magneticum.org} \citep{dolagip}, performed assuming the WMAP-7 cosmology as $\Omega_0=0.272$ and $h=0.704$ as given by \citet{komatsu11}. The volume of the \textit{Box2b/hr} box is 
$(909\,\mathrm{cMpc} )^{3}$, allowing for a broad range of galaxy cluster masses up to virial masses $M_\mathrm{vir}$ of 
several times $10^{15}\mathrm{M}_\odot$.
The galaxy cluster sample,
comprised of 116 clusters with $M_\mathrm{vir}\geq10^{14}\mathrm{M}_\odot$ at $z=0.252$, is described in detail 
in
\citet{kimmig23}. We 
select an additional 75 galaxy clusters at $z=0.518$ to account for redshift trends. The second sample is taken from a different box of equal resolution, {\it Box2/hr} of size $(500\,\mathrm{cMpc})^{3}$, to avoid double counting the 
halos at different redshifts.

Every cluster is projected from 100 random viewing angles. We then determine the projected center-of-mass via the shrinking sphere method \citep{power03} and around this center the peak of the X-ray emission between $0.5-2$keV. The intrinsic mis-centering is then the 
projected
displacement 
between the X-ray peak and the center of the halo, given by the most-bound particle.

We find that the mis-centering is well described by a single Rayleigh distribution with mean $\left \langle R_\text{mis}^\text{intr} \right\rangle = \sigma_\text{intr} R_\text{500c}$, with $\sigma_\text{intr} = 0.104 \pm 0.016$, and with no further resolved mass or redshift trends. As shown below in Section~\ref{sec:wl_biassactter} and Fig.~\ref{fig:bwlposteriors}, this uncertainty is not relevant when taken together with other systematic effects, especially photometric source redshift and hydro-dynamical modelling uncertainties. Valid concerns about the fidelity of the Magneticum prediction of the surface brightness in inner cluster regions are therefore unlikely to impact our results. In future analyses, we plan to validate the centering choice more carefully, for instance through the blind comparison of the results based on different centers \citep[as done for instance by][]{bocquet+ipa}.

\subsubsection{Observational mis-centering}\label{sec:obs_miscent}
For the observational mis-centering, we use the offset $\theta_\text{sep} $ between input 
and output positions of clusters in the ``eROSITA all-sky survey twin'' \citep{comparat19, comparat20, seppi22}, which 
produces and analyses simulations of
the X-ray signature from a realistic active galactic nuclei and cluster population in eRASS1 observational conditions 
that include 
exposure time variations and background. Furthermore, the X-ray cluster finding algorithm, as described in \citet{merloni2023, bulbul23} is run on the event file generated from this simulation, ensuring that PSF and shot noise of the real observations are 
well mapped. We select the input clusters matched with simulated sources that have detection likelihood $\mathcal{L}_\text{det}>3$, extent likelihood $\mathcal{L}_\texttt{EXT}>6$, and extent $4\leq \texttt{EXT} \leq 60$, to match the real cosmology sample. For each such object, we compute the separation $\theta_\text{sep}$ between input and output position, using the entries described in Table A.1 by \citet{seppi22}. 

We model the distribution of mis-centering with a mixture of a Rayleigh distribution and a Gamma distribution with shape parameter 2, as
\begin{equation}\label{eq:p_mis_obs}
\begin{split}
    P\left(\theta_\text{sep}|\left \langle \theta_\text{sep}\right \rangle, f, \lambda \right) =& \frac{1}{1+f} \frac{\theta_\text{sep}}{ \left \langle \theta_\text{sep}\right \rangle^2} \exp\left( -\frac{1}{2} \frac{\theta_\text{sep}^2}{\left \langle \theta_\text{sep}\right \rangle^2}\right) \\
    &+ \frac{f}{1+f}\frac{\lambda^2 \theta_\text{sep}}{ \left \langle \theta_\text{sep}\right \rangle^2} \exp\left( - \frac{\lambda \theta_\text{sep}}{\left \langle \theta_\text{sep}\right \rangle}\right), \\
\end{split}
\end{equation}
with the mean observational mis-centering given by 
\begin{equation}
    \left \langle \theta_\text{sep}\right \rangle =\left \langle \theta_\text{sep}\big| \alpha,\sigma_0,k\right\rangle =  \left( \frac{\mathcal{L}_\text{det}}{38.7} \right)^{-\alpha} \sqrt{ \sigma_0^2 + k^2 \texttt{EXT}^2},
\end{equation}
where $\mathcal{L}_\text{det}$ is the detection likelihood, $\texttt{EXT}$ the extent, as defined by the best-fit detection beta model, and $\boldsymbol{p}_\text{mct}=(\alpha,\,\sigma_0,\,k,\,\ln f,\,\ln \lambda)$ are free parameters. $f$ is the fraction of mis-centered objects in the heavy tailed component, $\alpha$ the detection likelihood trend of the mean observational mis-centering, $\sigma_0$ the extrapolated mean observational mis-centering for point source ($\texttt{EXT}=0$), $k$ the slope of the mean mis-centering w.r.t. the sources extent, and $\lambda$ the length of the mis-centering tail in units of the mean mis-centering. 

We fit these by sampling the likelihood 
\begin{equation}
    \ln \mathcal{L} = \sum_v \ln P\left(\theta_\text{sep}^v\big|\left\langle \theta_\text{sep}^v \big| \alpha,\sigma_0,k\right\rangle, \ln f, \ln\lambda\right), 
\end{equation}
assuming flat priors on all parameters, with $v$ running over the clusters selected from the digital twin simulation. All parameters are well constrained, their means and standard deviations are reported in Table~\ref{tab:clmcont_params}. Noticeably, the mean 
dependence on the detection likelihood
$\alpha = 0.43 \pm 0.02$ is quite close to the $\alpha=0.5$ trend one would expect for matched filter extraction in 
the
presence of Gaussian noise \citep{story+11, song+12}. Our constraint on $\ln f$ translates into a relative weight of $0.318 \pm 0.043 $ for the flatter Gamma distribution 
component, albeit only a part of that distributions populates the high mis-centering tail. Our other fit parameters result in a typical mis-centering  $\left \langle \theta_\text{sep}\right \rangle \sim 11$~arcsec.

We also check
that once the individual mis-centerings are corrected for the typical mis-centering of sources of that detection significance and extent, the residuals show no correlation with exposure time. This ensures that our parametrisation captures the main trends of the observational mis-centering, and that it can be safely used despite the exposure time varying over the survey footprint.
To further validate our model, we draw mock catalogs, by selecting a random posterior point ${\boldsymbol{p}}_\text{mct}=(\alpha,\,\sigma_0,\,k,\,\ln f,\,\ln \lambda)$. For each simulated cluster we then compute the typical mis-centering $\left \langle \theta_\text{sep}\right \rangle$ and draw a mock separation angle $\theta_\text{sep}$ from Eq.~\ref{eq:p_mis_obs}. With this procedure we create 1000 mock cluster samples
that follow our model, by drawing 1000 independent posterior points. In Fig.~\ref{fig:micentering-distr} we plot in blue the mis-centering distribution of the actual data, while in orange data points the distribution of mock data, sampled 
from our posterior, are shown. The two are indistinguishable, indicating that the actual mis-centering distribution from the digital twin looks like data drawn from our mis-centering model. Comparisons of the maximum likelihood of the data and the mocks
show that this holds also at the likelihood level (see insert of Fig.~\ref{fig:micentering-distr}). Our observational mis-centering model is 
therefore a good fit.

\subsection{Extraction model}\label{sec:extraction_mod}

When performing the mass measurement, we need to specify a mass extraction model
and coherently apply it to both the real and the synthetic data. In this analysis, we choose a simplified mis-centered model, corrected for the mean cluster member contamination, as suggested by \citet{grandis+21}. This model faithfully captures the main biases induced by mis-centering and cluster member contamination, while not adding numerical complexity compared to plain 2d-projected NFW model.
This approach is 
well suited for the repeated calls in the Monte Carlo Markov chains used for parameter inference. \citet{bocquet+ipa} make the same choices for the same reasons.

For the model, we assume that a cluster of radius $R_\text{500c}$ and mean observational mis-centering $\left \langle \theta_\text{sep}\right \rangle$ is effectively mis-centered by a physical distance 
\begin{equation}\label{eq:mean_miscent_extract_rad}
    R^\text{extr}_\text{mis} = \sqrt{\big( \sigma_\text{intr} R_\text{500c}\big)^2 + \left( D_\text{l} \left \langle \theta_\text{sep}\right \rangle \right)^2 },
\end{equation}
evaluated at the mean of the observational mis-centering parameters determined in section~\ref{sec:obs_miscent}. The radius $R_\text{500c}$ is derived from the input mass of the extraction model.
The cluster surface mass distribution is then well approximated by 
\begin{equation}
    \Sigma(R | M) = \begin{cases}
		      \Sigma_\text{NFW}(R^\text{extr}_\text{mis} | M) &\text{for } R < R^\text{extr}_\text{mis}, \, \text{and}\\
            \Sigma_\text{NFW}(R | M) & \text{otherwise},
		 \end{cases}
\end{equation}
where $\Sigma_\text{NFW}(R | M)$ is the 2d projected NFW profile, evaluated with the mean concentration mass relation reported by \citet{ragagnin21}. Deviations of this concentration mass relation from the one in the simulations are absorbed in the WL bias, while scatter in concentration at fixed mass contributes to the WL scatter. As shown 
in
\citet{bocquet+ipa}, this surface mass profile results in a radial density contrast $\Delta\Sigma(R | M)=0$ for $R<R^\text{extr}_\text{mis}$, and
\begin{equation}
    \Delta\Sigma(R | M) = \Delta\Sigma_\text{NFW}(R | M) - \left( \frac{R^{\text{extr}}_\text{mis}}{R}\right)^2 \Delta\Sigma_\text{NFW}(R^\text{extr}_\text{mis} | M),
\end{equation}
at larger projected radius.
This expression is numerically of the same complexity as evaluating the 2d density contrast of NFW profile $\Delta\Sigma_\text{NFW}(R | M)$, but is significantly more accurate. 

We then consider the average lensing efficiency 
\begin{equation}
    \Sigma_\text{crit,l}^{-1}=\left\langle \Sigma_\text{crit,ls}^{-1} \right\rangle_{P_\text{l}(z_\text{s})}
\end{equation}
of the lens l,
obtained by averaging the lensing efficiency of the individual source, $\Sigma_\text{crit,ls}^{-1}$, Eq.~\ref{eq:sigmacritinv_ls}, with source redshift distribution $P_\text{l}(z_\text{s})$. Using also the cluster member contamination fraction $f_\text{cl}(R | \lambda, z)$, evaluated at the mean cluster member contamination parameters, 
for the cluster richness $\lambda$ and the cluster redshift $z_\text{cl}$, we define the reduced shear model
\begin{equation}\label{eq:extraction_model}
    g_\text{t}^\text{mod}(R|M) = 
    \frac{\Sigma_\text{crit,l}^{-1}\Delta\Sigma(R | M)}{1-\Sigma_\text{crit,l}^{-1}\Sigma(R | M)} \left( 1-f_\text{cl}(R | \lambda, z_\text{cl})\right).
\end{equation}
This extraction model depends also on the clusters detection likelihood $\mathcal{L}_\text{det}$ and extent \texttt{EXT}, and the mean parameters of the mis-centering through the mean extraction mis-centering, eq.~\ref{eq:mean_miscent_extract_rad}. All these quantities are readily available for each cluster.

Any mis-matches between the real profiles and the extraction model are captured by considering that the best-fit mass assuming this model, the \emph{WL mass}
$M_\text{WL}$, 
will be biased and scattered around the true halo mass.



\begin{figure*}
  \includegraphics[width=\textwidth]{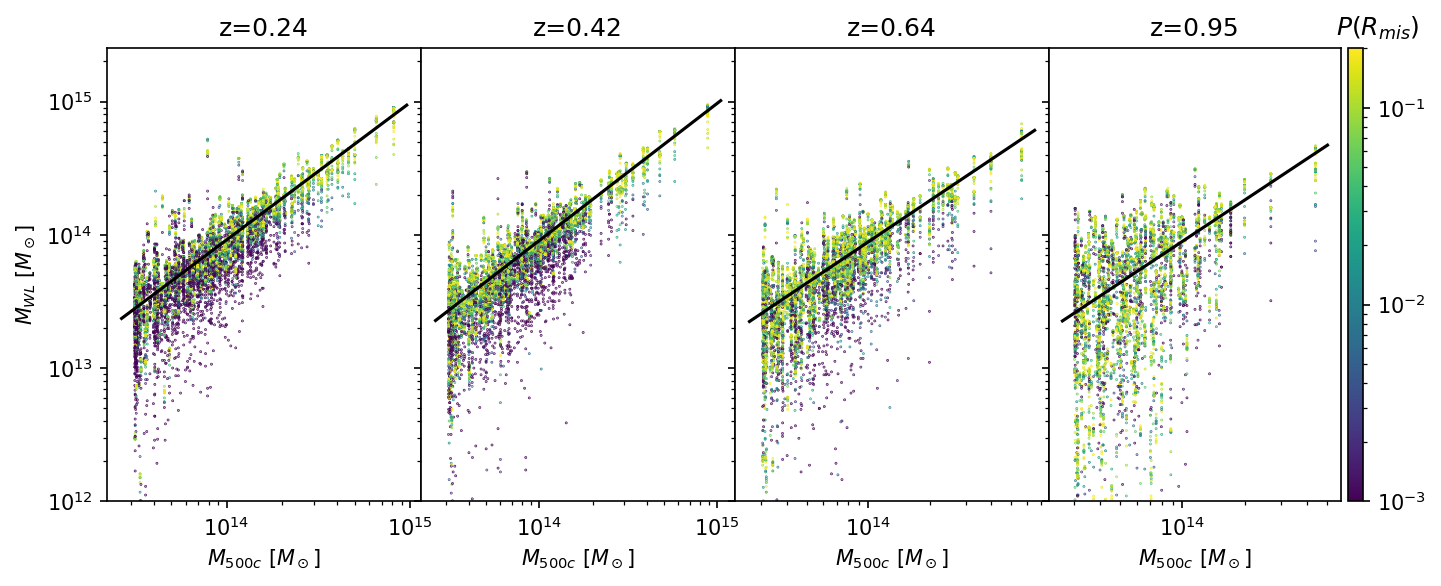}
  \caption{Scatter plot of the input halo mass versus the output WL mass from one realisation of the synthetic shear profiles, color coded by the value of the mis-centering distribution for that specific halos' mis-centering, and as solid black lines, the mis-centering-distribution-weighted scaling between WL and halo mass, which results in an estimate for the WL bias. The scatter around this relation results in the WL scatter. Generally speaking, extreme outliers have also highly improbable mis-centering. 
  }
  \label{fig:MWLvsM}
\end{figure*}

\subsection{Synthetic shear profiles}\label{sec:synth_prof}

The creation of the synthetic shear profiles proceeds as follows. 
\begin{itemize}
    \item The redshift of the synthetic clusters is fixed to the redshift of the hydro-simulation 
    outputs.
    \item Assuming the mean concentration $c_\text{200c}$--mass relation by \citet{child}, we convert the masses $M_\text{200c}$ from the simulation to $M_\text{500c}$, using the customary transformations \citep[see for instance][App.~A]{ettori11}. Note that the concentration--mass relation employed here was calibrated on gravity-only simulations, which matches the fact that our halo masses are also gravity-only masses.
    \item We assign a synthetic source redshift distribution $ P^\text{snth}_l(z_\text{s} )$ based on the tomographic bin weights $w_b(z_\text{l})$, as defined in Eq.~\ref{eq:tomo_weights}, and the realisations of their redshift distribution $P_\mathcal{H}(z_s | b)$, as provided by \citetalias{myles21}, discussed in Section~\ref{sec:des_photoz_shape_unc}, and shown in Fig.~\ref{fig:source_distr}. The normalisation of these distributions takes account of the multiplicative shear bias.
    \item Each cluster 
    is assigned a richness $\lambda$ based on the richness--mass relation calibrated 
    in \citet{chiu22a}. For later use, we remind the reader that this relation is defined by the parameters of the richness--mass relation $\boldsymbol{p}_{\lambda \text{M}}$. The richness is required both as an input for the extraction model, as well as to simulate the cluster member contamination, $f_\text{cl}(R | \lambda, z_\text{cl}, \boldsymbol{p}_\text{cl})$, on the synthetic shear profile, where $\boldsymbol{p}_\text{cl}$ are the parameters of the cluster member contamination.
    \item Querying the halos with $|\Delta\log_{10} M_\text{500c} / M_\odot|<0.1$ and $|\Delta z|<0.05$ from the eROSITA digital twin, we select the detection likelihood $\mathcal{L}_\text{det}$ and extent $\texttt{EXT}$ of a random pick. These two observables are needed to define 
    the 
    mean mis-centering 
    needed by 
    the extraction model and to assign a correct mis-centering distribution $P(R_\text{mis} | M_\text{500c}, z, \mathcal{L}_\text{det}, \texttt{EXT}, \boldsymbol{p}_\text{mct})$, resulting from the convolution of the observational and intrinsic mis-centering discussed in Section~\ref{sec:miscentering}. Note that this step depends on the parameters of the mis-centering $\boldsymbol{p}_\text{mct}$, determined in the same section.
    \item Improving on previous work, we now compute the reduced shear not only for each polar position in the map, but also for each source redshift $z_\text{s}$, as presented also in \citet{bocquet+ipa}. The azimuthal and source redshift average are computed after the computation of the reduced shear, as follows
    \begin{equation}
        g_\text{t}^\text{snth}(R| R_\text{mis}) = \int \text{d} z_\text{s} P^\text{snth}(z_\text{s} ) \int \frac{\text{d}\phi}{2 \pi} \frac{\Sigma_\text{crit,ls}^{-1} \Gamma_t(R,\,\phi | R_\text{mis})}{1-\Sigma_\text{crit,ls}^{-1} \Sigma (R,\,\phi | R_\text{mis}) }.
   \end{equation}
   Deviating from this 
   integration order leads to biases in the synthetic profiles of the order of 0.01, which is not acceptable given our accuracy
   requirements.
   \item We add cluster member contamination
   following the the fit we derived in Section~\ref{sec:cluster_mbr_cnt}, specifically the number density fit. We also add a non-linear shear bias and extra noise from the uncorrelated large scale structure along the line of sight, as discussed in \citet{grandis+21}.
   \item The extraction of the WL mass via the extraction model and the fit for the WL bias and scatter weighted by the mis-centering distribution follow exactly the method outlined in \citet{grandis+21}.
\end{itemize}

The output data product of such synthetic shear profile production and subsequent fit with the extraction model is shown in Fig~\ref{fig:MWLvsM}. There, we plot the input halo masses versus the WL masses,
color coded by the mis-centering distribution probability for the respective synthetic halo. Note that extreme outliers in WL mass are associated predominantly with highly mis-centered synthetic clusters, 
and are thus 
quite rare given the mis-centering distribution. As such, they contribute only marginally to the fit.

\begin{figure*}
    \centering
    \stackinset{r}{-.03\textwidth}{t}{-.04\textwidth}
  {\includegraphics[width=0.6\textwidth]{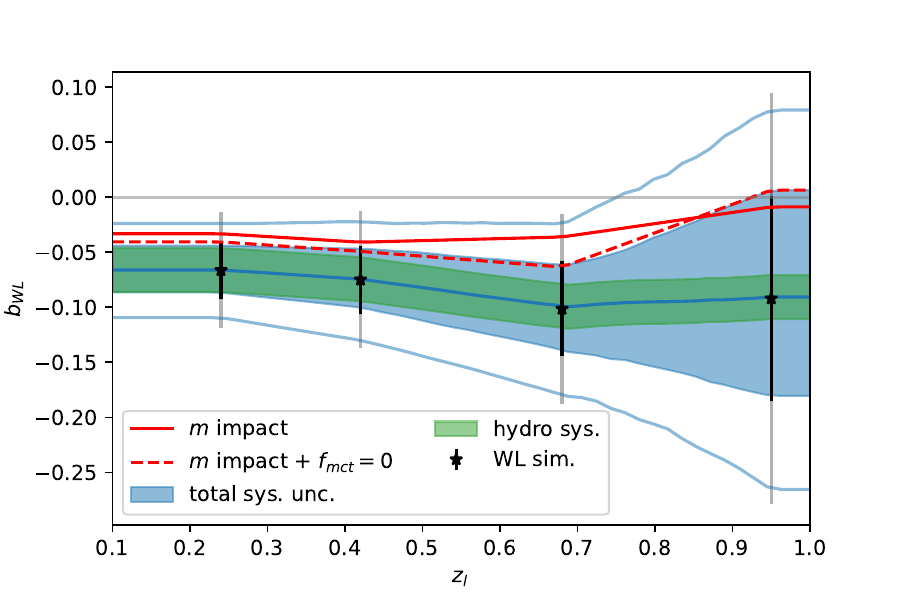}}
  {\includegraphics[width=\textwidth]{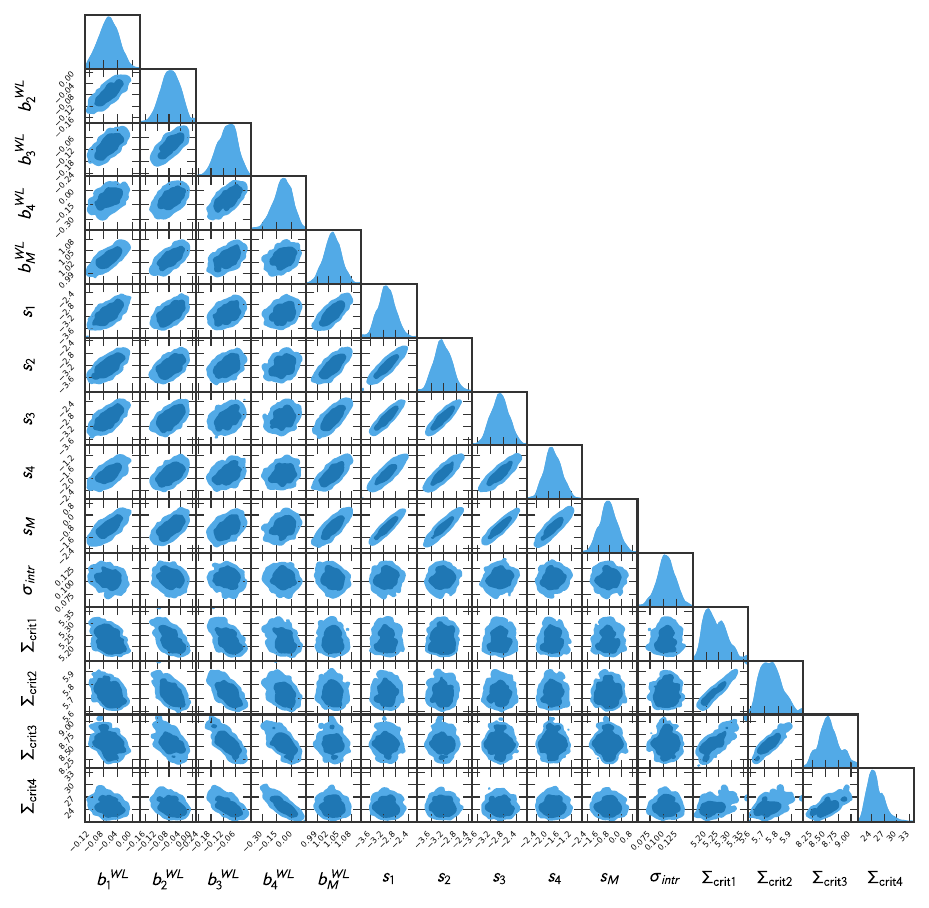}}
  \caption{\textit{Main panel}: Joint distribution on the WL bias and scatter parameters $(b^\text{WL}_l,\,s_l = \ln  (\sigma^\text{WL}_l)^2)$, and their mass slopes ($b_\text{M}^\text{WL}$, $\sigma_\text{M})$, together with the principal sources of systematic uncertainty: the lensing efficiencies $ \Sigma_{\text{crit},l}$ to the snapshots $l$, in units of $10^{15} h \,M_\odot $ Mpc$^{-2}$, and the scale of the intrinsic mis-centering $\sigma_\text{intr} $. The high redshift biases anticorrelate with the respective lensing efficiencies, indicating that in this regime we are limited by the DES Y3 photo-$z$ accuracy. Conversely, the low redshift biases, and the scatter values, correlate among themselves, as they are dominated by the hydro uncertainty floor we added. \textit{Insert}: Redshift interpolation of the WL lensing bias in cyan (filled region 1 sigma, faded lines 2 sigma), together with the 1 sigma uncertainty due to hydro modelling (green), explorations of the mean impact of the multiplicative shear bias $m$ modelling (red), and the additional omission of the high mis-centering tail $f=0$ (dashed red), and the posteriors of the WL bias in the four snap-shots. hydro-dynamical modelling uncertainties are dominant at low redshift. When considering our multiplicative shear bias modelling choices, the bias is consistent with zero, while the presence of the highly mis-centered tail does not affect the WL bias.
  }\label{fig:bwlposteriors}
\end{figure*}

\begin{table}
\caption{\label{tab:clmcont_params}
Calibration parameters with their priors, for the Monte Carlo marginalisation of the WL bias and scatter determination.}
\begin{tabular}{cc}
\multicolumn{2}{c}{Cluster Member Contamination (Section~\ref{sec:cluster_mbr_cnt})}  \\
\hline
$\log_{10} c = 4.83 \pm 0.22$ & $B_\text{cl}=0.467\pm 0.017$ \\
$\ln \sigma_\text{reg}=-0.61\pm 0.43$ & $A_0=-4.75\pm 0.11$ \\
$A_1 = -4.38\pm 0.05$ & $A_2= -3.68 \pm 0.04$ \\
$A_3 = -3.25\pm 0.04$ & $A_4 = -3.04 \pm 0.04$ \\
\hline
\hline
\multicolumn{2}{c}{Mis-centering Parameters (Section~\ref{sec:miscentering})} \\
\hline
$\sigma_0 [\text{arcseec}] = 4.83 \pm 0.22 $ & $k=0.308 \pm 0.007$  \\
$\alpha=0.443 \pm 0.012$ & $\ln f = -0.76 \pm 0.11$ \\
$\ln \beta = -0.18 \pm 0.04$ & $\sigma_\text{intr}=0.104 \pm 0.016$ \\
$\ln \lambda = -0.18 \pm 0.04$ & \\
\hline
\hline
\multicolumn{2}{c}{Richness Mass relation \citep[app. A]{chiu22a}} \\
\hline
$A_\lambda = 36.2 \pm 3.6$ & $B_\lambda = 0.881 \pm 0.083$ \\
$C_\lambda = -0.46 \pm 0.52$ & $\sigma_\lambda = 0.274 \pm 0.066$ \\
\hline
\hline
\multicolumn{2}{c}{photo-$z$ and shape calibrations} \\
\hline
$\mathcal{H} \sim \text{Int}(0, 999)$$^*$ & $ \alpha_\text{NL}=0.6\pm 0.4$ $^\dagger$ \\
\hline
\hline
\multicolumn{2}{c}{hydro-dynamical modelling uncertainties} \\
\hline
$\Delta b=0.02$, $ \Delta b_\text{M}=0.018$ & $\Delta s=0.25$, $ \Delta s_\text{M}=0.59$ \\
\multicolumn{2}{p{\linewidth-12pt} }{\small Note that the parameters for the cluster member contamination and the mis-centering are actually drawn from the respective posterior fits to conserve correlations among the parameters. Further notes: $^*$ $\text{Int}(a,b)$ is a uniform distribution over the integers $a\leq i \leq b $ (cf. Section~\ref{sec:des_photoz_shape_unc}); $^\dagger$ non-linear multiplicative shear bias \citep[Section 2.1.7]{grandis+21} }
\end{tabular}
\end{table}

\begin{table}
\caption{\label{tab:bWL--des}
Calibration of WL bias and scatter for DES Y3 WL on eRASS1 clusters.}
\begin{tabular}{lcccc}
$z$ & $0.24$& $0.42$& $0.68$& $0.95$ \\  
\hline
$\mu_\text{b}$ & $-0.066$ & $-0.075$ & $-0.101$ & $-0.092$ \\  
$\delta_\text{b1}$ & $-0.013$ & $-0.018$ & $-0.034$ & $-0.092$ \\  
$\delta_\text{b2}$ & $-0.019$ & $-0.022$ & $-0.023$ & $0.015$ \\   
$\mu_\text{s}$ & $-3.094$ & $-3.232$ & $-2.925$ & $-1.833$ \\  
$\delta_\text{s}$ & $0.272$ & $0.270$ & $0.275$ & $0.283$ \\ 
\hline
 &  \multicolumn{2}{c}{$b_\text{M}=1.030 \pm 0.021$} &  \multicolumn{2}{c}{$s_\text{M}=-0.919 \pm 0.608$} \\ 
\multicolumn{5}{p{\linewidth-12pt} }{\small Numerical values for the WL bias ($\mu_\text{b}$, $\delta_\text{b1}$, $\delta_\text{b2}$) and scatter ($\mu_\text{s}$, $\delta_{s}$) calibration at different redshift $z$, and their global mass trends $b_\text{M}$, and $s_\text{M}$, respectively. }
\end{tabular}
\end{table}

\subsection{WL bias and scatter}\label{sec:wl_biassactter}

At the end of the synthetic shear profile production and WL extraction, we get a bias and scatter $(b^\text{WL}_l,\,\sigma^\text{WL}_l)$ for each snap shot $l$, as well as a mass trend for both the bias ($b_\text{M}^\text{WL}$), and the scatter ($s_\text{M}$). They are obtained by fitting the relation  
\begin{align}
\bigg< \ln \frac{M_\text{WL}}{M_0} \bigg| M, z_\text{cl} \bigg> =b^\text{WL}_l + b_\text{M}^\text{WL} \ln \left( \frac{M}{M_0} \right)
\nonumber \\
\ln \sigma_\text{WL}^2 = s_l + s_\text{M} \ln \left( \frac{M}{M_0} \right),
\label{eq:WLbias}
\end{align}
with pivot mass $M_0 = 2 \times 10^{14}~M_\odot$, and $s_l = \ln  (\sigma^\text{WL}_l)^2$.
The fit of the relation 
includes cluster-by-cluster weights corresponding to the probability of the mis-centering of the individual synthetic halo, as discussed in \citet{grandis+21}. The mean relations resulting from the fit at each redshift are shown in Fig~\ref{fig:MWLvsM} as black lines. 

The uncertainty on these quantities is obtained by re-running the synthetic shear profile creation and WL mass extraction $\mathcal{O}(1000)$ times with slightly perturbed input parameters. Specifically, we vary the hyper parameter of the DES Y3 source redshift distributions $\mathcal{H}\in(0, 999)$ \citepalias[][and as outlined in Section~\ref{sec:des_photoz_shape_unc}]{myles21}, the parameters of the richness--mass relation $\boldsymbol{p}_{\lambda \text{M}}$  within the posterior reported by \citet{chiu22a}, the parameters of the cluster member contamination $\boldsymbol{p}_\text{cl}$ within the posterior determine in Section~\ref{sec:cluster_mbr_cnt}, the parameters of the mis-centering distribution $\boldsymbol{p}_\text{mct}$ within the posterior determine in Section~\ref{sec:miscentering}, as well as a wide prior on the non-linear shear biases. To these uncertainties we add a hydro-dynamical modelling uncertainty of $2\%$ on the WL bias, and corresponding values for the other parameters, as determined by \citet[][Table~2]{grandis+21}. 

The resulting 
parameter posterior distributions are 
presented in the contour plots 
in Fig.~\ref{fig:bwlposteriors}. We omit parameters that do not (anti)correlate with the WL bias and scatter parameters. 
As can be seen, the WL bias in the highest redshift slot $b^\text{WL}_4$ anticorrelates tightly with the critical surface density $\Sigma_{\text{crit},4} = \left\langle \Sigma_{\text{crit},s ,l=z_4}^{-1} \right\rangle_{P^\text{snth}_{z_4}(z_\text{s} )}^{-1}$, defined as the inverse lensing efficiency,  reported in units of $10^{15}$ $h$M$_\odot$ Mpc$^{-2}$. 
This indicates that the uncertainty on this bias parameter is dominated by the uncertainty on the lensing efficiency, as induced by the photo-$z$ uncertainty. 

Given these correlations, we compress 
the posterior on the WL bias into mean values $\mu_{\text{b},l}$, and two principal components $\delta_{\text{b1},l}$ and $\delta_{\text{b2},l}$, scaled by their variances. This means that a realisation of the WL bias at an arbitrary redshift $z_\text{cl}$ can be constructed by interpolating the mean and principal components and drawing two 
random standard normal variates $A_\text{WL}$, $B_\text{WL}$, such that
\begin{equation}\label{eq:bWL_w_params}
    b(z_\text{cl}) = \mathcal{I}(z_\text{cl}| z_l,\,\mu_{\text{b},l}) + A_\text{WL}\mathcal{I}(z_\text{cl}| z_l,\,\delta_{\text{b1},l}) + B_\text{WL}\mathcal{I}(z_\text{cl}| z_l,\,\delta_{\text{b2},l}),
\end{equation}
where $\mathcal{I}(z_\text{cl}| z_l,\,\mu_{\text{b},l})$ is the interpolation of the data vectors $( z_l,\,\mu_{\text{b},l})$ 
at redshift $z_\text{cl}$. The resulting posterior on the WL bias as a function of cluster redshift is shown in the insert of Fig.~\ref{fig:bwlposteriors} in blue bands for the 1\,$\sigma$ region, and faded lines for the 2\,$\sigma$ regions, together with the results for the 4 snapshots as black data points. For the natural logarithm of the variance, whose posterior is dominated by the uncertainty from hydro-dynamical modelling, we employ a single principal component (see below, Eq.~\ref{eq:sigmaWL}).

The numerical value of our WL bias is somewhat low, ranging from $-0.05$ to $-0.1$. A contributing factor is the response of the WL bias to our treatment of the multiplicative shear bias. While we include the multiplicative shear bias in the synthetic shear profiles, we do not include this bias in the model so as to reduce the number of independent calibration product we need to communicate to the cosmological number counts analysis \citep{ghirardini23}. To gauge the impact of this modelling choice on the numerical value of the bias, we perform a synthetic shear and mass extraction run at the mean values of all parameters, but artificially set the multiplicative shear bias $m=0$, that is we normalise the synthetic source redshfit distributions to 1. This results in a shift, shown as the red line in the insert of Fig.~\ref{fig:bwlposteriors} of the order of $-0.03$. It is two sigma away from our inferred bias values (black data points). This indicates that our extraction model is a reasonable representation of the synthetic data, once corrected for the above described modelling mis-match. Slightly negative WL biases can be explained by the fact that the density contrast of a halo is reduced by the projection of nearby correlated structure, resulting in a WL mass that is lower than the halo mass \citep{oguri+11, becker11, bahe12}.

We also test the sensitivity of the WL bias to the presence of the highly mis-centered tail in the eROSITA digital twin (cf. Section~\ref{sec:miscentering}). A similar large mis-centering tail is found by \citet{bulbul23}, Section~5 and Fig.~9, left panel, when comparing the centers obtained from the source detection algorithm and the X-ray post-processing. While the origin of this tail is still under investigation, we explore the hypothesis that the large mis-centering tail is completely suppressed by the X-ray post-processing. Setting the fraction $f=0$ fully suppresses the high exponential tail, but only mildly alters the WL bias, as can be seen by the difference between the red solid and dashed line. Our WL bias numbers, and therefore the mass anchoring of the number counts experiments, are stable w.r.t. the presence of this highly mis-centered tail, which might be an artefact of the eROSITA digital twin processing, and be corrected by the X-ray post-processing.

In summary, the two main contributions to uncertainty of the WL bias are the systematic floor resulting from the comparison of different hydro-dynamical simulations established by \citet{grandis+21}, and, at high redshift $z_\text{l}>0.6$ also the uncertainty of the DES Y3 photometric redshift distributions, shown in Fig.~\ref{fig:source_distr}. 

\section{Mass constraints}\label{sec:mass_calibration}

WL measurements around galaxy clusters provide a highly accurate (see Section~\ref{sec:4}) mass proxy.
They have, however, a quite low precision at the individual object basis, considering that the approximately 2.2k clusters we consider collectively only reach a signal-to-noise of 65 (cf. Section~\ref{sec:cosmo_rdy_prods}). 
A natural approach to counter-act this is to perform the mass measurement on stacked lensing signals \citepalias[e.g.][]{McClintock19, bellagamba19}. While this simplifies the mass measurement task, it significantly complicates the modelling in a cosmological context by introducing scale dependent selection bias effects \citep{sunayama20, desy1_clusters, wu22, sunayama23}. In this work, we therefore follow the Bayesian Population Modelling approach pioneered by \citet{mantz15, bocquet19, grandis19}, which predicts the distribution of expected shear profiles given the selection observable, as a function of the parameters governing the scaling between halo mass and selection observables. Complete derivations of this analysis framework can be found in \citet{ghirardini23, bocquet+ipa} in the context of eROSITA--selected clusters and SPT--selected clusters, respectively. Our results here are based on the implementation described in \citet{ghirardini23}, Section 4.

\subsection{Constraints of the count rate--mass relation}

In the following we discuss the likelihood setup 
used for the mass calibration
as well as the resulting mass calibration constraints. In general, WL mass calibration relies on statistically determining the scaling relation between the observables and halo mass \citep[for instance]{mantz16, dietrich19, schrabback21, zohren22}. In our case, we expand this standard treatment, to accommodate for the residual contamination of our sample by random superpositions
of X-ray candidates and optical systems, and, more importantly, mis-classified active galactic nuclei (see  \citet{bulbul23, kluge23, ghirardini23}). Both mis-classified AGNs and random superpositions are expect to produce far less WL signal compared to clusters, which inhabit massive halos. We therefore employ a \emph{mixture model}, a weighted sum of the different sub-populations, to perform the mass calibration in presence of non-negligible contamination.

\subsubsection{Mixture model}\label{sec:mixture_model}

For each individual cluster, we construct the probability density function (PDF) of its shear profile $g_\text{t}$ given the cluster observables count rate $\hat C_\text{R}$, redshift $z_\text{cl}$ and sky position $\hat{\mathcal{H}}$ as
\begin{equation}
    \begin{split}
P(\hat{g}_\text{t} | \hat{C}_R, z_\text{cl}, \hat{\mathcal{H}}) &= 
f_\text{C}(\hat{C}_R, z_\text{cl}, \hat{\mathcal{H}}) \,   P(\hat{g}_\text{t}  | \hat{C}_R, z_\text{cl}, \hat{\mathcal{H}}, C)  \\\
& + f_\text{RS}(\hat{C}_R, z_\text{cl}, \hat{\mathcal{H}}) \, P(\hat{g}_\text{t}  | \text{RS}) \\
& + f_\text{AGN}(\hat{C}_R, z_\text{cl}, \hat{\mathcal{H}}) \,  P(\hat{g}_\text{t}  | \text{AGN}, z_\text{cl}), 
\end{split}
\end{equation}
a weighted sum of the 
pdf of the shear profile for clusters (C), random 
superpositions
 (RS), 
and mis-classified active galactic nuclei (AGN), with their respective frequencies $f_\tau(\hat{C}_R, z_\text{cl}, \hat{\mathcal{H}})$ as a function of count rate, redshift and sky position, 
where $\tau\in$ (C, RS, AGN).
Naturally, $\sum_\tau f_\tau(\hat{C}_R, z_\text{cl}, \hat{\mathcal{H}}) =1 $, where the sum runs over $\tau\in$ (C, RS, AGN). 
As outlined in \citet{bulbul23, kluge23, ghirardini23}, the shape of the AGN and 
the random superpositions
contributions are calibrated partially on X-ray image simulations, partially 
with optical follow up of random positions and point sources, and have amplitudes ($f_\text{RS}$, $f_\text{AGN}$) that are fitted for using the richness distribution of the sample at a given count rate, redshift and sky position. They find $f_\text{RS}=0.0061\pm0.0023$ and $f_\text{AGN}=0.0462\pm 0.0038$, which we adopt as priors.

For the WL modelling, we assume that the density contrast of a random source is zero, resulting in their shear profile pdf being

\begin{equation}
     \ln P(\hat{g}_\text{t}  | \text{RS}) = -\frac{1}{2} \sum_k \frac{\hat g_{\text{t},k}^2}{ \delta g_{\mathrm{t},k}^2}, 
\end{equation}
where the sum runs over the radial bins $k$, with the measured shear profile $\hat g_{\text{t},k}$ defined in Eq.~\ref{eq:cosmo_rdy_gt}, and its shape noise error $\delta g_{\mathrm{t},k}$ defined in Eq.~\ref{eq:cosmo_rdy_dgt}. 
In practice, all objects in our lens sample have a richness $\hat\lambda>3$. At the lens redshift, the random source will thus have a few red galaxies along their lines of sight. Their density contrast is however expected to be small compared to cluster density contrasts.

For AGNs we assume that their density contrast $\Delta\Sigma_\text{AGN}(R)$ follows the profile measured by \citet{comparat23} for AGN in eFEDS using HSC WL data. We multiply this with the average lensing efficiency $\Sigma_\text{crit,c}^{-1}=\langle \Sigma_\text{crit,cs}^{-1} \rangle_{P_\text{c}(z_\text{s})}$, evaluated as a function of cosmology for the source redshift distribution  $P_\text{c}(z_\text{s})$ constructed in Eq.~\ref{eq:cosmo_rdy_pz}. The pdf thus reads

\begin{equation}
     \ln P(\hat{g}_\text{t}  | \text{AGN}, z_\text{cl}) = -\frac{1}{2} \sum_k \frac{1}{ \delta g_{\mathrm{t},k}^2} \left( \hat g_{\text{t},k} - \Sigma_\text{crit,c}^{-1} \Delta\Sigma_\text{AGN}(R=D_c \theta_k) \right)^2, 
\end{equation}
where $D_c$ is the angular diameter distance to the cluster, and $\theta_k$ the average angular distance to the sources in the respective radial bin $k$, as computed in Eq.~\ref{eq:cosmo_rdy_theta}. In practice, the density contrast of AGN is much smaller than that of galaxy clusters, given the lower host halo mass.

For galaxy clusters, the pdf of the shear profile depends on our extraction model $g_\text{t}^\text{mod}$, and the WL mass $M_\text{WL}$ defined by this model (cf. Section~\ref{sec:extraction_mod}). It therefore reads
\begin{equation}
     \ln P(\hat{g}_\text{t}  | \text{C}, z_\text{cl}, M_\text{WL} ) = -\frac{1}{2} \sum_k \frac{1}{ \delta g_{\mathrm{t},k}^2} \left( \hat g_{\text{t},k} - g_\text{t}^\text{mod}(R= D_c \theta_k |M_\text{WL})  \right)^2,
\end{equation}
with the extraction model depending on a host of cluster-specific observables, as outlined above. Note that in all the three pdfs we omitted the normalization term, which is the same for all three, and just depends on $\delta g_{\mathrm{t},k}$, thus being constant w.r.t. the parameters of the inference. 

The pdf of the shear profile for clusters needs to be marginalised over the distribution of WL masses $M_\text{WL}$ compatible with the clusters count rate $\hat C_\text{R}$, redshift $z_\text{cl}$, and sky position  $\hat{\mathcal{H}}$, that is $P(M_\text{WL} | \hat{C}_R, z_\text{cl}, \hat{\mathcal{H}}, \boldsymbol{p}_\text{SR})$, whose construction and dependence on model parameters $\boldsymbol{p}_\text{SR}$ we shall discuss below. The pdf of the shear profile for the cluster component is thus

\begin{align}
    P(\hat{g}_\text{t}  | \hat{C}_R, z_\text{cl}, \hat{\mathcal{H}}, \text{C}) = \int\text{d} M_\text{WL} & P(\hat{g}_\text{t}  | \text{C}, z_\text{cl}, M_\text{WL} ) \nonumber\\ 
    & P(M_\text{WL} | \hat{C}_R, z_\text{cl}, \hat{\mathcal{H}}, \boldsymbol{p}_\text{SR}).
\end{align}

The distribution of WL masses given the count rate, redshift and sky position is proportional to

\begin{align}
P(M_\text{WL} | \hat{C}_R, z_\text{cl}, \hat{\mathcal{H}}, \boldsymbol{p}_\text{SR}) \propto & \iint  \text{d}M
 \text{d}C_R 
P(M_\text{WL}, C_R | M, z_\text{cl}, \boldsymbol{p}_\text{SR}) 
 \nonumber \\
& \quad  P(I | C_R, z_\text{cl}, \hat{\mathcal{H}}_i) P(\hat{C}_R | C_R)
 P(M, z_\text{cl})
 ,
\end{align}
with $P(\hat{C}_R | C_R)$  modelling the observational uncertainty on the count rate, $P(M, z_\text{cl})$ being proportional to the halo mass function, and $P(I | C_R, z_\text{cl}, \hat{\mathcal{H}}_i) $ the X-ray incompleteness due to the extent selection. This ensures that we correctly model Eddington bias, which in cluster studies manifests as more low mass objects scattering up to a given observable value, than high mass objects scattering down to same observable values. This happens even when the observable--mass scatter is modelled as symmetric on account of the fact that there are just many more low mass than high mass objects. This fact is quantitatively expressed by the halo mass function. The expression given above is normalized to be a pdf in $M_\text{WL}$ via integration in $M_\text{WL}$ and subsequent re-scaling. See \citet{ghirardini23} for an alternative, but equivalent, derivation of the same likelihood.

\subsubsection{Scaling relation}

Given their deep potential wells, cluster observables scale tightly with the host halos mass. Leveraging this strong physical prior, we model the distribution of WL mass $M_\text{WL}$ and intrinsic (that is noise-free) count rate $C_\text{R}$ at given halo mass $M$ and redshift $z$ as a bi-variate log-normal, reading
\begin{equation}
    P(M_\text{WL}, C_R | M, z_\text{cl}, \boldsymbol{p}_\text{SR}) = \mathcal{N}(\bar{\mu}, \bar{\Sigma}),
\end{equation}
with mean and variance 
\[
\bar{\mu} = \left[\left\langle\ln C_R | M,z_\text{cl} \right\rangle, \left\langle\ln M_\text{WL} | M, z_\text{cl}\right\rangle\right]\]
\[
\bar{\Sigma} =
    \begin{bmatrix}
    \sigma_\text{X}^2 & \rho_\text{WL,X} \sigma_\text{X} \sigma_\text{WL} \\ 
    \rho_\text{WL,X} \sigma_\text{X} \sigma_\text{WL} & \sigma_\text{WL}^2 
    \end{bmatrix}
\]
where $\left\langle\ln C_R | M,z_\text{cl} \right\rangle$ and $\left\langle\ln M_\text{WL} | M, z_\text{cl}\right\rangle$ are the scaling relations between true mass and count rate and WL mass, respectively, $\sigma_\text{X}$ and $ \sigma_\text{WL}$ their intrinsic scatters, and $\rho_\text{WL,X}$ the correlation coefficient between the two intrinsic scatters. Inclusion of the latter is crucial to marginalise over possible selection effects, like more concentrated halos \emph{simultaneously} having a higher count rate and WL mass compared to other objects at their mass and redshift, or any other effect that might similarly correlate the intrinsic scatters.

As calibrated on dedicated simulations in Section~\ref{sec:4}, the mean WL mass -- halo mass relation is given by 
\begin{equation}
\bigg< \ln \frac{M_\text{WL}}{M_\text{p}} \bigg| M, z_\text{cl} \bigg> = b(z_\text{cl}) + b_\text{M} \ln \left( \frac{M}{M_\text{p}} \right),
\label{eq:WLbias_given_M_z}
\end{equation}
with the pivot value for the mass  $M_\text{p} = 2 \cdot 10 ^{14} M_\odot$, the redshift dependent amplitude of the WL bias given by equation~\ref{eq:bWL_w_params}, and the mass trend $b_\text{M} =  \mu_{b_\text{M}} + C_\text{WL} \cdot \sigma_{b_\text{M}}$, with mean and standard deviation derived in Section~\ref{sec:wl_biassactter} and reported in Table~\ref{tab:bWL--des}. 

The scatter is similarly modelled as
\begin{equation}\label{eq:sigmaWL}
    \sigma_\text{WL}^2 = \exp \left\{ \mathcal{I}(z_\text{cl}| z_l,\,\mu_{\text{s},l}) + D_\text{WL}\mathcal{I}(z_\text{cl}| z_l,\,\delta_{\text{s},l}) \right\},
\end{equation}
where $\mu_{\text{s},l}$ and $\delta_{\text{s},l}$ are the mean and standard deviations of the WL scatter in the simulation snapshot $l$, derived in Section~\ref{sec:wl_biassactter} and reported in Table~\ref{tab:bWL--des}, and $\mathcal{I}(z_\text{cl}| z_l,\,\mu_{\text{s},l})$ is the interpolation of the data vectors $( z_l,\,\mu_{\text{s},l})$ to the redshift $z_\text{cl}$. Given the tight correlations between the systematic uncertainties in the WL scatters of the different redshift seen in Fig.~\ref{fig:bwlposteriors}, we opt to model the WL scatter with a single principal component. We also set the mass trend of the scatter to $s_\text{M}=0$ as in this work we are not interested to explore mass trends in the intrinsic scatter of the other observables.

The count rate--mass relation is modelled as
\begin{equation} \label{eq:CR_given_M_z}
\bigg<\ln \frac{C_R}{C_{R,\text{p}}} \bigg| M, z_\text{cl}\bigg> = \ln A_\text{X} + b_\text{X}(M, z_\text{cl}) \cdot \ln \frac{M}{M_\text{p}} + e_\text{X}(z_\text{cl})
\end{equation}
where $C_{R,p} = 0.1 \text{ s}^{-1}$ is the pivot value for the count rate, $M_\text{p} = 2 \cdot 10 ^{14} M_\odot$ is the pivot value for the mass. $b_\text{X}(M, z)$ expresses the mass-redshift dependent slope of the scaling relation, given by
\begin{equation}
b_\text{X}(M, z_\text{cl}) = \bigg( B_\text{X} + C_\text{X} \cdot \ln \frac{M}{M_\text{p}} + F_\text{X} \cdot \ln \frac{1+z_\text{cl}}{1+z_\text{p}} \bigg)
\end{equation}
where $B_\text{X}$ is the classic standard single value mass slope, $C_\text{X}$ allows the slope to be mass dependent, and $F_\text{X}$ allows the slope to be redshift dependent, and $z_\text{p} = 0.35$ is the pivot value for the redshift. As shown in \citet{grandis19}, Appendix~B, $F_\text{X}$ needs to be sampled to retrieve unbiased cosmological results from the number counts. \citet{chiu22b} explored the impact of allowing for a mass dependence in the slope in a cosmological analysis of the eFEDS sample, finding that it is not necessary to give the scaling relation that freedom. We thus set $C_\text{X}=0$.

The redshift evolution $e_\text{X}(z_\text{cl})$ of the X-ray scaling relation is
\begin{equation}
e_\text{X}(z_\text{cl}) = D_\text{X} \cdot \ln \frac{d_\text{L}(z_\text{cl})}{d_\text{L}(z_\text{p})} + E_\text{X} \cdot \ln \frac{E(z_\text{cl})}{E(z_\text{p})} + G_\text{X} \cdot \ln \frac{1+z_\text{cl}}{1+z_\text{p}}
\end{equation}
where $E_\text{X} = 2$ and $D_\text{X}=-2$ are the values predicted by the self-similar model. The last term, with $G_\text{X}$, quantifies the deviation from the self-similar model. $E(z_\text{cl}) = H(z_\text{cl})/H_0$ is the dimensionless Hubble factor. 
 The intrinsic scatter of our scaling relation is represented by a single value, $\sigma_\text{X}$, independent of mass and redshift.

\subsubsection{Likelihood setup and priors}

\begin{table}[]
    \caption{Priors on the parameters used in the fitting of the count rate--mass relation.}
    \label{tab:parameter priors}
    \centering
    \begin{tabular}{ c c c }
    \hline
    \hline
    Parameter & Units & Prior \\
    \hline
    \multicolumn{3}{c}{Cosmology}\\
    \hline
        $\Omega_{\mathrm{M}}$ & - & $\mathcal{N}(0.331, 0.038)$ \\
        $\log_{10} A_\text{S}$ & - & Fixed to $-8.635$ \\
        $H_0$ & $\frac{\frac{\text{km}}{\text{s}}}{\text{Mpc}}$ & Fixed to 68.06 \\
        $\Omega_{\textrm{b}, 0}$ & - & Fixed to 0.048595 \\
        $n_\text{S}$ & - & Fixed to 0.9707 \\
        $w_0$ & - & Fixed to $-1$ \\
        $w_\text{a}$ & - & Fixed to 0  \\
        $\sum m_\nu$ & eV & Fixed to 0\\
        $\Omega_{\textrm{k}, 0}$ & - & Fixed to 0 \\
    \hline
    \multicolumn{3}{c}{X-ray scaling relation}\\
    \hline
        $A_\text{X}$ & - & $\mathcal{U}(0.01, 3)$ \\
        $B_\text{X}$ & - & $\mathcal{U}(0.1, 5)$ \\
        $C_\text{X}$ & - & Fixed to 0 \\
        $D_\text{X}$ & - & Fixed to $-2$ \\
        $E_\text{X}$ & - & Fixed to 2 \\
        $F_\text{X}$ & - & $\mathcal{U}(-5, 5)$ \\
        $G_\text{X}$ & - & $\mathcal{U}(-5, 5)$ \\
        $\sigma_\text{X}$ & - & $\mathcal{U}(0.05, 2)$ \\
    \hline
    \multicolumn{3}{c}{WL mass calibration}\\
    \hline
        $A_\text{WL, DES}$ & - & $\mathcal{N}(0, 1)$ \\
        $B_\text{WL, DES}$ & - & $\mathcal{N}(0, 1)$ \\
        $C_\text{WL, DES}$ & - & $\mathcal{N}(0, 1)$ \\
        $D_\text{WL, DES}$ & - & $\mathcal{N}(0, 1)$ \\
        $\rho_\text{WL,X}$ & - & $\mathcal{U}(-0.9, 0.9)$ \\ 
    \hline
    \multicolumn{3}{c}{Contamination modeling}\\
    \hline
    $f_\text{AGN}$ & - & $\mathcal{TN}(0.0412, 0.0040, 0, \infty)$  \\
    $f_\text{RS}$ & - &  $\mathcal{TN}(0.0064, 0.0024, 0, \infty)$ \\
    \hline
    \hline
    \end{tabular}
    \tablefoot{With $\mathcal{U}(\text{min}, \text{max})$ we indicate a uniform distribution between `min' and `max'. With $\mathcal{N}(\mu, \sigma)$ we indicate a normal distribution centered on $\mu$ and with standard deviation $\sigma$. With $\mathcal{TN}(\mu, \sigma, \text{min}, \text{max})$ we indicate a truncated normal distribution centered on $\mu$ and with standard deviation $\sigma$ bounded between `min' and `max'.}
\end{table}

As outlined in \citet{ghirardini23}, the total likelihood of our shear profiles $\hat g_{\text{t},\varkappa}$ of several clusters, given their  count rate $\hat C_{\text{R},\varkappa}$, redshift $z_{\text{cl},\varkappa}$ and sky position $\hat{\mathcal{H}}_\varkappa$, where $\varkappa$ runs over the cluster catalog, is given by
\begin{eqnarray}
    \ln \mathcal{L}_\text{WL} = \sum_\varkappa \ln P(\hat{g}_{\text{t},\varkappa} |\hat C_{\text{R},\varkappa}, z_{\text{cl},\varkappa}, \hat{\mathcal{H}}_\varkappa).
\end{eqnarray}

In light of the derivations above, this is a function of the parameters of the WL scaling relation $\boldsymbol{p}_\text{WL} = (A_\text{WL},\,B_\text{WL},\,C_\text{WL},\,D_\text{WL})$. To represent the WL anchoring performed in Section~\ref{sec:4}, we place independent Gaussian priors on these. The parameters of the count rate--mass relation $\boldsymbol{p}_\text{X} = (A_\text{X},\,B_\text{X},\,F_\text{X},\,G_\text{X}, \sigma_\text{X})$, instead, are the true target of our analysis. These are sampled within wide flat priors, reported in Table~\ref{tab:parameter priors}. Also the correlation coefficient between the WL and count rate intrinsic scatters is let free, but within the bounds $\rho_\text{WL,X}\in(-0.9,\,0.9)$. We purposefully do not sample all the way to $\pm$1, as the covariance of intrinsic scatters becomes singular in that limit. 

We place priors on the random source fractions and AGN ($f_\text{RS}$, $f_\text{AGN}$) from our optical follow-up observations. We do not expect the WL to significantly inform these fractions, as the individual shear profiles are hardly significantly different from zero themselves. As such, at an individual object basis, WL just lacks the statistical power to tell clusters apart from AGN or random sources.

\begin{figure*}
    \centering
   \includegraphics[width=\textwidth]{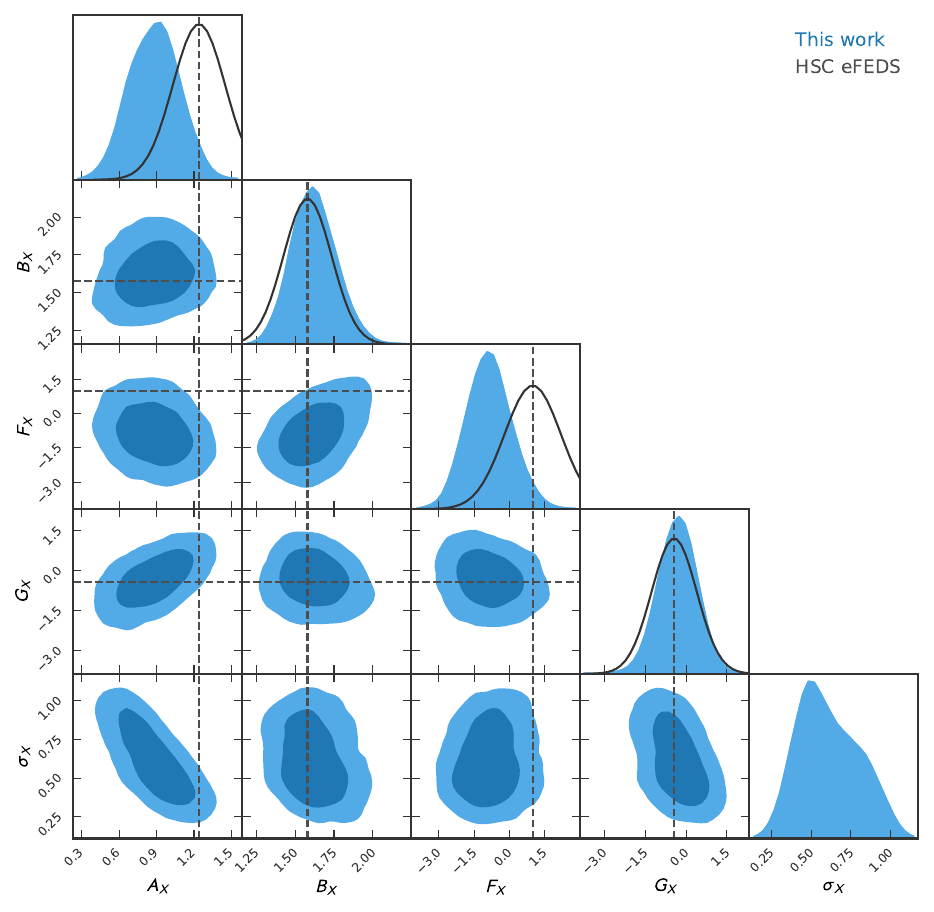}
  \caption{1- and 2-d marginal posterior on the parameters of the count rate--mass scaling relations from this work in blue, compared to the 1-d marginals of the analysis of HSC WL around eFEDS clusters and groups \citep{chiu22a} in black. We find good agreement on the mass trend $B_\text{X}$, and the deviation from self-similar redshift scaling $G_\text{X}$, while constraints on the amplitude $A_\text{X}$, and the redshift evolution of the mass slope $F_\text{X}$ are shifted by 1- to 2-sigmas, consistent with statistical fluctuations. Contrary to the eFEDS-HSC analysis, we do not place an informative prior of the intrinsic scatter $\sigma_\text{X}$, and get only weak constraints on it.
  }\label{fig:posterior_SR}
\end{figure*}

We fix all cosmological parameters to their Planck values values, except $\Omega_\text{M} = 0.331 \pm 0.038$ from the first Dark Energy Survey Supernovae Type Ia results \citep{des_sn}. Several elements of our likelihood, like angular diameter distances and the extraction model, depend on cosmology via the background evolution. The exact cosmological dependence of the WL calibrated number counts is accounted for by re-running the mass calibration presented here together with the number counts. In this work, we wish to provide mass calibration results marginalised over a reasonable range of cosmologies. Our $\Omega_\text{M}$-prior allows us to do this in a physically motivated way, while remaining independent of CMB results, with which the cosmological number counts analysis might be in tension.

\begin{figure}
  \includegraphics[width=\columnwidth]{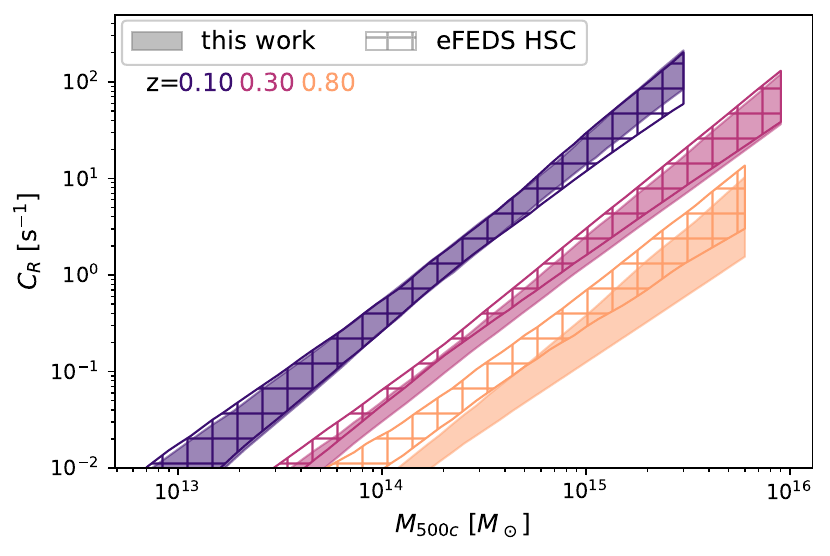}
  \caption{Posterior predictive distribution of the mean count rate--mass relation (the band ranges from the 16th to the 84th percentile at each mass), for three different redshifts (color coded). Filled bands represent the analysis in this work, hatched bands represent the eFEDS HSC result from \citet{chiu22a}. The predictions diverge for low mass, high redshift systems. 
}
  \label{fig:post_pred_SR}
\end{figure}

\subsubsection{Scaling relation parameters constraints}

We sample the total likelihood for the measured shear profiles with the priors using \texttt{ultranest} \citep{ultranest_is_the_best} to create a sample of the resulting posterior. We find that all parameters of interest, the parameters of the count rate--mass relation, are well constrained by our analysis (shown as blue contours in Fig.~\ref{fig:posterior_SR}), while all the calibration and other nuisance parameters have posteriors consistent with their priors. The find a constraint for the amplitude of the scaling relation $A_\text{X}=0.88\pm 0.20$, when considering the mean and standard deviation computed on the posterior sample. The mass slope of the relation is constrained to $B_\text{X}=1.62\pm 0.14$, the redshift trend of the mass slope to $F_\text{X}=-0.85\pm 0.93$. No significant deviation from the self-similar redshift evolution is detected, as $G_\text{X}=-0.32 \pm 0.69$ is consistent with 0. For the intrinsic scatter we find $\sigma_\text{X}=0.61\pm 0.19$. Noticeably, the intrinsic scatter can be constrained, despite the low precision of the individual lensing data, in line the predictions by \citet{grandis19}. While the intrinsic scatter values is somewhat large, it matches the count rate mass scatter we find in the digital twin simulations \citep{comparat20, seppi22}. To better visualize our scaling relation results, we create a posterior predictive distribution for the mean count rate as a function of mass and redshift. The 1 sigma region of the predicted relation at three redshifts is shown in Fig.~\ref{fig:post_pred_SR} in filled bands.

\begin{figure*}
    \centering
   \includegraphics[width=\textwidth]{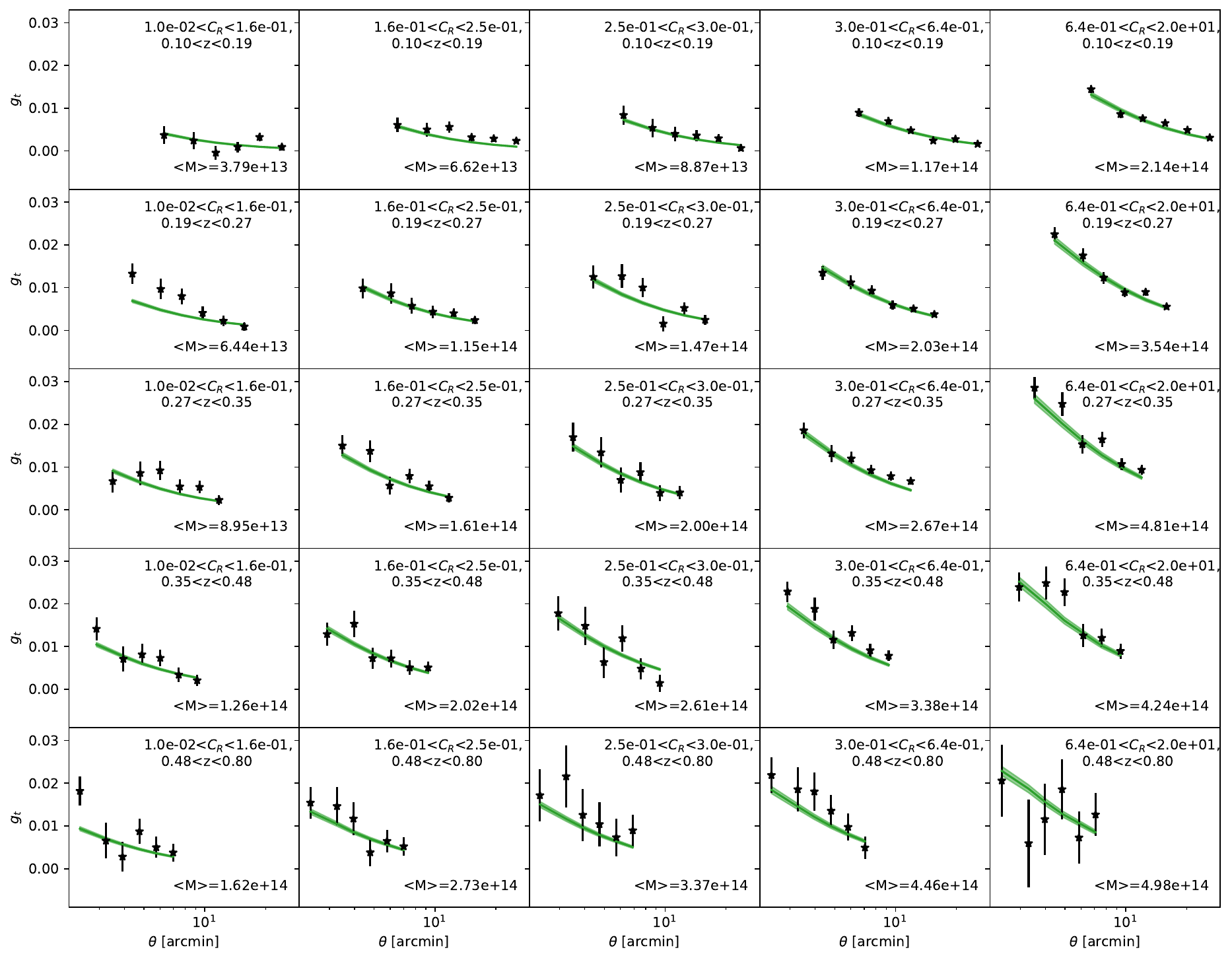}
  \caption{Stacked measured reduced tangential shear profiles with measurement uncertainty in black, for different count rate (rows) and redshift (columns) bins. As a green line the best-fit shear profile model and with filled green regions the propagation of the posterior uncertainty on the mean model. The chosen model fits the data well at all count rates and redshifts. We report also the median WL mass of each bin, in units of $M_\odot$.
  }\label{fig:goodness_of_fit}
\end{figure*}

\subsection{Goodness of fit}

To assess the goodness of fit of our mass calibration, we resort to binning the cluster sample in redshift, count rate bins, with edges $(0.10, 0.19, 0.27, 0.35, 0.48, 0.8)$, and $(0.01 0.16, 0.25, 0.30, 0.64, 20)$, in redshift and count rate respectively. For all clusters $\varkappa$ falling in each of the bins, we stack the tangential reduced shear profiles by defining the stacking weights $W_\varkappa =\sum_{b=2,3,4}  w^b \sum_{i \in b_\varkappa}   w^\mathrm{s}_i \mathcal{R}_{i} $, computing
\begin{equation}
    \hat g_\text{t}^\text{stack} = \frac{ \sum_\varkappa W_\varkappa\hat{g}_{\text{t},\varkappa} }{\sum_\varkappa W_\varkappa},
\end{equation}
where $i \in b_\varkappa$ symbolizes the sources $i$ associated to the cluster $\varkappa$ in the tomographic bin $b$. Combining this expression with Eq.~\ref{eq:cosmo_rdy_gt} shows that it is equivalent to the raw tangential reduced shear estimator in Eq.~\ref{eq:raw_shear}.
Using the same weighting we also derive the error on the stacked reduced shear
\begin{equation}
     \left( \delta g_\text{t}^\text{stack} \right)^2 = \frac{ \sum_\varkappa W_\varkappa^2 \delta g_{\text{t},\varkappa}^2 }{\sum_\varkappa W_\varkappa^2 },
\end{equation}
and the mean angular separation of the source galaxies
\begin{equation}
    \theta^\text{stack} = \frac{ \sum_\varkappa W_\varkappa \theta_{\varkappa} }{\sum_\varkappa W_\varkappa }.
\end{equation}
The resulting stacked tangential reduced shear profiles are shown as black points with their error-bars in Fig.~\ref{fig:goodness_of_fit} given by the square root of the variance calculated above.

For the model prediction, we draw $N_\text{MC}=1000$ points $\boldsymbol{p}_\text{SR}^v$ from the posterior sample. For each cluster $\varkappa$, and for each posterior point $v$, we compute its best-guess WL mass 
\begin{equation}
    M_{\text{WL},\varkappa}^{v} = \underset{M_\text{WL}}{\mathrm{argmax}} ~ P(M_\text{WL} | \hat{C}_{R,\varkappa}, z_\varkappa, \hat{\mathcal{H}}_\varkappa, \boldsymbol{p}_\text{SR}^v),
\end{equation}
as the maximum of the pdf of WL masses, given the count rate, redshift and sky position, evaluated for our specific posterior draw $\boldsymbol{p}_\text{SR}^v$. We report the median mass of the clusters in each stacking bin in Fig.~\ref{fig:goodness_of_fit}. For each WL mass $ M_{\text{WL},\varkappa}^{v} $ we evaluate our extraction model $g^{\text{mod},v}_{\text{t}, \varkappa}$, given in Eq.~\ref{eq:extraction_model} using the source redshift distribution computed on the real data, Eq.~\ref{eq:cosmo_rdy_pz}, and the mean positions of the sources $\theta_\alpha$ from Eq.~\ref{eq:cosmo_rdy_theta}. For each cluster and posterior point, we thus construct the predicted reduced shear profile as
\begin{equation}
    g^{\text{pred},v}_{\text{t},\varkappa} = (1 - f_{\text{RS}, \varkappa}^v - f_{\text{AGN}, \varkappa}^v ) g^{\text{mod},v}_{\text{t}, \varkappa} +  f_{\text{AGN}, \varkappa}^v g^{\text{AGN}}_{\text{t}, \varkappa}, 
\end{equation}
with $f_{\text{RS},\varkappa}^v$ and $f_{\text{AGN},\varkappa}^v$ the random source and AGN fractions at the count rate, redshift and sky position of the cluster $\varkappa$ given the model parameters $\boldsymbol{p}_\text{SR}^v$.

The average reduced shear model in the bin is then given by stacking the predicted reduced shear profile with the same weights as the data, and taking a mean over the posterior points, 
\begin{equation}
    g^{\text{pred}}_{\text{t}} = \frac{1}{N_\text{MC}} \sum_v \frac{ \sum_\varkappa W_\varkappa  g^{\text{pred},v}_{\text{t},\varkappa} }{\sum_\varkappa W_\varkappa},
\end{equation}
and variance 
\begin{equation}
    \left( \delta g^{\text{pred}}_{\text{t}}\right)^2 = \frac{1}{N_\text{MC}} \sum_v \frac{ \sum_\varkappa W_\varkappa  \left( g^{\text{pred},v}_{\text{t},\varkappa} - g^{\text{pred}}_{\text{t},\varkappa} \right)^2}{\sum_\varkappa W_\varkappa},
\end{equation}
propagating the uncertainty in the model parameters expressed by their posterior to the reduced shear profile prediction. The mean shear profile and its systematic uncertainty are shown as green lines in Fig.~\ref{fig:goodness_of_fit}. The visual impression of a good fit is corroborated by considering that the chi-squared between stacked data and predicted model is $\chi^2 = 180.0^{+45.8}_{-30.4}$ for 150 data points and 5 parameters that are effectively constrained. The upper and lower errors for the $\chi^2$ result from the difference induced when evaluating it with the model $g^{\text{pred}}_{\text{t}} \pm  \delta g^{\text{pred}}_{\text{t}}$, instead of the mean model. Within the errors propagated from the posterior, we attain a very good chi-squared.

\section{Discussion}\label{sec:discussion}

\subsection{Comparison to previous work}

The only previous work that calibrated the eROSITA count rate -- halo mass relation was performed by \citet{chiu22a}, measuring, calibrating and analysing the WL signal around 313 eFEDS selected clusters and groups with HSC data. Their results are shown as black lines in Fig.~\ref{fig:posterior_SR}.  We agree on all parameters within 2-sigma, with larger than 1-sigma deviations on the amplitude, the redshift trend of the slope and the scatter. When considering the predicted mean count rate as a function of mass and redshift, their results predict higher count rates for low mass, high redshift objects (hatched bands in Fig.~\ref{fig:post_pred_SR}). The difference is however not statistically significant. The constraining power of this work is only marginally better than the calibration derived by \citet{chiu22a}.

The similar constraining power of eFEDS-HSC analysis to this work, in spite of the significantly smaller number of objects (313 compared to 2201) is explained by the higher source density in HSC, as well as the cleaner background selection. Indeed, the signal-to-noise ratio of an individual cluster is proportional to
\begin{equation}
    \frac{\text{S}}{\text{N}} \propto \beta \sqrt{n_\text{eff}}\sqrt{1-f_\text{cl}}, \text{ with } \beta = \Bigg\langle \frac{\min (0, D_\text{ls})}{D_\text{s}} \Bigg\rangle_{P(z_\text{s})},
\end{equation}
where $\beta$ is the lensing efficiency, which is 0 for sources at the lens redshift, and converges to 1 for infinitely distant sources. $n_\text{eff}$ is the effective source density, and $f_\text{cl}$ the cluster member contamination. When considering these elements, each cluster in the HSC analysis has a WL signal-to-noise approximately a factor of 2.3 larger then the clusters considered in this work, assuming they have the same mass, and ignoring cluster member contamination. Considering the overall signal-to-noise gain stemming from the total number of clusters, $\sqrt{2201/313}=2.6$, we find that it just about compensates the much lower individual clusters signal-to-noise. Note also that in the eFEDS-HSC analysis, a tighter $\Omega_\text{M}$, and an informative $\sigma_\text{X}$ prior where used. Taken together, this explains the similar constraining power of the two works. 

\subsection{Cluster line of sight anomalies}

We find several cluster line of sight anomalies, that we shall discuss here in a cohesive manner. Firstly, it is somewhat anomalous that the cluster member contamination estimates based on the effective number density of sources display a
larger contaminant fraction than the estimate based on the decomposition of the source redshift distribution. The latter should be impervious to masking and obstruction effects, as argued by \citet{gruen14, varga19}. Estimations of the cluster member contamination via the effective number density of sources are biased low because of masking and obstruction by foreground and bright cluster sources, as shown by \citet{kleinebreil23}. It is therefore surprising that we find the number density estimate to be larger than the source redshift distribution estimate, despite it potentially being biased low. A possible source of error here is our very lenient background selection and the possible performance loss of the source redshift decomposition method if the cluster redshift lies too close to the peak of source redshift distribution of the field. This might be hinted at by the imperfect subtraction signatures discussed in Section~\ref{sec:cluster_mbr_cnt} and seen in Fig.~\ref{fig:Pz-valid}.

Also interesting are the decrement of the effective shape noise and the increment of the smooth shear response towards the cluster center. The effect is rather constant in lens redshift, as opposed to the cluster member contamination. Indeed, we see this anomaly also in the lowest source redshift bin, $0.1<z_\text{l}<0.19$, where we have practically no cluster member contamination. This makes it implausible that cluster member contaminants have significantly different responses or intrinsic shapes. It is worthwhile noting in Eq.~\ref{eq:sigma_eff} that a larger shear response leads to a smaller estimated effective shape noise. Also, the shear response used in this work is a smooth function of the source's signal-to-noise and its size w.r.t. the PSF. Noticeably, both these quantities are magnified by the cluster's potential -- a secondary effect of WL. Such magnification effects might thus alter the inferred smooth shear response. In general, shape measurement and photo-$z$ calibration in cluster lines of sight might be effected to second order not only by magnification, but also by increased blending \citep{hernandez-martin20} and the intra-cluster light \citep{zhang19}. Exploitation of future wide optical and NIR surveys might necessitate dedicated assessments of the performance of shape and photo-$z$ measurement techniques in cluster lines of sight.

\begin{figure}
  \includegraphics[width=\columnwidth]{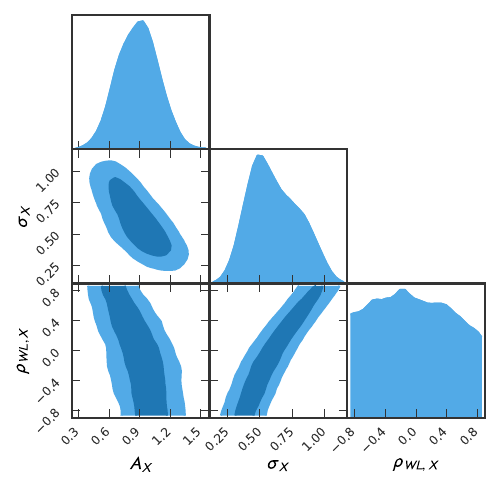}
  \caption{1-d and 2-d marginal contours of the amplitude of the count rate--mass scaling relation $A_\text{X}$, the intrinsic scatter around that relation $\sigma_\text{X}$, and the correlation coefficient between the WL and count rate scatter $\rho_{M_\text{WL}, C_R}$, labelled as $\rho_\text{WL,X}$ for brevity in the plot. While the correlation coefficient remains unconstrained, it leads to significant degeneracies with the amplitude and scatter. Astrophysical processes like cluster age, merger state, environment, orientation, triaxiality, etc., might impact both observables and lead to selection biases. We agnostically marginalize over the entire range of such effects, thereby weakening our amplitude and scatter measurements, but making the mass calibration impervious to such astrophysical effects.}
  \label{fig:selection_effects}
\end{figure}

\subsection{Accounting for cluster selection biases}

Much of the astrophysically driven conversation about the viability of galaxy clusters as a cosmological probe, and the limits of our ability to effectively calibrate their masses, focus on so called \emph{selection biases}. When taking a forward modelling approach using Bayesian population models, these concerns can be naturally accounted for. We shall discuss in the following the most (in)famous "selection bias" as an example: the "cool core bias" \citep{rossetti17}. The cooling time in the center of massive galaxy clusters can fall below the Hubble time, especially if they have been undisturbed for a long time. Effective cooling boosts the central luminosity, resulting in very peaked X-ray surface brightness profiles. Conversely, the SZe selection, sensitive to the pressure, is not effected by this. As discussed already above, taking a population modelling approach, it immediately results that any sample selected on an observable that scatters around the halo mass will preferentially contain up-scattered objects compared to down-scattered objects, on account of the much higher frequency of the former. If a specific physical property drives the scatter in some observable at fixed mass, then a sample selected in that observable would show more objects with that physical property, compared to a sample selected on another observable. X-ray selected samples will preferentially contain objects that scattered up in X-rays, for whatever astrophysical reason.

For cosmological inference the complication arises if the scatter in the follow-up observable used for the mass calibration correlates with scatter in the selection observable. A glaring case which is being much investigated, is clusters selected via the number of photometric member galaxies, and the WL signal of these objects. The scatter of both depends on the concentration, orientation and triaxiality of the halo, as well as the neighboring environment, resulting in strong selection biases hampering cosmological analyses on those samples \citep{desy1_clusters, wu22}. Analogously, X-ray-selected clusters have been reported to live preferentially in nodes of the cosmic web, as compared to filaments or sheets, as found by \citet{popesso+23} when comparing eFEDS-selected halos with spectroscopically selected ones. This means that at fixed halo mass, the cluster properties may depend on secondary halo properties like environment or concentration -- an effect analogous to the \emph{assembly bias} of galaxies \citep{wu08, wechsler18}. Our scale cuts on the tangential reduced shear are designed  to ensure that we measure the WL signal only in the 1-halo regime \citep{grandis+21}, that is scales dominated by the halo's own density profile, and not by neighboring structures. This already limits significantly the impact of assembly bias. Still, the eRASS1 selection  might correlate with some property that impact also the WL mass, such as concentration, triaxiality, or dynamical state. To account for any such possible effect, we introduce the correlation coefficient $\rho_\text{WL,X}$ among the scatters in count rate and WL mass. Fig.~\ref{fig:selection_effects} shows the posterior on the correlation coefficient, together with its degeneracies with the amplitude of the scaling relation and intrinsic scatter at fixed halo mass and redshift. The prior range for the correlation coefficient is naturally given by $(-1, 1)$, though we sample $(-0.9, 0.9)$ to avoid numerical instabilities for singular covariance matrix. Marginalising over the correlation coefficient makes our analysis impervious to selection bias effects, but deteriorates our ability to calibrate the observable--mass relation. Given that the numerical value of the amplitude of the scaling relation directly impacts our cosmological results, and crucially their agreement with external experiments, foregoing the use of strong astrophysical priors on the selection effects at the cost of a less stringent mass calibration is undoubtedly the more prudent choice.

\section{Conclusions}

In this work, we perform the measurement and calibration of the WL signal around eRASS1 clusters and groups in the DES Y3 data. To this end, we define a background sample of DES galaxies by weighting tomographic redshift bins defined and calibrated for the cosmic shear experiment \citepalias{myles21, gatti21} as a function of the photometric redshift of our clusters. After querying sources around our lenses, out to an angular distance corresponding to $15 h^{-1}$Mpc, we validate and calibrate that measurement. We fit the lens richness, redshift and source-lens distance dependence of the effective source density and the source redshift distribution to determine the contamination of our source sample. The  different methods employed agree with each other within the systematic uncertainties of the source redshift and shape measurement.

We fit a lens redshift trend to the effective shape noise, allowing us to define a semi-analytical covariance metric for the shape noise. The profiles of the B-mode-like cross component are consistent with zero, as expected for a well-calibrated WL signal sourced by gravity. Investigation of the selection and shear response of our tangential reduced shear estimators show that the former can be ignored, as it induces a correction much smaller than the systematic uncertainty induced by source redshift and shape measurements. On small cluster-centric distances we find interesting anomalies in the shear response and effective shape noise, that might point at cluster line of sight specific calibration issues, which arguably are too small to impact our subsequent analysis steps.

We derive cosmology-ready data products, used as inputs for the mass calibration of eRASS1 selected clusters, both in this work, and self-consistently also in the subsequent cosmological analysis by \citet{ghirardini23}. These products are:
\begin{itemize}
    \item the tangential reduced shear profile in angular bins, corrected for the shear response,
    \item a semi-analytical estimate of the shape noise on that profile, and 
    \item an estimate of the field source redshift distribution from the local background,
\end{itemize}
for each source with DES Y3 lensing data. This results in WL data for 2201 eRASS1 selected clusters, with a raw signal-to-noise ratio of 92, reduced to 65 after the our radial scale cuts are applied.

Exploitation of these data for mass calibration and number counts cosmology requires the anchoring of the mapping between halo mass and WL signal via $\mathcal{O}(1000)$ realisations of synthetic shear profile simulations, following the scheme presented by \citet{grandis+21}. These simulations include:
\begin{itemize}
    \item realistic 2d projected surface mass density maps around massive halos, extracted from the cosmological hydro-dynamical TNG300 simulations \citep{Pillepich2018MNRAS.475..648P, Marinacci2018MNRAS.480.5113M, Springel2018MNRAS.475..676S, Nelson2018MNRAS.475..624N, Naiman2018MNRAS.477.1206N, Nelson2019ComAC...6....2N},
    \item a distribution of the off-set between the observational centers used for the tangential reduced shear profile measurements, constructed from X-ray images of the \textsc{Magneticum Pathfinder} simulations  (Dolag et al., in prep.), and the `eROSITA digital sky twin'  \citep{seppi22}, 
    \item a painting of the richness using the richness--mass relation calibrated by \citet{chiu22a},
    \item dilution of the tangential reduced shear signal based on the degree of cluster member contamination we calibrated,
    \item a noise contribution from the uncorrelated large-scale structure, and
    \item realistic DES Y3 source distributions reflecting our background selection based on weighted sums of the DES Y3 tomographic redshift bins \citepalias{myles21}, together with a prescription for non-linear shear bias.
\end{itemize}

These synthetic shear profiles are analysed with the same shear profile model that we use in the subsequent mass calibration. It features a simplified mis-centered NFW with fixed concentration--mass relation, and accounts for the mean cluster member contamination, as advocated by \citet{grandis+21}. It thus has one free parameter, the \emph{WL mass}. Extracting the WL mass on our synthetic shear profile simulations allows us to establish the relation between WL and halo mass, summarized in the WL bias and scatter. Monte-Carlo realisations of the synthetic shear profile simulation, obtained by varying all the inputs within our expectation, result in an uncertainty estimate on the WL bias and scatter, that acts as a calibration prior for the subsequent mass calibration. We find that at low redshift our mass extraction uncertainty is dominated by the effect of hydro-dynamical modelling, leading to a systematics floor of $2\%$. At higher redshift, DES photo-$z$ uncertainties become incrementally more important.

We calibrate the count rate -- halo mass relation by using a Bayesian population model to forward-model the probability density function of each cluster's shear profile given its count rate and redshift. Improving on previous work \citep{mantz16, dietrich19, bocquet19, chiu22a}, we take explicit account of contaminants via a mixture model. We find that we are able to measure 5 parameters of the count rate--mass relation: the amplitude, the slope and its redshift dependence, the deviation from a self-similar redshift evolution, and the intrinsic scatter around the relation. We assess the quality of this population model based fit on stacked signals in count rate, redshift bins, and find excellent agreement among the data and the model. Our analysis finds comparable, and marginally tighter constraints than the calibration of the eROSITA count rate--mass relation performed by \citet{chiu22a} with HSC data on eFEDS-selected clusters.

This analysis will serve, along with similar analyses on HSC and KiDS, as the main mass calibration for the number counts of eRASS1 selected clusters \citep{ghirardini23}, which constrain different cosmological parameters like the present-day matter density, the amplitude of matter fluctuations, the equation of state of Dark Energy, and the sum of the neutrino masses. Methodologically, we demonstrate how photometric data from wide surveys, calibrated specifically for cosmic shear experiments, can be effectively used the perform a WL mass calibration of cluster number counts. These two experiments are highly complementary. Cosmic shear probes the linear and mildly non-linear scale of the large-scale structure, while WL calibrated number counts probe the most non-linear regions, massive galaxy clusters. These probes have very different sensitivities to astrophysical and observational systematics, only sharing a joint dependence on the source photo-$z$ and shape calibration. In light of upcoming deep and wide surveys, tailored mainly for cosmic shear, like \textit{Euclid}\footnote{\url{http://www.euclid-ec.org/}}, LSST\footnote{\url{https://www.lsst.org/}}, or Roman\footnote{\url{https://roman.gsfc.nasa.gov/}}, this work provides a demonstration on how to leverage these data to inform cluster number counts, enabling independent and competitive cosmological constraints.

\begin{acknowledgements}
This work is based on data from eROSITA, the soft X-ray instrument aboard SRG, a joint Russian-German science mission supported by the Russian Space Agency (Roskosmos), in the interests of the Russian Academy of Sciences represented by its Space Research Institute (IKI), and the Deutsches Zentrum f{\"{u}}r Luft und Raumfahrt (DLR). The SRG spacecraft was built by Lavochkin Association (NPOL) and its subcontractors and is operated by NPOL with support from the Max Planck Institute for Extraterrestrial Physics (MPE).

The development and construction of the eROSITA X-ray instrument was led by MPE, with contributions from the Dr. Karl Remeis Observatory Bamberg \& ECAP (FAU Erlangen-Nuernberg), the University of Hamburg Observatory, the Leibniz Institute for Astrophysics Potsdam (AIP), and the Institute for Astronomy and Astrophysics of the University of T{\"{u}}bingen, with the support of DLR and the Max Planck Society. The Argelander Institute for Astronomy of the University of Bonn and the Ludwig Maximilians Universit{\"{a}}t Munich also participated in the science preparation for eROSITA.

The eROSITA data shown here were processed using the {\tt eSASS} software system developed by the German eROSITA consortium.
\\

Funding for the DES Projects has been provided by the U.S. Department of Energy, the U.S. National Science Foundation, the Ministry of Science and Education of Spain, the Science and Technology Facilities, the Council of the United Kingdom, the Higher Education Funding Council for England, the National Center for Supercomputing Applications at the University of Illinois at Urbana-Champaign, the Kavli Institute of Cosmological Physics at the University of Chicago, the Center for Cosmology and Astro-Particle Physics at the Ohio State University, the Mitchell Institute for Fundamental Physics and Astronomy at Texas A\&M University, Financiadora de Estudos e Projetos, Funda{\c c}{\~a}o Carlos Chagas Filho de Amparo {\`a} Pesquisa do Estado do Rio de Janeiro, Conselho Nacional de Desenvolvimento Cient{\'i}fico e Tecnol{\'o}gico and the Minist{\'e}rio da Ci{\^e}ncia, Tecnologia e Inova{\c c}{\~a}o, the Deutsche Forschungsgemeinschaft, and the Collaborating Institutions in the Dark Energy Survey.

The Collaborating Institutions are Argonne National Laboratory, the University of California at Santa Cruz, the University of Cambridge, Centro de Investigaciones Energ{\'e}ticas, Medioambientales y Tecnol{\'o}gicas-Madrid, the University of Chicago, University College London, the DES-Brazil Consortium, the University of Edinburgh, the Eidgen{\"o}ssische Technische Hochschule (ETH) Z{\"u}rich,  Fermi National Accelerator Laboratory, the University of Illinois at Urbana-Champaign, the Institut de Ci{\`e}ncies de l'Espai (IEEC/CSIC), the Institut de F{\'i}sica d'Altes Energies, Lawrence Berkeley National Laboratory, the Ludwig-Maximilians Universit{\"a}t M{\"u}nchen and the associated Excellence Cluster Universe, the University of Michigan, the National Optical Astronomy Observatory, the University of Nottingham, The Ohio State University, the OzDES Membership Consortium, the University of Pennsylvania, the University of Portsmouth, SLAC National Accelerator Laboratory, Stanford University, the University of Sussex, and Texas A\&M University.

Based in part on observations at Cerro Tololo Inter-American Observatory, National Optical Astronomy Observatory, which is operated by the Association of Universities for Research in Astronomy (AURA) under a cooperative agreement with the National Science Foundation.

E.B., A.L., V.G., X.Z., and  acknowledges financial support from the European Research Council (ERC) Consolidator Grant under the European Union’s Horizon 2020 research and innovation programme (grant agreement CoG DarkQuest No 101002585). LK acknowledges support by the COMPLEX project from the European Research Council (ERC) under the European Union’s Horizon 2020 research and innovation program grant agreement ERC-2019-AdG 882679. N.O. acknowledges JSPS KAKENHI Grant Number JP19KK0076.

The Innsbruck group acknowledges support from the German Research Foundation (DFG) under grant 415537506, as well as the Austrian Research Promotion Agency (FFG) and the Federal Ministry of the Republic of
Austria for Climate Action, Environment, Mobility, Innovation and
Technology (BMK) via grants 899537 and 900565.
\\
      
\end{acknowledgements}

%
\bibliographystyle{aa} 
\bibliography{bibliography} 

\begin{thebibliography}{95}
\expandafter\ifx\csname natexlab\endcsname\relax\def\natexlab#1{#1}\fi

\bibitem[{{Allen} {et~al.}(2011){Allen}, {Evrard}, \& {Mantz}}]{allen11}
{Allen}, S.~W., {Evrard}, A.~E., \& {Mantz}, A.~B. 2011, \araa, 49, 409

\bibitem[{{Angulo} {et~al.}(2012){Angulo}, {Springel}, {White}, {Jenkins},
  {Baugh}, \& {Frenk}}]{angulo12}
{Angulo}, R.~E., {Springel}, V., {White}, S.~D.~M., {et~al.} 2012, \mnras, 426,
  2046

\bibitem[{{Bah{\'e}} {et~al.}(2012){Bah{\'e}}, {McCarthy}, \& {King}}]{bahe12}
{Bah{\'e}}, Y.~M., {McCarthy}, I.~G., \& {King}, L.~J. 2012, \mnras, 421, 1073

\bibitem[{{Becker} \& {Kravtsov}(2011)}]{becker11}
{Becker}, M.~R. \& {Kravtsov}, A.~V. 2011, \apj, 740, 25

\bibitem[{{Bellagamba} {et~al.}(2019){Bellagamba}, {Sereno}, {Roncarelli},
  {Maturi}, {Radovich}, {Bardelli}, {Puddu}, {Moscardini}, {Getman},
  {Hildebrandt}, \& {Napolitano}}]{bellagamba19}
{Bellagamba}, F., {Sereno}, M., {Roncarelli}, M., {et~al.} 2019, \mnras, 484,
  1598

\bibitem[{{Ben{\'\i}tez}(2000)}]{benitez00}
{Ben{\'\i}tez}, N. 2000, \apj, 536, 571

\bibitem[{{Bleem} {et~al.}(2020){Bleem}, {Bocquet}, {Stalder}, {Gladders},
  {Ade}, {Allen}, {Anderson}, {Annis}, {Ashby}, {Austermann}, {Avila}, {Avva},
  {Bayliss}, {Beall}, {Bechtol}, {Bender}, {Benson}, {Bertin}, {Bianchini},
  {Blake}, {Brodwin}, {Brooks}, {Buckley-Geer}, {Burke}, {Carlstrom}, {Rosell},
  {Carrasco Kind}, {Carretero}, {Chang}, {Chiang}, {Citron}, {Moran},
  {Costanzi}, {Crawford}, {Crites}, {da Costa}, {de Haan}, {De Vicente},
  {Desai}, {Diehl}, {Dietrich}, {Dobbs}, {Eifler}, {Everett}, {Flaugher},
  {Floyd}, {Frieman}, {Gallicchio}, {Garc{\'\i}a-Bellido}, {George}, {Gerdes},
  {Gilbert}, {Gruen}, {Gruendl}, {Gschwend}, {Gupta}, {Gutierrez}, {Halverson},
  {Harrington}, {Henning}, {Heymans}, {Holder}, {Hollowood}, {Holzapfel},
  {Honscheid}, {Hrubes}, {Huang}, {Hubmayr}, {Irwin}, {James}, {Jeltema},
  {Joudaki}, {Khullar}, {Klein}, {Knox}, {Kuropatkin}, {Lee}, {Li}, {Lidman},
  {Lowitz}, {MacCrann}, {Mahler}, {Maia}, {Marshall}, {McDonald}, {McMahon},
  {Melchior}, {Menanteau}, {Meyer}, {Miquel}, {Mocanu}, {Mohr}, {Montgomery},
  {Nadolski}, {Natoli}, {Nibarger}, {Noble}, {Novosad}, {Padin}, {Palmese},
  {Parkinson}, {Patil}, {Paz-Chinch{\'o}n}, {Plazas}, {Pryke}, {Ramachandra},
  {Reichardt}, {Remolina Gonz{\'a}lez}, {Romer}, {Roodman}, {Ruhl}, {Rykoff},
  {Saliwanchik}, {Sanchez}, {Saro}, {Sayre}, {Schaffer}, {Schrabback},
  {Serrano}, {Sharon}, {Sievers}, {Smecher}, {Smith}, {Soares-Santos}, {Stark},
  {Story}, {Suchyta}, {Tarle}, {Tucker}, {Vanderlinde}, {Veach}, {Vieira},
  {Wang}, {Weller}, {Whitehorn}, {Wu}, {Yefremenko}, \& {Zhang}}]{bleem20}
{Bleem}, L.~E., {Bocquet}, S., {Stalder}, B., {et~al.} 2020, \apjs, 247, 25

\bibitem[{{Bocquet} {et~al.}(2019){Bocquet}, {Dietrich}, {Schrabback}, {Bleem},
  {Klein}, {Allen}, {Applegate}, {Ashby}, {Bautz}, {Bayliss}, {Benson},
  {Brodwin}, {Bulbul}, {Canning}, {Capasso}, {Carlstrom}, {Chang}, {Chiu},
  {Cho}, {Clocchiatti}, {Crawford}, {Crites}, {de Haan}, {Desai}, {Dobbs},
  {Foley}, {Forman}, {Garmire}, {George}, {Gladders}, {Gonzalez}, {Grandis},
  {Gupta}, {Halverson}, {Hlavacek-Larrondo}, {Hoekstra}, {Holder}, {Holzapfel},
  {Hou}, {Hrubes}, {Huang}, {Jones}, {Khullar}, {Knox}, {Kraft}, {Lee}, {von
  der Linden}, {Luong-Van}, {Mantz}, {Marrone}, {McDonald}, {McMahon}, {Meyer},
  {Mocanu}, {Mohr}, {Morris}, {Padin}, {Patil}, {Pryke}, {Rapetti},
  {Reichardt}, {Rest}, {Ruhl}, {Saliwanchik}, {Saro}, {Sayre}, {Schaffer},
  {Shirokoff}, {Stalder}, {Stanford}, {Staniszewski}, {Stark}, {Story},
  {Strazzullo}, {Stubbs}, {Vanderlinde}, {Vieira}, {Vikhlinin}, {Williamson},
  \& {Zenteno}}]{bocquet19}
{Bocquet}, S., {Dietrich}, J.~P., {Schrabback}, T., {et~al.} 2019, \apj, 878,
  55

\bibitem[{{Bocquet} {et~al.}(2023){Bocquet}, {Grandis}, {Bleem}, {Klein},
  {Mohr}, {Aguena}, {Alarcon}, {Allam}, {Allen}, {Alves}, {Amon},
  {Ansarinejad}, {Bacon}, {Bayliss}, {Bechtol}, {Becker}, {Benson},
  {Bernstein}, {Brodwin}, {Brooks}, {Campos}, {Canning}, {Carlstrom}, {Carnero
  Rosell}, {Carrasco Kind}, {Carretero}, {Cawthon}, {Chang}, {Chen}, {Choi},
  {Cordero}, {Costanzi}, {da Costa}, {Pereira}, {Davis}, {de Haan}, {DeRose},
  {Desai}, {Diehl}, {Dodelson}, {Doel}, {Doux}, {Drlica-Wagner}, {Eckert},
  {Elvin-Poole}, {Everett}, {Ferrero}, {Fert{\'e}}, {Flores}, {Frieman},
  {Garc{\'\i}a-Bellido}, {Gatti}, {Giannini}, {Gladders}, {Gruen}, {Gruendl},
  {Harrison}, {Hartley}, {Herner}, {Hinton}, {Hollowood}, {Holzapfel},
  {Honscheid}, {Huang}, {Huff}, {James}, {Jarvis}, {K{\'e}ruzor{\'e}},
  {Khullar}, {Kim}, {Kraft}, {Kuehn}, {Kuropatkin}, {Lee}, {Leget}, {MacCrann},
  {Mahler}, {Mantz}, {Marshall}, {McCullough}, {McDonald},
  {Mena-Fern{\'a}ndez}, {Miquel}, {Myles}, {Navarro-Alsina}, {Ogando},
  {Palmese}, {Pandey}, {Pieres}, {Plazas Malag{\'o}n}, {Prat}, {Raveri},
  {Reichardt}, {Roberson}, {Rollins}, {Romer}, {Romero}, {Roodman}, {Ross},
  {Rykoff}, {Salvati}, {S{\'a}nchez}, {Sanchez}, {Sanchez Cid}, {Saro},
  {Schrabback}, {Schubnell}, {Secco}, {Sevilla-Noarbe}, {Sharon}, {Sheldon},
  {Shin}, {Smith}, {Somboonpanyakul}, {Stalder}, {Stark}, {Strazzullo},
  {Suchyta}, {Swanson}, {Tarle}, {To}, {Troxel}, {Tutusaus}, {Varga}, {von der
  Linden}, {Weaverdyck}, {Weller}, {Wiseman}, {Yanny}, {Yin}, {Young}, {Zhang},
  \& {Zuntz}}]{bocquet+ipa}
{Bocquet}, S., {Grandis}, S., {Bleem}, L.~E., {et~al.} 2023, arXiv e-prints,
  arXiv:2310.12213

\bibitem[{{Buchner}(2021)}]{ultranest_is_the_best}
{Buchner}, J. 2021, The Journal of Open Source Software, 6, 3001

\bibitem[{{Bulbul} {et~al.}(A\&A,~subm.){Bulbul}, {Liu}, {Kluge}, {Zhang}, \&
  {X}}]{bulbul23}
{Bulbul}, E., {Liu}, A., {Kluge}, M., {Zhang}, X., \& {X}. A\&A,~subm.

\bibitem[{{Castro} {et~al.}(2021){Castro}, {Borgani}, {Dolag}, {Marra},
  {Quartin}, {Saro}, \& {Sefusatti}}]{castro+21}
{Castro}, T., {Borgani}, S., {Dolag}, K., {et~al.} 2021, \mnras, 500, 2316

\bibitem[{{Child} {et~al.}(2018){Child}, {Habib}, {Heitmann}, {Frontiere},
  {Finkel}, {Pope}, \& {Morozov}}]{child}
{Child}, H.~L., {Habib}, S., {Heitmann}, K., {et~al.} 2018, \apj, 859, 55

\bibitem[{{Chiu} {et~al.}(2022){Chiu}, {Ghirardini}, {Liu}, {Grandis},
  {Bulbul}, {Bahar}, {Comparat}, {Bocquet}, {Clerc}, {Klein}, {Liu}, {Li},
  {Miyatake}, {Mohr}, {More}, {Oguri}, {Okabe}, {Pacaud}, {Ramos-Ceja},
  {Reiprich}, {Schrabback}, \& {Umetsu}}]{chiu22a}
{Chiu}, I.~N., {Ghirardini}, V., {Liu}, A., {et~al.} 2022, \aap, 661, A11

\bibitem[{{Chiu} {et~al.}(2023){Chiu}, {Klein}, {Mohr}, \& {Bocquet}}]{chiu22b}
{Chiu}, I.~N., {Klein}, M., {Mohr}, J., \& {Bocquet}, S. 2023, \mnras, 522,
  1601

\bibitem[{{Clerc} {et~al.}(A\&A,~subm.){Clerc}, {Comparat}, {Seppi}, \&
  {Artis}}]{clerc23}
{Clerc}, N., {Comparat}, J., {Seppi}, R., \& {Artis}, E. A\&A,~subm.

\bibitem[{{Comparat} {et~al.}(2020){Comparat}, {Eckert}, {Finoguenov},
  {Schmidt}, {Sanders}, {Nagai}, {Lau}, {K{\"a}}, {fer}, {Pacaud}, {Clerc},
  {Reiprich}, {Bulbul}, {Chitham}, {Chiang}, {Ghirardini}, {Gonzalez-Perez},
  {Gozaliasl}, {Fitzpatrick}, {Klypin}, {Merloni}, {Nandra}, {Liu}, {Prada},
  {Ramos-Ceja}, {Salvato}, {Seppi}, {Tempel}, \& {Yepes}}]{comparat20}
{Comparat}, J., {Eckert}, D., {Finoguenov}, A., {et~al.} 2020, The Open Journal
  of Astrophysics, 3, 13

\bibitem[{{Comparat} {et~al.}(2023){Comparat}, {Luo}, {Merloni}, {More},
  {Salvato}, {Krumpe}, {Miyaji}, {Brandt}, {Georgakakis}, {Akiyama}, {Buchner},
  {Dwelly}, {Kawaguchi}, {Liu}, {Nagao}, {Nandra}, {Silverman}, {Toba},
  {Anderson}, \& {Kollmeier}}]{comparat23}
{Comparat}, J., {Luo}, W., {Merloni}, A., {et~al.} 2023, \aap, 673, A122

\bibitem[{{Comparat} {et~al.}(2019){Comparat}, {Merloni}, {Salvato}, {Nandra},
  {Boller}, {Georgakakis}, {Finoguenov}, {Dwelly}, {Buchner}, {Del Moro},
  {Clerc}, {Wang}, {Zhao}, {Prada}, {Yepes}, {Brusa}, {Krumpe}, \&
  {Liu}}]{comparat19}
{Comparat}, J., {Merloni}, A., {Salvato}, M., {et~al.} 2019, \mnras, 487, 2005

\bibitem[{{Cordero} {et~al.}(2022){Cordero}, {Harrison}, {Rollins},
  {Bernstein}, {Bridle}, {Alarcon}, {Alves}, {Amon}, {Andrade-Oliveira},
  {Camacho}, {Campos}, {Choi}, {DeRose}, {Dodelson}, {Eckert}, {Eifler},
  {Everett}, {Fang}, {Friedrich}, {Gruen}, {Gruendl}, {Hartley}, {Huff},
  {Krause}, {Kuropatkin}, {MacCrann}, {McCullough}, {Myles}, {Pandey},
  {Raveri}, {Rosenfeld}, {Rykoff}, {S{\'a}nchez}, {S{\'a}nchez},
  {Sevilla-Noarbe}, {Sheldon}, {Troxel}, {Wechsler}, {Yanny}, {Yin}, {Zhang},
  {Aguena}, {Allam}, {Bertin}, {Brooks}, {Burke}, {Carnero Rosell}, {Carrasco
  Kind}, {Carretero}, {Castander}, {Cawthon}, {Costanzi}, {da Costa}, {da Silva
  Pereira}, {De Vicente}, {Diehl}, {Dietrich}, {Doel}, {Elvin-Poole},
  {Ferrero}, {Flaugher}, {Fosalba}, {Frieman}, {Garcia-Bellido}, {Gerdes},
  {Gschwend}, {Gutierrez}, {Hinton}, {Hollowood}, {Honscheid}, {Hoyle},
  {James}, {Kuehn}, {Lahav}, {Maia}, {March}, {Menanteau}, {Miquel}, {Morgan},
  {Muir}, {Palmese}, {Paz-Chinchon}, {Pieres}, {Plazas Malag{\'o}n},
  {S{\'a}nchez}, {Scarpine}, {Serrano}, {Smith}, {Soares-Santos}, {Suchyta},
  {Swanson}, {Tarle}, {Thomas}, {To}, {Varga}, \& {DES
  Collaboration}}]{cordero22}
{Cordero}, J.~P., {Harrison}, I., {Rollins}, R.~P., {et~al.} 2022, \mnras, 511,
  2170

\bibitem[{{Dark Energy Survey Collaboration} {et~al.}(2020){Dark Energy Survey
  Collaboration}, {Abbott}, {Aguena}, {Alarcon}, {Allam}, {Allen}, {Annis},
  {Avila}, {Bacon}, {Bechtol}, {Bermeo}, {Bernstein}, {Bertin}, {Bhargava},
  {Bocquet}, {Brooks}, {Brout}, {Buckley-Geer}, {Burke}, {Carnero Rosell},
  {Carrasco Kind}, {Carretero}, {Castander}, {Cawthon}, {Chang}, {Chen},
  {Choi}, {Costanzi}, {Crocce}, {da Costa}, {Davis}, {De Vicente}, {DeRose},
  {Desai}, {Diehl}, {Dietrich}, {Dodelson}, {Doel}, {Drlica-Wagner}, {Eckert},
  {Eifler}, {Elvin-Poole}, {Estrada}, {Everett}, {Evrard}, {Farahi}, {Ferrero},
  {Flaugher}, {Fosalba}, {Frieman}, {Garc{\'\i}a-Bellido}, {Gatti},
  {Gaztanaga}, {Gerdes}, {Giannantonio}, {Giles}, {Grandis}, {Gruen},
  {Gruendl}, {Gschwend}, {Gutierrez}, {Hartley}, {Hinton}, {Hollowood},
  {Honscheid}, {Hoyle}, {Huterer}, {James}, {Jarvis}, {Jeltema}, {Johnson},
  {Johnson}, {Kent}, {Krause}, {Kron}, {Kuehn}, {Kuropatkin}, {Lahav}, {Li},
  {Lidman}, {Lima}, {Lin}, {MacCrann}, {Maia}, {Mantz}, {Marshall}, {Martini},
  {Mayers}, {Melchior}, {Mena-Fern{\'a}ndez}, {Menanteau}, {Miquel}, {Mohr},
  {Nichol}, {Nord}, {Ogando}, {Palmese}, {Paz-Chinch{\'o}n}, {Plazas}, {Prat},
  {Rau}, {Romer}, {Roodman}, {Rooney}, {Rozo}, {Rykoff}, {Sako}, {Samuroff},
  {S{\'a}nchez}, {Sanchez}, {Saro}, {Scarpine}, {Schubnell}, {Scolnic},
  {Serrano}, {Sevilla-Noarbe}, {Sheldon}, {Smith}, {Smith}, {Suchyta},
  {Swanson}, {Tarle}, {Thomas}, {To}, {Troxel}, {Tucker}, {Varga}, {von der
  Linden}, {Walker}, {Wechsler}, {Weller}, {Wilkinson}, {Wu}, {Yanny}, {Zhang},
  {Zhang}, {Zuntz}, \& {DES Collaboration}}]{desy1_clusters}
{Dark Energy Survey Collaboration}, {Abbott}, T.~M.~C., {Aguena}, M., {et~al.}
  2020, \prd, 102, 023509

\bibitem[{{Dark Energy Survey Collaboration} {et~al.}(2022){Dark Energy Survey
  Collaboration}, {Abbott}, {Aguena}, {Alarcon}, {Allam}, {Alves}, {Amon},
  {Andrade-Oliveira}, {Annis}, {Avila}, {Bacon}, {Baxter}, {Bechtol}, {Becker},
  {Bernstein}, {Bhargava}, {Birrer}, {Blazek}, {Brandao-Souza}, {Bridle},
  {Brooks}, {Buckley-Geer}, {Burke}, {Camacho}, {Campos}, {Carnero Rosell},
  {Carrasco Kind}, {Carretero}, {Castander}, {Cawthon}, {Chang}, {Chen},
  {Chen}, {Choi}, {Conselice}, {Cordero}, {Costanzi}, {Crocce}, {da Costa}, {da
  Silva Pereira}, {Davis}, {Davis}, {De Vicente}, {DeRose}, {Desai}, {Di
  Valentino}, {Diehl}, {Dietrich}, {Dodelson}, {Doel}, {Doux}, {Drlica-Wagner},
  {Eckert}, {Eifler}, {Elsner}, {Elvin-Poole}, {Everett}, {Evrard}, {Fang},
  {Farahi}, {Fernandez}, {Ferrero}, {Fert{\'e}}, {Fosalba}, {Friedrich},
  {Frieman}, {Garc{\'\i}a-Bellido}, {Gatti}, {Gaztanaga}, {Gerdes},
  {Giannantonio}, {Giannini}, {Gruen}, {Gruendl}, {Gschwend}, {Gutierrez},
  {Harrison}, {Hartley}, {Herner}, {Hinton}, {Hollowood}, {Honscheid}, {Hoyle},
  {Huff}, {Huterer}, {Jain}, {James}, {Jarvis}, {Jeffrey}, {Jeltema}, {Kovacs},
  {Krause}, {Kron}, {Kuehn}, {Kuropatkin}, {Lahav}, {Leget}, {Lemos}, {Liddle},
  {Lidman}, {Lima}, {Lin}, {MacCrann}, {Maia}, {Marshall}, {Martini},
  {McCullough}, {Melchior}, {Mena-Fern{\'a}ndez}, {Menanteau}, {Miquel},
  {Mohr}, {Morgan}, {Muir}, {Myles}, {Nadathur}, {Navarro-Alsina}, {Nichol},
  {Ogando}, {Omori}, {Palmese}, {Pandey}, {Park}, {Paz-Chinch{\'o}n},
  {Petravick}, {Pieres}, {Plazas Malag{\'o}n}, {Porredon}, {Prat}, {Raveri},
  {Rodriguez-Monroy}, {Rollins}, {Romer}, {Roodman}, {Rosenfeld}, {Ross},
  {Rykoff}, {Samuroff}, {S{\'a}nchez}, {Sanchez}, {Sanchez}, {Sanchez Cid},
  {Scarpine}, {Schubnell}, {Scolnic}, {Secco}, {Serrano}, {Sevilla-Noarbe},
  {Sheldon}, {Shin}, {Smith}, {Soares-Santos}, {Suchyta}, {Swanson}, {Tabbutt},
  {Tarle}, {Thomas}, {To}, {Troja}, {Troxel}, {Tucker}, {Tutusaus}, {Varga},
  {Walker}, {Weaverdyck}, {Wechsler}, {Weller}, {Yanny}, {Yin}, {Zhang},
  {Zuntz}, \& {DES Collaboration}}]{DES_Y3_3x2pt}
{Dark Energy Survey Collaboration}, {Abbott}, T.~M.~C., {Aguena}, M., {et~al.}
  2022, \prd, 105, 023520

\bibitem[{{Dark Energy Survey Collaboration} {et~al.}(2019){Dark Energy Survey
  Collaboration}, {Abbott}, {Allam}, {Andersen}, {Angus}, {Asorey}, {Avelino},
  {Avila}, {Bassett}, {Bechtol}, {Bernstein}, {Bertin}, {Brooks}, {Brout},
  {Brown}, {Burke}, {Calcino}, {Carnero Rosell}, {Carollo}, {Carrasco Kind},
  {Carretero}, {Casas}, {Castander}, {Cawthon}, {Challis}, {Childress},
  {Clocchiatti}, {Cunha}, {D'Andrea}, {da Costa}, {Davis}, {Davis}, {De
  Vicente}, {DePoy}, {Desai}, {Diehl}, {Doel}, {Drlica-Wagner}, {Eifler},
  {Evrard}, {Fernandez}, {Filippenko}, {Finley}, {Flaugher}, {Foley},
  {Fosalba}, {Frieman}, {Galbany}, {Garc{\'\i}a-Bellido}, {Gaztanaga},
  {Giannantonio}, {Glazebrook}, {Goldstein}, {Gonz{\'a}lez-Gait{\'a}n},
  {Gruen}, {Gruendl}, {Gschwend}, {Gupta}, {Gutierrez}, {Hartley}, {Hinton},
  {Hollowood}, {Honscheid}, {Hoormann}, {Hoyle}, {James}, {Jeltema}, {Johnson},
  {Johnson}, {Kasai}, {Kent}, {Kessler}, {Kim}, {Kirshner}, {Kovacs}, {Krause},
  {Kron}, {Kuehn}, {Kuhlmann}, {Kuropatkin}, {Lahav}, {Lasker}, {Lewis}, {Li},
  {Lidman}, {Lima}, {Lin}, {Macaulay}, {Maia}, {Mandel}, {March}, {Marriner},
  {Marshall}, {Martini}, {Menanteau}, {Miller}, {Miquel}, {Miranda}, {Mohr},
  {Morganson}, {Muthukrishna}, {M{\"o}ller}, {Neilsen}, {Nichol}, {Nord},
  {Nugent}, {Ogando}, {Palmese}, {Pan}, {Plazas}, {Pursiainen}, {Romer},
  {Roodman}, {Rozo}, {Rykoff}, {Sako}, {Sanchez}, {Scarpine}, {Schindler},
  {Schubnell}, {Scolnic}, {Serrano}, {Sevilla-Noarbe}, {Sharp}, {Smith},
  {Soares-Santos}, {Sobreira}, {Sommer}, {Spinka}, {Suchyta}, {Sullivan},
  {Swann}, {Tarle}, {Thomas}, {Thomas}, {Troxel}, {Tucker}, {Uddin}, {Walker},
  {Wester}, {Wiseman}, {Wolf}, {Yanny}, {Zhang}, {Zhang}, \& {DES
  Collaboration}}]{des_sn}
{Dark Energy Survey Collaboration}, {Abbott}, T.~M.~C., {Allam}, S., {et~al.}
  2019, \apjl, 872, L30

\bibitem[{{De Vicente} {et~al.}(2016){De Vicente}, {S{\'a}nchez}, \&
  {Sevilla-Noarbe}}]{devicente16}
{De Vicente}, J., {S{\'a}nchez}, E., \& {Sevilla-Noarbe}, I. 2016, \mnras, 459,
  3078

\bibitem[{{Dey} {et~al.}(2019){Dey}, {Schlegel}, {Lang}, {Blum}, {Burleigh},
  {Fan}, {Findlay}, {Finkbeiner}, {Herrera}, {Juneau}, {Landriau}, {Levi},
  {McGreer}, {Meisner}, {Myers}, {Moustakas}, {Nugent}, {Patej}, {Schlafly},
  {Walker}, {Valdes}, {Weaver}, {Y{\`e}che}, {Zou}, {Zhou}, {Abareshi},
  {Abbott}, {Abolfathi}, {Aguilera}, {Alam}, {Allen}, {Alvarez}, {Annis},
  {Ansarinejad}, {Aubert}, {Beechert}, {Bell}, {BenZvi}, {Beutler}, {Bielby},
  {Bolton}, {Brice{\~n}o}, {Buckley-Geer}, {Butler}, {Calamida}, {Carlberg},
  {Carter}, {Casas}, {Castander}, {Choi}, {Comparat}, {Cukanovaite}, {Delubac},
  {DeVries}, {Dey}, {Dhungana}, {Dickinson}, {Ding}, {Donaldson}, {Duan},
  {Duckworth}, {Eftekharzadeh}, {Eisenstein}, {Etourneau}, {Fagrelius},
  {Farihi}, {Fitzpatrick}, {Font-Ribera}, {Fulmer}, {G{\"a}nsicke},
  {Gaztanaga}, {George}, {Gerdes}, {Gontcho}, {Gorgoni}, {Green}, {Guy},
  {Harmer}, {Hernandez}, {Honscheid}, {Huang}, {James}, {Jannuzi}, {Jiang},
  {Joyce}, {Karcher}, {Karkar}, {Kehoe}, {Kneib}, {Kueter-Young}, {Lan},
  {Lauer}, {Le Guillou}, {Le Van Suu}, {Lee}, {Lesser}, {Perreault Levasseur},
  {Li}, {Mann}, {Marshall}, {Mart{\'\i}nez-V{\'a}zquez}, {Martini}, {du Mas des
  Bourboux}, {McManus}, {Meier}, {M{\'e}nard}, {Metcalfe},
  {Mu{\~n}oz-Guti{\'e}rrez}, {Najita}, {Napier}, {Narayan}, {Newman}, {Nie},
  {Nord}, {Norman}, {Olsen}, {Paat}, {Palanque-Delabrouille}, {Peng},
  {Poppett}, {Poremba}, {Prakash}, {Rabinowitz}, {Raichoor}, {Rezaie},
  {Robertson}, {Roe}, {Ross}, {Ross}, {Rudnick}, {Safonova}, {Saha},
  {S{\'a}nchez}, {Savary}, {Schweiker}, {Scott}, {Seo}, {Shan}, {Silva},
  {Slepian}, {Soto}, {Sprayberry}, {Staten}, {Stillman}, {Stupak}, {Summers},
  {Sien Tie}, {Tirado}, {Vargas-Maga{\~n}a}, {Vivas}, {Wechsler}, {Williams},
  {Yang}, {Yang}, {Yapici}, {Zaritsky}, {Zenteno}, {Zhang}, {Zhang}, {Zhou}, \&
  {Zhou}}]{dey19}
{Dey}, A., {Schlegel}, D.~J., {Lang}, D., {et~al.} 2019, \aj, 157, 168

\bibitem[{{Dietrich} {et~al.}(2019){Dietrich}, {Bocquet}, {Schrabback},
  {Applegate}, {Hoekstra}, {Grandis}, {Mohr}, {Allen}, {Bayliss}, {Benson},
  {Bleem}, {Brodwin}, {Bulbul}, {Capasso}, {Chiu}, {Crawford}, {Gonzalez}, {de
  Haan}, {Klein}, {von der Linden}, {Mantz}, {Marrone}, {McDonald},
  {Raghunathan}, {Rapetti}, {Reichardt}, {Saro}, {Stalder}, {Stark}, {Stern},
  \& {Stubbs}}]{dietrich19}
{Dietrich}, J.~P., {Bocquet}, S., {Schrabback}, T., {et~al.} 2019, \mnras, 483,
  2871

\bibitem[{{Dolag} {et~al.}(in prep.){Dolag}, X., \& et~al.}]{dolagip}
{Dolag}, K., X., ., \& et~al. in prep.

\bibitem[{{Dyson} {et~al.}(1920){Dyson}, {Eddington}, \& {Davidson}}]{dyson20}
{Dyson}, F.~W., {Eddington}, A.~S., \& {Davidson}, C. 1920, Philosophical
  Transactions of the Royal Society of London Series A, 220, 291

\bibitem[{{Ettori} {et~al.}(2011){Ettori}, {Gastaldello}, {Leccardi},
  {Molendi}, {Rossetti}, {Buote}, \& {Meneghetti}}]{ettori11}
{Ettori}, S., {Gastaldello}, F., {Leccardi}, A., {et~al.} 2011, {Mass profiles
  and c - M$_{DM}$ relation in X-ray luminous galaxy clusters (Corrigendum)},
  Astronomy \& Astrophysics, Volume 526, id.C1, 1 pp.

\bibitem[{{Flaugher} {et~al.}(2015){Flaugher}, {Diehl}, {Honscheid}, {Abbott},
  {Alvarez}, {Angstadt}, {Annis}, {Antonik}, {Ballester}, {Beaufore},
  {Bernstein}, {Bernstein}, {Bigelow}, {Bonati}, {Boprie}, {Brooks},
  {Buckley-Geer}, {Campa}, {Cardiel-Sas}, {Castander}, {Castilla}, {Cease},
  {Cela-Ruiz}, {Chappa}, {Chi}, {Cooper}, {da Costa}, {Dede}, {Derylo},
  {DePoy}, {de Vicente}, {Doel}, {Drlica-Wagner}, {Eiting}, {Elliott}, {Emes},
  {Estrada}, {Fausti Neto}, {Finley}, {Flores}, {Frieman}, {Gerdes},
  {Gladders}, {Gregory}, {Gutierrez}, {Hao}, {Holland}, {Holm}, {Huffman},
  {Jackson}, {James}, {Jonas}, {Karcher}, {Karliner}, {Kent}, {Kessler},
  {Kozlovsky}, {Kron}, {Kubik}, {Kuehn}, {Kuhlmann}, {Kuk}, {Lahav}, {Lathrop},
  {Lee}, {Levi}, {Lewis}, {Li}, {Mandrichenko}, {Marshall}, {Martinez},
  {Merritt}, {Miquel}, {Mu{\~n}oz}, {Neilsen}, {Nichol}, {Nord}, {Ogando},
  {Olsen}, {Palaio}, {Patton}, {Peoples}, {Plazas}, {Rauch}, {Reil}, {Rheault},
  {Roe}, {Rogers}, {Roodman}, {Sanchez}, {Scarpine}, {Schindler}, {Schmidt},
  {Schmitt}, {Schubnell}, {Schultz}, {Schurter}, {Scott}, {Serrano}, {Shaw},
  {Smith}, {Soares-Santos}, {Stefanik}, {Stuermer}, {Suchyta}, {Sypniewski},
  {Tarle}, {Thaler}, {Tighe}, {Tran}, {Tucker}, {Walker}, {Wang}, {Watson},
  {Weaverdyck}, {Wester}, {Woods}, {Yanny}, \& {DES
  Collaboration}}]{flaugher15}
{Flaugher}, B., {Diehl}, H.~T., {Honscheid}, K., {et~al.} 2015, \aj, 150, 150

\bibitem[{{Friedrich} {et~al.}(2021){Friedrich}, {Andrade-Oliveira}, {Camacho},
  {Alves}, {Rosenfeld}, {Sanchez}, {Fang}, {Eifler}, {Krause}, {Chang},
  {Omori}, {Amon}, {Baxter}, {Elvin-Poole}, {Huterer}, {Porredon}, {Prat},
  {Terra}, {Troja}, {Alarcon}, {Bechtol}, {Bernstein}, {Buchs}, {Campos},
  {Carnero Rosell}, {Carrasco Kind}, {Cawthon}, {Choi}, {Cordero}, {Crocce},
  {Davis}, {DeRose}, {Diehl}, {Dodelson}, {Doux}, {Drlica-Wagner}, {Elsner},
  {Everett}, {Fosalba}, {Gatti}, {Giannini}, {Gruen}, {Gruendl}, {Harrison},
  {Hartley}, {Jain}, {Jarvis}, {MacCrann}, {McCullough}, {Muir}, {Myles},
  {Pandey}, {Raveri}, {Roodman}, {Rodriguez-Monroy}, {Rykoff}, {Samuroff},
  {S{\'a}nchez}, {Secco}, {Sevilla-Noarbe}, {Sheldon}, {Troxel}, {Weaverdyck},
  {Yanny}, {Aguena}, {Avila}, {Bacon}, {Bertin}, {Bhargava}, {Brooks}, {Burke},
  {Carretero}, {Costanzi}, {da Costa}, {Pereira}, {De Vicente}, {Desai},
  {Evrard}, {Ferrero}, {Frieman}, {Garc{\'\i}a-Bellido}, {Gaztanaga}, {Gerdes},
  {Giannantonio}, {Gschwend}, {Gutierrez}, {Hinton}, {Hollowood}, {Honscheid},
  {James}, {Kuehn}, {Lahav}, {Lima}, {Maia}, {Menanteau}, {Miquel}, {Morgan},
  {Palmese}, {Paz-Chinch{\'o}n}, {Plazas}, {Sanchez}, {Scarpine}, {Serrano},
  {Soares-Santos}, {Smith}, {Suchyta}, {Tarle}, {Thomas}, {To}, {Varga},
  {Weller}, {Wilkinson}, {Wilkinson}, \& {DES Collaboration}}]{friedrich21}
{Friedrich}, O., {Andrade-Oliveira}, F., {Camacho}, H., {et~al.} 2021, \mnras,
  508, 3125

\bibitem[{{Gatti} {et~al.}(2021){Gatti}, {Sheldon}, {Amon}, {Becker}, {Troxel},
  {Choi}, {Doux}, {MacCrann}, {Navarro-Alsina}, {Harrison}, {Gruen},
  {Bernstein}, {Jarvis}, {Secco}, {Fert{\'e}}, {Shin}, {McCullough}, {Rollins},
  {Chen}, {Chang}, {Pandey}, {Tutusaus}, {Prat}, {Elvin-Poole}, {Sanchez},
  {Plazas}, {Roodman}, {Zuntz}, {Abbott}, {Aguena}, {Allam}, {Annis}, {Avila},
  {Bacon}, {Bertin}, {Bhargava}, {Brooks}, {Burke}, {Carnero Rosell}, {Carrasco
  Kind}, {Carretero}, {Castander}, {Conselice}, {Costanzi}, {Crocce}, {da
  Costa}, {Davis}, {De Vicente}, {Desai}, {Diehl}, {Dietrich}, {Doel},
  {Drlica-Wagner}, {Eckert}, {Everett}, {Ferrero}, {Frieman},
  {Garc{\'\i}a-Bellido}, {Gerdes}, {Giannantonio}, {Gruendl}, {Gschwend},
  {Gutierrez}, {Hartley}, {Hinton}, {Hollowood}, {Honscheid}, {Hoyle}, {Huff},
  {Huterer}, {Jain}, {James}, {Jeltema}, {Krause}, {Kron}, {Kuropatkin},
  {Lima}, {Maia}, {Marshall}, {Miquel}, {Morgan}, {Myles}, {Palmese},
  {Paz-Chinch{\'o}n}, {Rykoff}, {Samuroff}, {Sanchez}, {Scarpine}, {Schubnell},
  {Serrano}, {Sevilla-Noarbe}, {Smith}, {Soares-Santos}, {Suchyta}, {Swanson},
  {Tarle}, {Thomas}, {To}, {Tucker}, {Varga}, {Wechsler}, {Weller}, {Wester},
  \& {Wilkinson}}]{gatti21}
{Gatti}, M., {Sheldon}, E., {Amon}, A., {et~al.} 2021, \mnras, 504, 4312

\bibitem[{{Ghirardini} {et~al.}(A\&A,~subm.){Ghirardini}, {Esra}, {Artis},
  {Clerc}, \& {Garrel}}]{ghirardini23}
{Ghirardini}, V., {Esra}, B., {Artis}, A., {Clerc}, N., \& {Garrel}, C.
  A\&A,~subm.

\bibitem[{{Grandis} {et~al.}(2021{\natexlab{a}}){Grandis}, {Bocquet}, {Mohr},
  {Klein}, \& {Dolag}}]{grandis+21}
{Grandis}, S., {Bocquet}, S., {Mohr}, J.~J., {Klein}, M., \& {Dolag}, K.
  2021{\natexlab{a}}, \mnras, 507, 5671

\bibitem[{{Grandis} {et~al.}(2020){Grandis}, {Klein}, {Mohr}, {Bocquet},
  {Paulus}, {Abbott}, {Aguena}, {Allam}, {Annis}, {Benson}, {Bertin},
  {Bhargava}, {Brooks}, {Burke}, {Carnero Rosell}, {Carrasco Kind},
  {Carretero}, {Capasso}, {Costanzi}, {da Costa}, {De Vicente}, {Desai},
  {Dietrich}, {Doel}, {Eifler}, {Evrard}, {Flaugher}, {Fosalba}, {Frieman},
  {Garc{\'\i}a-Bellido}, {Gaztanaga}, {Gerdes}, {Gruen}, {Gruendl}, {Gschwend},
  {Gutierrez}, {Hartley}, {Hinton}, {Hollowood}, {Honscheid}, {James},
  {Jeltema}, {Kuehn}, {Kuropatkin}, {Lima}, {Maia}, {Marshall}, {Melchior},
  {Menanteau}, {Miquel}, {Ogando}, {Palmese}, {Paz-Chinch{\'o}n}, {Plazas},
  {Romer}, {Roodman}, {Sanchez}, {Saro}, {Scarpine}, {Schubnell}, {Serrano},
  {Sheldon}, {Smith}, {Stark}, {Suchyta}, {Swanson}, {Tarle}, {Thomas},
  {Tucker}, {Varga}, {Weller}, \& {Wilkinson}}]{grandis20}
{Grandis}, S., {Klein}, M., {Mohr}, J.~J., {et~al.} 2020, \mnras, 498, 771

\bibitem[{{Grandis} {et~al.}(2021{\natexlab{b}}){Grandis}, {Mohr}, {Costanzi},
  {Saro}, {Bocquet}, {Klein}, {Aguena}, {Allam}, {Annis}, {Ansarinejad},
  {Bacon}, {Bertin}, {Bleem}, {Brooks}, {Burke}, {Carnero Rosel}, {Carrasco
  Kind}, {Carretero}, {Castander}, {Choi}, {da Costa}, {De Vincente}, {Desai},
  {Diehl}, {Dietrich}, {Doel}, {Eifler}, {Everett}, {Ferrero}, {Floyd},
  {Fosalba}, {Frieman}, {Garc{\'\i}a-Bellido}, {Gaztanaga}, {Gruen}, {Gruendl},
  {Gschwend}, {Gupta}, {Gutierrez}, {Hinton}, {Hollowood}, {Honscheid},
  {James}, {Jeltema}, {Kuehn}, {Lahav}, {Lidman}, {Lima}, {Maia}, {March},
  {Marshall}, {Melchior}, {Menanteau}, {Miquel}, {Morgan}, {Myles}, {Ogando},
  {Palmese}, {Paz-Chinch{\'o}n}, {Plazas}, {Reichardt}, {Romer}, {Sanchez},
  {Scarpine}, {Serrano}, {Sevilla-Noarbe}, {Singh}, {Smith}, {Suchyta},
  {Swanson}, {Tarle}, {Thomas}, {To}, {Weller}, {Wilkinson}, \&
  {Wu}}]{grandis21b}
{Grandis}, S., {Mohr}, J.~J., {Costanzi}, M., {et~al.} 2021{\natexlab{b}},
  \mnras, 504, 1253

\bibitem[{{Grandis} {et~al.}(2019){Grandis}, {Mohr}, {Dietrich}, {Bocquet},
  {Saro}, {Klein}, {Paulus}, \& {Capasso}}]{grandis19}
{Grandis}, S., {Mohr}, J.~J., {Dietrich}, J.~P., {et~al.} 2019, \mnras, 488,
  2041

\bibitem[{{Gruen} {et~al.}(2014){Gruen}, {Seitz}, {Brimioulle}, {Kosyra},
  {Koppenhoefer}, {Lee}, {Bender}, {Riffeser}, {Eichner}, {Weidinger}, \&
  {Bierschenk}}]{gruen14}
{Gruen}, D., {Seitz}, S., {Brimioulle}, F., {et~al.} 2014, \mnras, 442, 1507

\bibitem[{{Haiman} {et~al.}(2001){Haiman}, {Mohr}, \& {Holder}}]{haiman01}
{Haiman}, Z., {Mohr}, J.~J., \& {Holder}, G.~P. 2001, \apj, 553, 545

\bibitem[{{Hennig} {et~al.}(2017){Hennig}, {Mohr}, {Zenteno}, {Desai},
  {Dietrich}, {Bocquet}, {Strazzullo}, {Saro}, {Abbott}, {Abdalla}, {Bayliss},
  {Benoit-L{\'e}vy}, {Bernstein}, {Bertin}, {Brooks}, {Capasso}, {Capozzi},
  {Carnero}, {Carrasco Kind}, {Carretero}, {Chiu}, {D'Andrea}, {daCosta},
  {Diehl}, {Doel}, {Eifler}, {Evrard}, {Fausti-Neto}, {Fosalba}, {Frieman},
  {Gangkofner}, {Gonzalez}, {Gruen}, {Gruendl}, {Gupta}, {Gutierrez},
  {Honscheid}, {Hlavacek-Larrondo}, {James}, {Kuehn}, {Kuropatkin}, {Lahav},
  {March}, {Marshall}, {Martini}, {McDonald}, {Melchior}, {Miller}, {Miquel},
  {Neilsen}, {Nord}, {Ogando}, {Plazas}, {Reichardt}, {Romer}, {Rozo},
  {Rykoff}, {Sanchez}, {Santiago}, {Schubnell}, {Sevilla-Noarbe}, {Smith},
  {Soares-Santos}, {Sobreira}, {Stalder}, {Stanford}, {Suchyta}, {Swanson},
  {Tarle}, {Thomas}, {Vikram}, {Walker}, \& {Zhang}}]{hennig17}
{Hennig}, C., {Mohr}, J.~J., {Zenteno}, A., {et~al.} 2017, \mnras, 467, 4015

\bibitem[{{Hern{\'a}ndez-Mart{\'\i}n}
  {et~al.}(2020){Hern{\'a}ndez-Mart{\'\i}n}, {Schrabback}, {Hoekstra},
  {Martinet}, {Hlavacek-Larrondo}, {Bleem}, {Gladders}, {Stalder}, {Stark}, \&
  {Bayliss}}]{hernandez-martin20}
{Hern{\'a}ndez-Mart{\'\i}n}, B., {Schrabback}, T., {Hoekstra}, H., {et~al.}
  2020, \aap, 640, A117

\bibitem[{{Hoekstra} {et~al.}(2012){Hoekstra}, {Mahdavi}, {Babul}, \&
  {Bildfell}}]{hoekstra12}
{Hoekstra}, H., {Mahdavi}, A., {Babul}, A., \& {Bildfell}, C. 2012, \mnras,
  427, 1298

\bibitem[{{Huff} \& {Mandelbaum}(2017)}]{huff&mandelbaum17}
{Huff}, E. \& {Mandelbaum}, R. 2017, arXiv e-prints, arXiv:1702.02600

\bibitem[{{Huterer}(2022)}]{huterer2022}
{Huterer}, D. 2022, arXiv e-prints, arXiv:2212.05003

\bibitem[{{Jarvis} {et~al.}(2021){Jarvis}, {Bernstein}, {Amon}, {Davis},
  {L{\'e}get}, {Bechtol}, {Harrison}, {Gatti}, {Roodman}, {Chang}, {Chen},
  {Choi}, {Desai}, {Drlica-Wagner}, {Gruen}, {Gruendl}, {Hernandez},
  {MacCrann}, {Meyers}, {Navarro-Alsina}, {Pandey}, {Plazas}, {Secco},
  {Sheldon}, {Troxel}, {Vorperian}, {Wei}, {Zuntz}, {Abbott}, {Aguena},
  {Allam}, {Avila}, {Bhargava}, {Bridle}, {Brooks}, {Carnero Rosell}, {Carrasco
  Kind}, {Carretero}, {Costanzi}, {da Costa}, {De Vicente}, {Diehl}, {Doel},
  {Everett}, {Flaugher}, {Fosalba}, {Frieman}, {Garc{\'\i}a-Bellido},
  {Gaztanaga}, {Gerdes}, {Gutierrez}, {Hinton}, {Hollowood}, {Honscheid},
  {James}, {Kent}, {Kuehn}, {Kuropatkin}, {Lahav}, {Maia}, {March}, {Marshall},
  {Melchior}, {Menanteau}, {Miquel}, {Ogando}, {Paz-Chinch{\'o}n}, {Rykoff},
  {Sanchez}, {Scarpine}, {Schubnell}, {Serrano}, {Sevilla-Noarbe}, {Smith},
  {Suchyta}, {Swanson}, {Tarle}, {Varga}, {Walker}, {Wester}, {Wilkinson}, \&
  {DES Collaboration}}]{jarvis21}
{Jarvis}, M., {Bernstein}, G.~M., {Amon}, A., {et~al.} 2021, \mnras, 501, 1282

\bibitem[{{Kimmig} {et~al.}(2023){Kimmig}, {Remus}, {Dolag}, \&
  {Biffi}}]{kimmig23}
{Kimmig}, L.~C., {Remus}, R.-S., {Dolag}, K., \& {Biffi}, V. 2023, \apj, 949,
  92

\bibitem[{{Kleinebreil} {et~al.}(in prep.){Kleinebreil}, {Grandis},
  {Schrabback}, \& {Ghirardini}}]{kleinebreil23}
{Kleinebreil}, F., {Grandis}, S., {Schrabback}, T., \& {Ghirardini}, V. in
  prep.

\bibitem[{{Kluge} {et~al.}(A\&A,~subm.){Kluge}, {Comparat}, et~{Liu}, {X}, \&
  {Y}}]{kluge23}
{Kluge}, M., {Comparat}, J., et~{Liu}, A., {X}, \& {Y}. A\&A,~subm.

\bibitem[{{Komatsu} {et~al.}(2011){Komatsu}, {Smith}, {Dunkley}, {Bennett},
  {Gold}, {Hinshaw}, {Jarosik}, {Larson}, {Nolta}, {Page}, {Spergel},
  {Halpern}, {Hill}, {Kogut}, {Limon}, {Meyer}, {Odegard}, {Tucker}, {Weiland},
  {Wollack}, \& {Wright}}]{komatsu11}
{Komatsu}, E., {Smith}, K.~M., {Dunkley}, J., {et~al.} 2011, \apjs, 192, 18

\bibitem[{{MacCrann} {et~al.}(2022){MacCrann}, {Becker}, {McCullough}, {Amon},
  {Gruen}, {Jarvis}, {Choi}, {Troxel}, {Sheldon}, {Yanny}, {Herner},
  {Dodelson}, {Zuntz}, {Eckert}, {Rollins}, {Varga}, {Bernstein}, {Gruendl},
  {Harrison}, {Hartley}, {Sevilla-Noarbe}, {Pieres}, {Bridle}, {Myles},
  {Alarcon}, {Everett}, {S{\'a}nchez}, {Huff}, {Tarsitano}, {Gatti}, {Secco},
  {Abbott}, {Aguena}, {Allam}, {Annis}, {Bacon}, {Bertin}, {Brooks}, {Burke},
  {Carnero Rosell}, {Carrasco Kind}, {Carretero}, {Costanzi}, {Crocce},
  {Pereira}, {De Vicente}, {Desai}, {Diehl}, {Dietrich}, {Doel}, {Eifler},
  {Ferrero}, {Fert{\'e}}, {Flaugher}, {Fosalba}, {Frieman},
  {Garc{\'\i}a-Bellido}, {Gaztanaga}, {Gerdes}, {Giannantonio}, {Gschwend},
  {Gutierrez}, {Hinton}, {Hollowood}, {Honscheid}, {James}, {Lahav}, {Lima},
  {Maia}, {March}, {Marshall}, {Martini}, {Melchior}, {Menanteau}, {Miquel},
  {Mohr}, {Morgan}, {Muir}, {Ogando}, {Palmese}, {Paz-Chinch{\'o}n}, {Plazas},
  {Rodriguez-Monroy}, {Roodman}, {Samuroff}, {Sanchez}, {Scarpine}, {Serrano},
  {Smith}, {Soares-Santos}, {Suchyta}, {Swanson}, {Tarle}, {Thomas}, {To},
  {Wilkinson}, {Wilkinson}, \& {DES Collaboration}}]{maccrann22}
{MacCrann}, N., {Becker}, M.~R., {McCullough}, J., {et~al.} 2022, \mnras, 509,
  3371

\bibitem[{{Majumdar} \& {Mohr}(2004)}]{majumdar04}
{Majumdar}, S. \& {Mohr}, J.~J. 2004, \apj, 613, 41

\bibitem[{{Mantz} {et~al.}(2016){Mantz}, {Allen}, {Morris}, {von der Linden},
  {Applegate}, {Kelly}, {Burke}, {Donovan}, \& {Ebeling}}]{mantz16}
{Mantz}, A.~B., {Allen}, S.~W., {Morris}, R.~G., {et~al.} 2016, \mnras, 463,
  3582

\bibitem[{{Mantz} {et~al.}(2015){Mantz}, {von der Linden}, {Allen},
  {Applegate}, {Kelly}, {Morris}, {Rapetti}, {Schmidt}, {Adhikari}, {Allen},
  {Burchat}, {Burke}, {Cataneo}, {Donovan}, {Ebeling}, {Shandera}, \&
  {Wright}}]{mantz15}
{Mantz}, A.~B., {von der Linden}, A., {Allen}, S.~W., {et~al.} 2015, \mnras,
  446, 2205

\bibitem[{{Marinacci} {et~al.}(2018){Marinacci}, {Vogelsberger}, {Pakmor},
  {Torrey}, {Springel}, {Hernquist}, {Nelson}, {Weinberger}, {Pillepich},
  {Naiman}, \& {Genel}}]{Marinacci2018MNRAS.480.5113M}
{Marinacci}, F., {Vogelsberger}, M., {Pakmor}, R., {et~al.} 2018, \mnras, 480,
  5113

\bibitem[{{McClintock} {et~al.}(2019){McClintock}, {Varga}, {Gruen}, {Rozo},
  {Rykoff}, {Shin}, {Melchior}, {DeRose}, {Seitz}, {Dietrich}, {Sheldon},
  {Zhang}, {von der Linden}, {Jeltema}, {Mantz}, {Romer}, {Allen}, {Becker},
  {Bermeo}, {Bhargava}, {Costanzi}, {Everett}, {Farahi}, {Hamaus}, {Hartley},
  {Hollowood}, {Hoyle}, {Israel}, {Li}, {MacCrann}, {Morris}, {Palmese},
  {Plazas}, {Pollina}, {Rau}, {Simet}, {Soares-Santos}, {Troxel}, {Vergara
  Cervantes}, {Wechsler}, {Zuntz}, {Abbott}, {Abdalla}, {Allam}, {Annis},
  {Avila}, {Bridle}, {Brooks}, {Burke}, {Carnero Rosell}, {Carrasco Kind},
  {Carretero}, {Castander}, {Crocce}, {Cunha}, {D'Andrea}, {da Costa}, {Davis},
  {De Vicente}, {Diehl}, {Doel}, {Drlica-Wagner}, {Evrard}, {Flaugher},
  {Fosalba}, {Frieman}, {Garc{\'\i}a-Bellido}, {Gaztanaga}, {Gerdes},
  {Giannantonio}, {Gruendl}, {Gutierrez}, {Honscheid}, {James}, {Kirk},
  {Krause}, {Kuehn}, {Lahav}, {Li}, {Lima}, {March}, {Marshall}, {Menanteau},
  {Miquel}, {Mohr}, {Nord}, {Ogando}, {Roodman}, {Sanchez}, {Scarpine},
  {Schindler}, {Sevilla-Noarbe}, {Smith}, {Smith}, {Sobreira}, {Suchyta},
  {Swanson}, {Tarle}, {Tucker}, {Vikram}, {Walker}, {Weller}, \& {DES
  Collaboration}}]{McClintock19}
{McClintock}, T., {Varga}, T.~N., {Gruen}, D., {et~al.} 2019, \mnras, 482, 1352

\bibitem[{{Merloni} {et~al.}(2024){Merloni}, {Lamer}, {Liu}, {Ramos-Ceja},
  {Brunner}, {Bulbul}, {Dennerl}, {Doroshenko}, {Freyberg}, {Friedrich},
  {Gatuzz}, {Georgakakis}, {Haberl}, {Igo}, {Kreykenbohm}, {Liu}, {Maitra},
  {Malyali}, {Mayer}, {Nandra}, {Predehl}, {Robrade}, {Salvato}, {Sanders},
  {Stewart}, {Tub{\'\i}n-Arenas}, {Weber}, {Wilms}, {Arcodia}, {Artis},
  {Aschersleben}, {Avakyan}, {Aydar}, {Bahar}, {Balzer}, {Becker}, {Berger},
  {Boller}, {Bornemann}, {Br{\"u}ggen}, {Brusa}, {Buchner}, {Burwitz},
  {Camilloni}, {Clerc}, {Comparat}, {Coutinho}, {Czesla}, {Dannhauer},
  {Dauner}, {Dauser}, {Dietl}, {Dolag}, {Dwelly}, {Egg}, {Ehl}, {Freund},
  {Friedrich}, {Gaida}, {Garrel}, {Ghirardini}, {Gokus}, {Gr{\"u}nwald},
  {Grandis}, {Grotova}, {Gruen}, {Gueguen}, {H{\"a}mmerich}, {Hamaus},
  {Hasinger}, {Haubner}, {Homan}, {Ider Chitham}, {Joseph}, {Joyce},
  {K{\"o}nig}, {Kaltenbrunner}, {Khokhriakova}, {Kink}, {Kirsch}, {Kluge},
  {Knies}, {Krippendorf}, {Krumpe}, {Kurpas}, {Li}, {Liu}, {Locatelli},
  {Lorenz}, {M{\"u}ller}, {Magaudda}, {Mannes}, {McCall}, {Meidinger},
  {Michailidis}, {Migkas}, {Mu{\~n}oz-Giraldo}, {Musiimenta}, {Nguyen-Dang},
  {Ni}, {Olechowska}, {Ota}, {Pacaud}, {Pasini}, {Perinati}, {Pires},
  {Pommranz}, {Ponti}, {Poppenhaeger}, {P{\"u}hlhofer}, {Rau}, {Reh},
  {Reiprich}, {Roster}, {Saeedi}, {Santangelo}, {Sasaki}, {Schmitt},
  {Schneider}, {Schrabback}, {Schuster}, {Schwope}, {Seppi}, {Serim},
  {Shreeram}, {Sokolova-Lapa}, {Starck}, {Stelzer}, {Stierhof}, {Suleimanov},
  {Tenzer}, {Traulsen}, {Tr{\"u}mper}, {Tsuge}, {Urrutia}, {Veronica},
  {Waddell}, {Willer}, {Wolf}, {Yeung}, {Zainab}, {Zangrandi}, {Zhang},
  {Zhang}, \& {Zheng}}]{merloni2023}
{Merloni}, A., {Lamer}, G., {Liu}, T., {et~al.} 2024, \aap, 682, A34

\bibitem[{{Mohr} \& {Evrard}(1997)}]{mohr97}
{Mohr}, J.~J. \& {Evrard}, A.~E. 1997, \apj, 491, 38

\bibitem[{{Myles} {et~al.}(2021){Myles}, {Alarcon}, {Amon}, {S{\'a}nchez},
  {Everett}, {DeRose}, {McCullough}, {Gruen}, {Bernstein}, {Troxel},
  {Dodelson}, {Campos}, {MacCrann}, {Yin}, {Raveri}, {Amara}, {Becker}, {Choi},
  {Cordero}, {Eckert}, {Gatti}, {Giannini}, {Gschwend}, {Gruendl}, {Harrison},
  {Hartley}, {Huff}, {Kuropatkin}, {Lin}, {Masters}, {Miquel}, {Prat},
  {Roodman}, {Rykoff}, {Sevilla-Noarbe}, {Sheldon}, {Wechsler}, {Yanny},
  {Abbott}, {Aguena}, {Allam}, {Annis}, {Bacon}, {Bertin}, {Bhargava},
  {Bridle}, {Brooks}, {Burke}, {Carnero Rosell}, {Carrasco Kind}, {Carretero},
  {Castander}, {Conselice}, {Costanzi}, {Crocce}, {da Costa}, {Pereira},
  {Desai}, {Diehl}, {Eifler}, {Elvin-Poole}, {Evrard}, {Ferrero}, {Fert{\'e}},
  {Flaugher}, {Fosalba}, {Frieman}, {Garc{\'\i}a-Bellido}, {Gaztanaga},
  {Giannantonio}, {Hinton}, {Hollowood}, {Honscheid}, {Hoyle}, {Huterer},
  {James}, {Krause}, {Kuehn}, {Lahav}, {Lima}, {Maia}, {Marshall}, {Martini},
  {Melchior}, {Menanteau}, {Mohr}, {Morgan}, {Muir}, {Ogando}, {Palmese},
  {Paz-Chinch{\'o}n}, {Plazas}, {Rodriguez-Monroy}, {Samuroff}, {Sanchez},
  {Scarpine}, {Secco}, {Serrano}, {Smith}, {Soares-Santos}, {Suchyta},
  {Swanson}, {Tarle}, {Thomas}, {To}, {Varga}, {Weller}, \& {Wester}}]{myles21}
{Myles}, J., {Alarcon}, A., {Amon}, A., {et~al.} 2021, \mnras, 505, 4249

\bibitem[{{Naiman} {et~al.}(2018){Naiman}, {Pillepich}, {Springel},
  {Ramirez-Ruiz}, {Torrey}, {Vogelsberger}, {Pakmor}, {Nelson}, {Marinacci},
  {Hernquist}, {Weinberger}, \& {Genel}}]{Naiman2018MNRAS.477.1206N}
{Naiman}, J.~P., {Pillepich}, A., {Springel}, V., {et~al.} 2018, \mnras, 477,
  1206

\bibitem[{{Navarro} {et~al.}(1996){Navarro}, {Frenk}, \& {White}}]{NFW}
{Navarro}, J.~F., {Frenk}, C.~S., \& {White}, S. D.~M. 1996, \apj, 462, 563

\bibitem[{{Nelson} {et~al.}(2018){Nelson}, {Pillepich}, {Springel},
  {Weinberger}, {Hernquist}, {Pakmor}, {Genel}, {Torrey}, {Vogelsberger},
  {Kauffmann}, {Marinacci}, \& {Naiman}}]{Nelson2018MNRAS.475..624N}
{Nelson}, D., {Pillepich}, A., {Springel}, V., {et~al.} 2018, \mnras, 475, 624

\bibitem[{{Nelson} {et~al.}(2019){Nelson}, {Springel}, {Pillepich},
  {Rodriguez-Gomez}, {Torrey}, {Genel}, {Vogelsberger}, {Pakmor}, {Marinacci},
  {Weinberger}, {Kelley}, {Lovell}, {Diemer}, \&
  {Hernquist}}]{Nelson2019ComAC...6....2N}
{Nelson}, D., {Springel}, V., {Pillepich}, A., {et~al.} 2019, Computational
  Astrophysics and Cosmology, 6, 2

\bibitem[{{Oguri} \& {Hamana}(2011)}]{oguri+11}
{Oguri}, M. \& {Hamana}, T. 2011, \mnras, 414, 1851

\bibitem[{{Paulus}(2021)}]{paulus21}
{Paulus}, M. 2021, PhD thesis, Ludwig-Maximilians University of Munich, Germany

\bibitem[{{Pillepich} {et~al.}(2018){Pillepich}, {Nelson}, {Hernquist},
  {Springel}, {Pakmor}, {Torrey}, {Weinberger}, {Genel}, {Naiman}, {Marinacci},
  \& {Vogelsberger}}]{Pillepich2018MNRAS.475..648P}
{Pillepich}, A., {Nelson}, D., {Hernquist}, L., {et~al.} 2018, \mnras, 475, 648

\bibitem[{{Popesso} {et~al.}(2023){Popesso}, {Biviano}, {Bulbul}, {Merloni},
  {Comparat}, {Clerc}, {Igo}, {Liu}, {Driver}, {Salvato}, {Brusa}, {Bahar},
  {Malavasi}, {Ghirardini}, {Ponti}, {Robotham}, {Liske}, \&
  {Grandis}}]{popesso+23}
{Popesso}, P., {Biviano}, A., {Bulbul}, E., {et~al.} 2023, arXiv e-prints,
  arXiv:2302.08405

\bibitem[{{Power} {et~al.}(2003){Power}, {Navarro}, {Jenkins}, {Frenk},
  {White}, {Springel}, {Stadel}, \& {Quinn}}]{power03}
{Power}, C., {Navarro}, J.~F., {Jenkins}, A., {et~al.} 2003, \mnras, 338, 14

\bibitem[{{Prat} {et~al.}(2022){Prat}, {Blazek}, {S{\'a}nchez}, {Tutusaus},
  {Pandey}, {Elvin-Poole}, {Krause}, {Troxel}, {Secco}, {Amon}, {DeRose},
  {Zacharegkas}, {Chang}, {Jain}, {MacCrann}, {Park}, {Sheldon}, {Giannini},
  {Bocquet}, {To}, {Alarcon}, {Alves}, {Andrade-Oliveira}, {Baxter}, {Bechtol},
  {Becker}, {Bernstein}, {Camacho}, {Campos}, {Carnero Rosell}, {Carrasco
  Kind}, {Cawthon}, {Chen}, {Choi}, {Cordero}, {Crocce}, {Davis}, {De Vicente},
  {Diehl}, {Dodelson}, {Doux}, {Drlica-Wagner}, {Eckert}, {Eifler}, {Elsner},
  {Everett}, {Fang}, {Farahi}, {Fert{\'e}}, {Fosalba}, {Friedrich}, {Gatti},
  {Gruen}, {Gruendl}, {Harrison}, {Hartley}, {Herner}, {Huang}, {Huff},
  {Huterer}, {Jarvis}, {Kuropatkin}, {Leget}, {Lemos}, {Liddle}, {McCullough},
  {Muir}, {Myles}, {Navarro-Alsina}, {Porredon}, {Raveri}, {Rodriguez-Monroy},
  {Rollins}, {Roodman}, {Rosenfeld}, {Ross}, {Rykoff}, {Sanchez},
  {Sevilla-Noarbe}, {Shin}, {Troja}, {Varga}, {Weaverdyck}, {Wechsler},
  {Yanny}, {Yin}, {Zuntz}, {Abbott}, {Aguena}, {Allam}, {Annis}, {Bacon},
  {Brooks}, {Burke}, {Carretero}, {Conselice}, {Costanzi}, {da Costa},
  {Pereira}, {Desai}, {Dietrich}, {Doel}, {Evrard}, {Ferrero}, {Flaugher},
  {Frieman}, {Garc{\'\i}a-Bellido}, {Gaztanaga}, {Gerdes}, {Giannantonio},
  {Gschwend}, {Gutierrez}, {Hinton}, {Hollowood}, {Honscheid}, {James},
  {Kuehn}, {Lahav}, {Lin}, {Maia}, {Marshall}, {Martini}, {Melchior},
  {Menanteau}, {Miller}, {Miquel}, {Mohr}, {Morgan}, {Ogando}, {Palmese},
  {Paz-Chinch{\'o}n}, {Petravick}, {Plazas Malag{\'o}n}, {Sanchez}, {Serrano},
  {Smith}, {Soares-Santos}, {Suchyta}, {Tarle}, {Thomas}, {Weller}, \& {DES
  Collaboration}}]{prat22}
{Prat}, J., {Blazek}, J., {S{\'a}nchez}, C., {et~al.} 2022, \prd, 105, 083528

\bibitem[{{Predehl} {et~al.}(2021){Predehl}, {Andritschke}, {Arefiev},
  {Babyshkin}, {Batanov}, {Becker}, {B{\"o}hringer}, {Bogomolov}, {Boller},
  {Borm}, {Bornemann}, {Br{\"a}uninger}, {Br{\"u}ggen}, {Brunner}, {Brusa},
  {Bulbul}, {Buntov}, {Burwitz}, {Burkert}, {Clerc}, {Churazov}, {Coutinho},
  {Dauser}, {Dennerl}, {Doroshenko}, {Eder}, {Emberger}, {Eraerds},
  {Finoguenov}, {Freyberg}, {Friedrich}, {Friedrich}, {F{\"u}rmetz},
  {Georgakakis}, {Gilfanov}, {Granato}, {Grossberger}, {Gueguen}, {Gureev},
  {Haberl}, {H{\"a}lker}, {Hartner}, {Hasinger}, {Huber}, {Ji}, {Kienlin},
  {Kink}, {Korotkov}, {Kreykenbohm}, {Lamer}, {Lomakin}, {Lapshov}, {Liu},
  {Maitra}, {Meidinger}, {Menz}, {Merloni}, {Mernik}, {Mican}, {Mohr},
  {M{\"u}ller}, {Nandra}, {Nazarov}, {Pacaud}, {Pavlinsky}, {Perinati},
  {Pfeffermann}, {Pietschner}, {Ramos-Ceja}, {Rau}, {Reiffers}, {Reiprich},
  {Robrade}, {Salvato}, {Sanders}, {Santangelo}, {Sasaki}, {Scheuerle},
  {Schmid}, {Schmitt}, {Schwope}, {Shirshakov}, {Steinmetz}, {Stewart},
  {Str{\"u}der}, {Sunyaev}, {Tenzer}, {Tiedemann}, {Tr{\"u}mper}, {Voron},
  {Weber}, {Wilms}, \& {Yaroshenko}}]{Predehl2021}
{Predehl}, P., {Andritschke}, R., {Arefiev}, V., {et~al.} 2021, \aap, 647, A1

\bibitem[{{Ragagnin} {et~al.}(2021){Ragagnin}, {Saro}, {Singh}, \&
  {Dolag}}]{ragagnin21}
{Ragagnin}, A., {Saro}, A., {Singh}, P., \& {Dolag}, K. 2021, \mnras, 500, 5056

\bibitem[{{Rossetti} {et~al.}(2017){Rossetti}, {Gastaldello}, {Eckert}, {Della
  Torre}, {Pantiri}, {Cazzoletti}, \& {Molendi}}]{rossetti17}
{Rossetti}, M., {Gastaldello}, F., {Eckert}, D., {et~al.} 2017, \mnras, 468,
  1917

\bibitem[{{Rozo} {et~al.}(2015){Rozo}, {Rykoff}, {Becker}, {Reddick}, \&
  {Wechsler}}]{rozo15}
{Rozo}, E., {Rykoff}, E.~S., {Becker}, M., {Reddick}, R.~M., \& {Wechsler},
  R.~H. 2015, \mnras, 453, 38

\bibitem[{{Rykoff} {et~al.}(2014){Rykoff}, {Rozo}, {Busha}, {Cunha},
  {Finoguenov}, {Evrard}, {Hao}, {Koester}, {Leauthaud}, {Nord}, {Pierre},
  {Reddick}, {Sadibekova}, {Sheldon}, \& {Wechsler}}]{rykoff14}
{Rykoff}, E.~S., {Rozo}, E., {Busha}, M.~T., {et~al.} 2014, \apj, 785, 104

\bibitem[{{Rykoff} {et~al.}(2016){Rykoff}, {Rozo}, {Hollowood},
  {Bermeo-Hernandez}, {Jeltema}, {Mayers}, {Romer}, {Rooney}, {Saro}, {Vergara
  Cervantes}, {Wechsler}, {Wilcox}, {Abbott}, {Abdalla}, {Allam}, {Annis},
  {Benoit-L{\'e}vy}, {Bernstein}, {Bertin}, {Brooks}, {Burke}, {Capozzi},
  {Carnero Rosell}, {Carrasco Kind}, {Castander}, {Childress}, {Collins},
  {Cunha}, {D'Andrea}, {da Costa}, {Davis}, {Desai}, {Diehl}, {Dietrich},
  {Doel}, {Evrard}, {Finley}, {Flaugher}, {Fosalba}, {Frieman}, {Glazebrook},
  {Goldstein}, {Gruen}, {Gruendl}, {Gutierrez}, {Hilton}, {Honscheid}, {Hoyle},
  {James}, {Kay}, {Kuehn}, {Kuropatkin}, {Lahav}, {Lewis}, {Lidman}, {Lima},
  {Maia}, {Mann}, {Marshall}, {Martini}, {Melchior}, {Miller}, {Miquel},
  {Mohr}, {Nichol}, {Nord}, {Ogando}, {Plazas}, {Reil}, {Sahl{\'e}n},
  {Sanchez}, {Santiago}, {Scarpine}, {Schubnell}, {Sevilla-Noarbe}, {Smith},
  {Soares-Santos}, {Sobreira}, {Stott}, {Suchyta}, {Swanson}, {Tarle},
  {Thomas}, {Tucker}, {Uddin}, {Viana}, {Vikram}, {Walker}, {Zhang}, \& {DES
  Collaboration}}]{rykoff16}
{Rykoff}, E.~S., {Rozo}, E., {Hollowood}, D., {et~al.} 2016, \apjs, 224, 1

\bibitem[{{Saro} {et~al.}(2015){Saro}, {Bocquet}, {Rozo}, {Benson}, {Mohr},
  {Rykoff}, {Soares-Santos}, {Bleem}, {Dodelson}, {Melchior}, {Sobreira},
  {Upadhyay}, {Weller}, {Abbott}, {Abdalla}, {Allam}, {Armstrong}, {Banerji},
  {Bauer}, {Bayliss}, {Benoit-L{\'e}vy}, {Bernstein}, {Bertin}, {Brodwin},
  {Brooks}, {Buckley-Geer}, {Burke}, {Carlstrom}, {Capasso}, {Capozzi},
  {Carnero Rosell}, {Carrasco Kind}, {Chiu}, {Covarrubias}, {Crawford},
  {Crocce}, {D'Andrea}, {da Costa}, {DePoy}, {Desai}, {de Haan}, {Diehl},
  {Dietrich}, {Doel}, {Cunha}, {Eifler}, {Evrard}, {Fausti Neto}, {Fernandez},
  {Flaugher}, {Fosalba}, {Frieman}, {Gangkofner}, {Gaztanaga}, {Gerdes},
  {Gruen}, {Gruendl}, {Gupta}, {Hennig}, {Holzapfel}, {Honscheid}, {Jain},
  {James}, {Kuehn}, {Kuropatkin}, {Lahav}, {Li}, {Lin}, {Maia}, {March},
  {Marshall}, {Martini}, {McDonald}, {Miller}, {Miquel}, {Nord}, {Ogando},
  {Plazas}, {Reichardt}, {Romer}, {Roodman}, {Sako}, {Sanchez}, {Schubnell},
  {Sevilla}, {Smith}, {Stalder}, {Stark}, {Strazzullo}, {Suchyta}, {Swanson},
  {Tarle}, {Thaler}, {Thomas}, {Tucker}, {Vikram}, {von der Linden}, {Walker},
  {Wechsler}, {Wester}, {Zenteno}, \& {Ziegler}}]{saro15}
{Saro}, A., {Bocquet}, S., {Rozo}, E., {et~al.} 2015, \mnras, 454, 2305

\bibitem[{{Schneider}(2006)}]{schneider05}
{Schneider}, P. 2006, in Saas-Fee Advanced Course 33: Gravitational Lensing:
  Strong, Weak and Micro, ed. G.~{Meylan}, P.~{Jetzer}, P.~{North},
  P.~{Schneider}, C.~S. {Kochanek}, \& J.~{Wambsganss}, 269--451

\bibitem[{{Schrabback} {et~al.}(2021){Schrabback}, {Bocquet}, {Sommer},
  {Zohren}, {van den Busch}, {Hern{\'a}ndez-Mart{\'\i}n}, {Hoekstra}, {Raihan},
  {Schirmer}, {Applegate}, {Bayliss}, {Benson}, {Bleem}, {Dietrich}, {Floyd},
  {Hilbert}, {Hlavacek-Larrondo}, {McDonald}, {Saro}, {Stark}, \&
  {Weissgerber}}]{schrabback21}
{Schrabback}, T., {Bocquet}, S., {Sommer}, M., {et~al.} 2021, \mnras, 505, 3923

\bibitem[{{Seppi} {et~al.}(2022){Seppi}, {Comparat}, {Bulbul}, {Nandra},
  {Merloni}, {Clerc}, {Liu}, {Ghirardini}, {Liu}, {Salvato}, {Sanders},
  {Wilms}, {Dwelly}, {Dauser}, {K{\"o}nig}, {Ramos-Ceja}, {Garrel}, \&
  {Reiprich}}]{seppi22}
{Seppi}, R., {Comparat}, J., {Bulbul}, E., {et~al.} 2022, \aap, 665, A78

\bibitem[{{Sevilla-Noarbe} {et~al.}(2021){Sevilla-Noarbe}, {Bechtol}, {Carrasco
  Kind}, {Carnero Rosell}, {Becker}, {Drlica-Wagner}, {Gruendl}, {Rykoff},
  {Sheldon}, {Yanny}, {Alarcon}, {Allam}, {Amon}, {Benoit-L{\'e}vy},
  {Bernstein}, {Bertin}, {Burke}, {Carretero}, {Choi}, {Diehl}, {Everett},
  {Flaugher}, {Gaztanaga}, {Gschwend}, {Harrison}, {Hartley}, {Hoyle},
  {Jarvis}, {Johnson}, {Kessler}, {Kron}, {Kuropatkin}, {Leistedt}, {Li},
  {Menanteau}, {Morganson}, {Ogando}, {Palmese}, {Paz-Chinch{\'o}n}, {Pieres},
  {Pond}, {Rodriguez-Monroy}, {Smith}, {Stringer}, {Troxel}, {Tucker}, {de
  Vicente}, {Wester}, {Zhang}, {Abbott}, {Aguena}, {Annis}, {Avila},
  {Bhargava}, {Bridle}, {Brooks}, {Brout}, {Castander}, {Cawthon}, {Chang},
  {Conselice}, {Costanzi}, {Crocce}, {da Costa}, {Pereira}, {Davis}, {Desai},
  {Dietrich}, {Doel}, {Eckert}, {Evrard}, {Ferrero}, {Fosalba},
  {Garc{\'\i}a-Bellido}, {Gerdes}, {Giannantonio}, {Gruen}, {Gutierrez},
  {Hinton}, {Hollowood}, {Honscheid}, {Huff}, {Huterer}, {James}, {Jeltema},
  {Kuehn}, {Lahav}, {Lidman}, {Lima}, {Lin}, {Maia}, {Marshall}, {Martini},
  {Melchior}, {Miquel}, {Mohr}, {Morgan}, {Neilsen}, {Plazas}, {Romer},
  {Roodman}, {Sanchez}, {Scarpine}, {Schubnell}, {Serrano}, {Smith}, {Suchyta},
  {Tarle}, {Thomas}, {To}, {Varga}, {Wechsler}, {Weller}, {Wilkinson}, \& {DES
  Collaboration}}]{sevilla-noarbe21}
{Sevilla-Noarbe}, I., {Bechtol}, K., {Carrasco Kind}, M., {et~al.} 2021, \apjs,
  254, 24

\bibitem[{{Sheldon} \& {Huff}(2017)}]{sheldon&huff17}
{Sheldon}, E.~S. \& {Huff}, E.~M. 2017, \apj, 841, 24

\bibitem[{{Shin} {et~al.}(2021){Shin}, {Jain}, {Adhikari}, {Baxter}, {Chang},
  {Pandey}, {Salcedo}, {Weinberg}, {Amsellem}, {Battaglia}, {Belyakov},
  {Dacunha}, {Goldstein}, {Kravtsov}, {Varga}, {Abbott}, {Aguena}, {Alarcon},
  {Allam}, {Amon}, {Andrade-Oliveira}, {Annis}, {Bacon}, {Bechtol}, {Becker},
  {Bernstein}, {Bertin}, {Bocquet}, {Bond}, {Brooks}, {Buckley-Geer}, {Burke},
  {Campos}, {Rosell}, {Kind}, {Carretero}, {Chen}, {Choi}, {Costanzi}, {da
  Costa}, {DeRose}, {Desai}, {De Vicente}, {Devlin}, {Diehl}, {Dietrich},
  {Dodelson}, {Doel}, {Doux}, {Drlica-Wagner}, {Eckert}, {Elvin-Poole},
  {Everett}, {Ferraro}, {Ferrero}, {Fert{\'e}}, {Flaugher}, {Frieman},
  {Gallardo}, {Gatti}, {Gaztanaga}, {Gerdes}, {Gruen}, {Gruendl}, {Gutierrez},
  {Harrison}, {Hartley}, {Hill}, {Hilton}, {Hinton}, {Hollowood}, {Hughes},
  {James}, {Jarvis}, {Jeltema}, {Koopman}, {Krause}, {Kuehn}, {Kuropatkin},
  {Lahav}, {Lima}, {Lokken}, {MacCrann}, {Madhavacheril}, {Maia}, {McCullough},
  {McMahon}, {Melchior}, {Menanteau}, {Miquel}, {Mohr}, {Moodley}, {Morgan},
  {Myles}, {Nati}, {Navarro-Alsina}, {Niemack}, {Ogando}, {Page}, {Palmese},
  {Partridge}, {Paz-Chinch{\'o}n}, {Pereira}, {Pieres}, {Malag{\'o}n}, {Prat},
  {Raveri}, {Rodriguez-Monroy}, {Rollins}, {Romer}, {Rykoff}, {Salatino},
  {S{\'a}nchez}, {Sanchez}, {Santiago}, {Scarpine}, {Schillaci}, {Secco},
  {Serrano}, {Sevilla-Noarbe}, {Sheldon}, {Sherwin}, {Sif{\'o}n}, {Smith},
  {Soares-Santos}, {Staggs}, {Suchyta}, {Swanson}, {Tarle}, {Thomas}, {To},
  {Troxel}, {Tutusaus}, {Vavagiakis}, {Weller}, {Wollack}, {Yanny}, {Yin}, \&
  {Zhang}}]{shin21}
{Shin}, T., {Jain}, B., {Adhikari}, S., {et~al.} 2021, \mnras, 507, 5758

\bibitem[{{Sommer} {et~al.}(2023){Sommer}, {Schrabback}, {Ragagnin}, \&
  {Rockenfeller}}]{sommer23}
{Sommer}, M.~W., {Schrabback}, T., {Ragagnin}, A., \& {Rockenfeller}, R. 2023,
  arXiv e-prints, arXiv:2306.13187

\bibitem[{{Song} {et~al.}(2012){Song}, {Zenteno}, {Stalder}, {Desai}, {Bleem},
  {Aird}, {Armstrong}, {Ashby}, {Bayliss}, {Bazin}, {Benson}, {Bertin},
  {Brodwin}, {Carlstrom}, {Chang}, {Cho}, {Clocchiatti}, {Crawford}, {Crites},
  {de Haan}, {Dobbs}, {Dudley}, {Foley}, {George}, {Gettings}, {Gladders},
  {Gonzalez}, {Halverson}, {Harrington}, {High}, {Holder}, {Holzapfel},
  {Hoover}, {Hrubes}, {Joy}, {Keisler}, {Knox}, {Lee}, {Leitch}, {Liu},
  {Lueker}, {Luong-Van}, {Marrone}, {McDonald}, {McMahon}, {Mehl}, {Meyer},
  {Mocanu}, {Mohr}, {Montroy}, {Natoli}, {Nurgaliev}, {Padin}, {Plagge},
  {Pryke}, {Reichardt}, {Rest}, {Ruel}, {Ruhl}, {Saliwanchik}, {Saro}, {Sayre},
  {Schaffer}, {Shaw}, {Shirokoff}, {{\v{S}}uhada}, {Spieler}, {Stanford},
  {Staniszewski}, {Stark}, {Story}, {Stubbs}, {van Engelen}, {Vanderlinde},
  {Vieira}, {Williamson}, \& {Zahn}}]{song+12}
{Song}, J., {Zenteno}, A., {Stalder}, B., {et~al.} 2012, \apj, 761, 22

\bibitem[{{Springel} {et~al.}(2018){Springel}, {Pakmor}, {Pillepich},
  {Weinberger}, {Nelson}, {Hernquist}, {Vogelsberger}, {Genel}, {Torrey},
  {Marinacci}, \& {Naiman}}]{Springel2018MNRAS.475..676S}
{Springel}, V., {Pakmor}, R., {Pillepich}, A., {et~al.} 2018, \mnras, 475, 676

\bibitem[{{Story} {et~al.}(2011){Story}, {Aird}, {Andersson}, {Armstrong},
  {Bazin}, {Benson}, {Bleem}, {Bonamente}, {Brodwin}, {Carlstrom}, {Chang},
  {Clocchiatti}, {Crawford}, {Crites}, {de Haan}, {Desai}, {Dobbs}, {Dudley},
  {Foley}, {George}, {Gladders}, {Gonzalez}, {Halverson}, {High}, {Holder},
  {Holzapfel}, {Hoover}, {Hrubes}, {Joy}, {Keisler}, {Knox}, {Lee}, {Leitch},
  {Lueker}, {Luong-Van}, {Marrone}, {McMahon}, {Mehl}, {Meyer}, {Mohr},
  {Montroy}, {Padin}, {Plagge}, {Pryke}, {Reichardt}, {Rest}, {Ruel}, {Ruhl},
  {Saliwanchik}, {Saro}, {Schaffer}, {Shaw}, {Shirokoff}, {Song}, {Spieler},
  {Stalder}, {Staniszewski}, {Stark}, {Stubbs}, {Vanderlinde}, {Vieira},
  {Williamson}, \& {Zenteno}}]{story+11}
{Story}, K., {Aird}, K.~A., {Andersson}, K., {et~al.} 2011, \apjl, 735, L36

\bibitem[{{Sunayama}(2023)}]{sunayama23}
{Sunayama}, T. 2023, \mnras, 521, 5064

\bibitem[{{Sunayama} {et~al.}(2020){Sunayama}, {Park}, {Takada}, {Kobayashi},
  {Nishimichi}, {Kurita}, {More}, {Oguri}, \& {Osato}}]{sunayama20}
{Sunayama}, T., {Park}, Y., {Takada}, M., {et~al.} 2020, \mnras, 496, 4468

\bibitem[{{Sunyaev} {et~al.}(2021){Sunyaev}, {Arefiev}, {Babyshkin},
  {Bogomolov}, {Borisov}, {Buntov}, {Brunner}, {Burenin}, {Churazov},
  {Coutinho}, {Eder}, {Eismont}, {Freyberg}, {Gilfanov}, {Gureyev}, {Hasinger},
  {Khabibullin}, {Kolmykov}, {Komovkin}, {Krivonos}, {Lapshov}, {Levin},
  {Lomakin}, {Lutovinov}, {Medvedev}, {Merloni}, {Mernik}, {Mikhailov},
  {Molodtsov}, {Mzhelsky}, {M{\"u}ller}, {Nandra}, {Nazarov}, {Pavlinsky},
  {Poghodin}, {Predehl}, {Robrade}, {Sazonov}, {Scheuerle}, {Shirshakov},
  {Tkachenko}, \& {Voron}}]{Sunyaev2021}
{Sunyaev}, R., {Arefiev}, V., {Babyshkin}, V., {et~al.} 2021, \aap, 656, A132

\bibitem[{{Umetsu}(2020)}]{umetsu20}
{Umetsu}, K. 2020, \aapr, 28, 7

\bibitem[{{Varga} {et~al.}(2019){Varga}, {DeRose}, {Gruen}, {McClintock},
  {Seitz}, {Rozo}, {Costanzi}, {Hoyle}, {MacCrann}, {Plazas}, {Rykoff},
  {Simet}, {von der Linden}, {Wechsler}, {Annis}, {Avila}, {Bertin}, {Brooks},
  {Buckley-Geer}, {Burke}, {Carnero Rosell}, {Carrasco Kind}, {Carretero},
  {Cunha}, {D'Andrea}, {da Costa}, {De Vicente}, {Desai}, {Diehl}, {Dietrich},
  {Doel}, {Evrard}, {Flaugher}, {Fosalba}, {Frieman}, {Garc{\'\i}a-Bellido},
  {Gaztanaga}, {Gerdes}, {Gruendl}, {Gschwend}, {Gutierrez}, {Hartley},
  {Hollowood}, {Honscheid}, {James}, {Jeltema}, {Kuehn}, {Kuropatkin}, {Lima},
  {Maia}, {March}, {Marshall}, {Melchior}, {Menanteau}, {Miller}, {Miquel},
  {Ogando}, {Romer}, {Sanchez}, {Scarpine}, {Schubnell}, {Serrano},
  {Sevilla-Noarbe}, {Smith}, {Sobreira}, {Suchyta}, {Swanson}, {Tarle},
  {Thomas}, {Tucker}, {Zhang}, \& {DES Collaboration}}]{varga19}
{Varga}, T.~N., {DeRose}, J., {Gruen}, D., {et~al.} 2019, \mnras, 489, 2511

\bibitem[{{Wechsler} \& {Tinker}(2018)}]{wechsler18}
{Wechsler}, R.~H. \& {Tinker}, J.~L. 2018, \araa, 56, 435

\bibitem[{{Wu} {et~al.}(2022){Wu}, {Costanzi}, {To}, {Salcedo}, {Weinberg},
  {Annis}, {Bocquet}, {da Silva Pereira}, {DeRose}, {Esteves}, {Farahi},
  {Grandis}, {Rozo}, {Rykoff}, {Varga}, {Wechsler}, {Zeng}, {Zhang}, {Zhang},
  \& {DES Collaboration}}]{wu22}
{Wu}, H.-Y., {Costanzi}, M., {To}, C.-H., {et~al.} 2022, \mnras, 515, 4471

\bibitem[{{Wu} {et~al.}(2008){Wu}, {Rozo}, \& {Wechsler}}]{wu08}
{Wu}, H.-Y., {Rozo}, E., \& {Wechsler}, R.~H. 2008, \apj, 688, 729

\bibitem[{{Zhang} {et~al.}(2019){Zhang}, {Yanny}, {Palmese}, {Gruen}, {To},
  {Rykoff}, {Leung}, {Collins}, {Hilton}, {Abbott}, {Annis}, {Avila}, {Bertin},
  {Brooks}, {Burke}, {Carnero Rosell}, {Carrasco Kind}, {Carretero}, {Cunha},
  {D'Andrea}, {da Costa}, {De Vicente}, {Desai}, {Diehl}, {Dietrich}, {Doel},
  {Drlica-Wagner}, {Eifler}, {Evrard}, {Flaugher}, {Fosalba}, {Frieman},
  {Garc{\'\i}a-Bellido}, {Gaztanaga}, {Gerdes}, {Gruendl}, {Gschwend},
  {Gutierrez}, {Hartley}, {Hollowood}, {Honscheid}, {Hoyle}, {James},
  {Jeltema}, {Kuehn}, {Kuropatkin}, {Li}, {Lima}, {Maia}, {March}, {Marshall},
  {Melchior}, {Menanteau}, {Miller}, {Miquel}, {Mohr}, {Ogando}, {Plazas},
  {Romer}, {Sanchez}, {Scarpine}, {Schubnell}, {Serrano}, {Sevilla-Noarbe},
  {Smith}, {Soares-Santos}, {Sobreira}, {Suchyta}, {Swanson}, {Tarle},
  {Thomas}, {Wester}, \& {DES Collaboration}}]{zhang19}
{Zhang}, Y., {Yanny}, B., {Palmese}, A., {et~al.} 2019, \apj, 874, 165

\bibitem[{{Zohren} {et~al.}(2022){Zohren}, {Schrabback}, {Bocquet}, {Sommer},
  {Raihan}, {Hern{\'a}ndez-Mart{\'\i}n}, {Marggraf}, {Ansarinejad}, {Bayliss},
  {Bleem}, {Erben}, {Hoekstra}, {Floyd}, {Gladders}, {Kleinebreil}, {McDonald},
  {Schirmer}, {Scognamiglio}, {Sharon}, \& {Wright}}]{zohren22}
{Zohren}, H., {Schrabback}, T., {Bocquet}, S., {et~al.} 2022, \aap, 668, A18

\end{thebibliography}
%

\appendix

\section{Effective shape noise and number density in presence of shear responses}\label{app:Nsigma_eff}

The aim of this appendix is to present a derivation for the expression of the effective shape variance, Eq.~\ref{eq:sigma_eff} and the effective number density, Eq.~\ref{eq:N_eff}, in presence of a shear response $\mathcal{R}$.

\citet{friedrich21} demonstrate that the variance of a reduced shear profile $\hat g_\text{t}$ is given by 
\begin{equation}
    \text{Var}\left[ \hat g_\text{t}\right] = \frac{\sigma^2_\text{eff}}{N_\text{pair}} \text{ with } \sigma^2_\text{eff}=\frac{1}{2 N_\text{s}} \sum |e|^2 w_\text{n}^2,
\end{equation}
where $N_\text{pair}$ is the number of source lens pairs, $N_\text{s}$ the number of sources, $|e|^2=e_\text{t}^2 + e_\text{x}^2$ the squared modulus of the ellipticity when interpreted as a complex number, and $w_\text{n}$ normalized weights.

To generalize this expression to un-normalized weight $w$, just consider that such un-normalized weights can always be rescaled to normalized weights with the ansatz $w_\text{n} = C w$. The normalisation condition then readily leads to an expression for $C$, reading
\begin{equation}
    \frac{1}{N_\text{s}} \sum w_n = 1 = \frac{C}{N_\text{s}} \sum w \Leftrightarrow C^{-1} = \frac{1}{N_\text{s}} \sum w.
\end{equation}

Moving into the regime of individual cluster lensing, where $N_\text{pair}=N_\text{s}$, combining all the above equations, we find 
\begin{equation}\label{eq:appA:3}
    \text{Var}\left[ \hat g_\text{t}\right] = \frac{1}{2} \frac{1}{\left( \sum w \right)^2} \sum |e|^2 w^2,
\end{equation}
for generic, un-normalized weights.

When considering shear data with shear responses $\mathcal{R}$, \citet{friedrich21}, Appendix~D, suggest to transform the above equations by re-scaling both the ellipticities $e$, as well as the weight $w$, as follows
\begin{equation}
    \tilde e = \frac{e}{\mathcal{R}} \text{ and } \tilde w = \mathcal{R} w,
\end{equation}
thus correcting the ellipticity for the shear response, and adjusting the weights for the necessary re-scaling. Note that the summand in the numerator of Eq.~\ref{eq:appA:3} is invariant under this re-scaling, $|\tilde e|^2 \tilde w^2 = |e|^2 w^2$, and we thus only have to modify the summand in the denominator. Given that the above expression is valid for un-normalized weights, no further corrections are required. 

Dropping the tildes, we can readily write the variance of a tangential reduced shear profile in presence of shear response as 
\begin{equation}
    \text{Var}\left[ \hat g_\text{t}\right] = \frac{1}{2} \frac{1}{\left( \sum  w \mathcal{R}\right)^2} \sum |e|^2 w^2.
\end{equation}
Similarly, this justifies also why the source redshift distribution needs to be shear response weighted. 

Interpreting this expression as the ratio between an effective shape dispersion $\sigma_{\text{eff}, \alpha}$ for $\alpha \in$ (t, x), and an effective number of source $N_\text{eff}$ we can recover the above expression by using
\begin{equation}
    N_\text{eff} = \frac{\left( \sum  w \mathcal{R}\right)^2 }{\sum  w^2 \mathcal{R}^2} \text{ and } \sigma^2_{\text{eff},\alpha} = \frac{ \sum e^2_\alpha w^2}{\sum  w^2 \mathcal{R}^2} \text{ for } \alpha \in(\text{t},\, \text{x}),
\end{equation}
which are the expressions that we also use for our shape noise modelling.

\section{Parametrisation of cluster member contamination}\label{app:fcl_model}
We shall justify in the following the parameterisation used for the cluster member contamination in Section~\ref{sec:cluster_mbr_cnt}. As shown by \citet{hennig17}, cluster member galaxies follow an NFW profile, with the total number of galaxies enclosed in the virial region correlating tightly with mass, while the concentration varies depending on the galaxy type. Red-sequence galaxies have a larger concentration, more similar to the concentration of the dark matter, than non-red-sequence galaxies. We therefore assume that the cluster member galaxies that contaminated our background sample also follow an NFW profile.

The fraction of red-sequence galaxies in the cluster member contamination is unknown, though it stands to argue that their more precise photometric redshifts make it easier to exclude them with a background selection. We therefore let the concentration $c$ of the cluster member contaminants profile as a free parameter.

Similarly, we choose an agnostic approach to the redshift trend of the cluster member contamination. We simply assume that the ratio between the number density of field galaxies $n_\text{field}(z)$ and cluster member contaminants is a smooth function of redshift. We therefore employ a non-parameteric fit of the amplitudes $A_{z_i}$ for pivot redshifts $z_i$.

To account for possible mass trends we use the richness $\lambda$ as a mass proxy, given that we have to fit for the cluster member contamination prior to performing a mass calibration, and therefore do not have WL calibrated mass estimates at our disposal. For ICM-selected cluster samples, richness is known to scale approximately linearly with halo mass, $\lambda\sim M$ \citep{saro15, bleem20, grandis20, grandis21b}. We therefore use the richness to define an approximate virial radius for the NFW profile of the cluster member contaminants as $R_\text{vir} \propto \left( \lambda / 20 \right)^{1/3}$. We also allow the amplitude of the cluster member contamination to vary with richness.

Taken together, the above consideration lead to the ansatz

\begin{equation}
\begin{split}
    & n_\text{fcl}(\lambda, z, R) = A(\lambda, z, R) \,n_\text{field}(z) \text{ with} \\
    & A(\lambda, z, R) = A_{j} \left( \frac{\lambda}{25} \right)^{B_\lambda} \Sigma^\text{norm}_\text{NFW}\left(R \Big| r_S = c^{-1} \left( \frac{\lambda}{20}  \right)^{1/3}\right).
\end{split}
\end{equation}

The total number density results from the sum of the cluster member contaminants number density and the field number density, as

\begin{equation}
    n_\text{tot} = n_\text{fcl}(\lambda, z, R) + n_\text{field}(z) = \left[ A(\lambda, z, R)+1 \right] n_\text{field}(z),
\end{equation}
while the fraction of cluster member contaminants is
\begin{equation}
    f_\text{cl}(\lambda, z, R) = \frac{n_\text{fcl}}{n_\text{tot}} = \frac{A(\lambda, z, R) }{ A(\lambda, z, R)+1 },
\end{equation}
thus justifying the parametrisation chosen in Eq.~\ref{eq:fcl_profile}. This parametrisation has the benefit, that for $A(\lambda, z, R)>0$, $0<f_\text{cl}(\lambda, z, R)<1$, as already noted by \citet{grandis21b} in the context of the richness-dependent contamination of photometrically selected cluster samples.

Finally, one can show that $1+A = (1-f_\text{cl})^{-1}$, thus also motivating Eq.~\ref{eq:neff_clmc}.

\section*{Affiliations}
\begin{enumerate}
\item Universit\"at Innsbruck, Institut für Astro- und Teilchenphysik, Technikerstr. 25/8, 6020 Innsbruck, Austria
\item Max Planck Institute for Extraterrestrial Physics, Giessenbachstrasse 1, 85748 Garching, Germany \item  Faculty of Physics, Ludwig-Maximilians-Universit\"at M\"unchen, Scheinerstr.\ 1, 81679 Munich, Germany
\item  Argelander-Institut für Astronomie (AIfA), Universität Bonn, Auf dem H\"ugel 71, 53121 Bonn, Germany
\item  Link\"opings Universitet, Institutionen f\"or Systemteknik,
Link\"opings Universitet, Link\"oping 581 83, Sweden
\item Kavli Institute for Cosmology, University of Cambridge, Mading-
ley Road, Cambridge CB3 0HA, United Kingdom 
\item Institute of Astronomy, University of Cambridge, Madingley
Road, Cambridge, CB3 0HA, United Kingdom
\item Physics Department, University of Wisconsin-Madison, Madison, WI 53706, USA
\item Argonne National Laboratory, 9700 South Cass Avenue, Lemont,
IL 60439, USA
\item Department of Physics and Astronomy, University of Pennsylvania,
Philadelphia, PA 19104, USA
\item Department of Physics, Carnegie Mellon University, Pittsburgh,
Pennsylvania 15312, USA
\item Instituto de Astrofisica de Canarias, E-38205 La Laguna, Tenerife,
Spain
\item Laboratório Interinstitucional de e-Astronomia - LIneA, Rua Gal.
José Cristino 77, Rio de Janeiro, RJ - 20921-400, Brazil
\item Universidad de La Laguna, Dpto. Astrofísica, E-38206 La Laguna,
Tenerife, Spain
\item Center for Astrophysical Surveys, National Center for Supercom-
puting Applications, 1205 West Clark St., Urbana, IL 61801, USA
\item Department of Astronomy, University of Illinois at Urbana-
Champaign, 1002 W. Green Street, Urbana, IL 61801, USA
\item Physics Department, William Jewell College, Liberty, MO, 64068, USA
\item Department of Astronomy and Astrophysics, University of
Chicago, Chicago, IL 60637, USA
\item Kavli Institute for Cosmological Physics, University of Chicago,
Chicago, IL 60637, USA
\item Department of Physics, Duke University Durham, NC
27708, USA
\item Department of Physics, National Cheng Kung University, 70101 Tainan, Taiwan
\item NASA Goddard Space Flight Center, 8800 Greenbelt Rd, Green-
belt, MD 20771, USA
\item IRAP, Université de Toulouse, CNRS, UPS, CNES, Toulouse, France
\item Jodrell Bank Center for Astrophysics, School of Physics and
Astronomy, University of Manchester, Oxford Road, Manchester,
M13 9PL, United Kingdom
\item School of Mathematics and Physics, University of Queensland, Brisbane, QLD 4072, Australia
\item Lawrence Berkeley National Laboratory, 1 Cyclotron Road, Berke-
ley, CA 94720, USA
\item Fermi National Accelerator Laboratory, P. O. Box 500, Batavia,
IL 60510, USA
\item NSF AI Planning Institute for Physics of the Future, Carnegie
Mellon University, Pittsburgh, PA 15213, USA
\item Universit\'e Grenoble Alpes, CNRS, LPSC-IN2P3, 38000
Grenoble, France
\item Department of Physics and Astronomy, University of Waterloo,
200 University Ave W, Waterloo, ON N2L 3G1, Canada
\item Jet Propulsion Laboratory, California Institute of Technology,
4800 Oak Grove Dr., Pasadena, CA 91109, USA
\item SLAC National Accelerator Laboratory, Menlo Park, CA
94025, USA
\item Institut de F\'isica d’Altes Energies (IFAE), The Barcelona
Institute of Science and Technology, Campus UAB, 08193
Bellaterra (Barcelona), Spain
\item Department of Physics and Astronomy, Pevensey Building, University of Sussex, Brighton, BN1 9QH, UK
\item School of Physics and Astronomy, Cardiff University, CF24 3AA, United Kingdom
\item Department of Astronomy, University of Geneva, ch. d’\'Ecogia 16,
CH-1290 Versoix, Switzerland
\item Kavli Institute for Particle Astrophysics \& Cosmology, P. O. Box
2450, Stanford University, Stanford, CA 94305, USA
\item Department of Applied Mathematics and Theoretical Physics,
University of Cambridge, Cambridge CB3 0WA, United Kingdom
\item Department of Physics, Stanford University, 382 Via Pueblo
Mall, Stanford, CA 94305, USA
\item Instituto de F\'isica Gleb Wataghin, Universidade Estadual de
Campinas, 13083-859, Campinas, SP, Brazil
\item Physics Program, Graduate School of Advanced Science and Engineering, Hiroshima University, 1-3-1 Kagamiyama, Higashi-Hiroshima, Hiroshima 739-8526, Japan
\item Hiroshima Astrophysical Science Center, Hiroshima University, 1-3-1 Kagamiyama, Higashi-Hiroshima, Hiroshima 739-8526, Japan
\item Core Research for Energetic Universe, Hiroshima University, 1-3-1, Kagamiyama, Higashi-Hiroshima, Hiroshima 739-8526, Japan
\item Department of Physics, University of Genova and INFN, Via
Dodecaneso 33, 16146, Genova, Italy
\item Center for Cosmology and Astro-Particle Physics, The Ohio
State University, Columbus, OH 43210, USA
\item Centro de Investigaciones Energ\'eticas, Medioambientales y
Tecnol\'ogicas (CIEMAT), Madrid, Spain
\item Brookhaven National Laboratory, Bldg 510, Upton, NY 11973,
USA
\item Department of Physics and Astronomy, Stony Brook University,
Stony Brook, NY 11794, USA
\item D\'epartement de Physique Th\'eorique and Center for Astroparticle Physics (CAP), University of Geneva, 24 quai Ernest Ansermet, 1211 Gen\`eve, Switzerland
\item Institut de Recherche en Astrophysique et Plan\'etologie (IRAP), Universit\'e de Toulouse, CNRS, UPS, CNES, 14 Av. Edouard Belin, 31400 Toulouse, France
\item Department of Physics, Boise State University, Boise, ID 83725, USA
\item McWilliams Center for Cosmology, Department of Physics,
Carnegie Mellon University, Pittsburgh, PA 15213, USA
\item Department of Physics, University of Michigan, Ann Arbor, MI
48109, USA
\item Astronomy Unit, Department of Physics, University of Trieste, via Tiepolo 11, I-34131 Trieste, Italy
\item INAF-Osservatorio Astronomico di Trieste, via G. B. Tiepolo 11, I-34143 Trieste, Italy
\item Institute for Fundamental Physics of the Universe, Via Beirut 2, 34014 Trieste, Italy
\item Santa Cruz Institute for Particle Physics, Santa Cruz, CA 95064, USA
\item Computer Science and Mathematics Division, Oak Ridge National Laboratory, Oak Ridge, TN 37831
\item Hamburger Sternwarte, Universit\"{a}t Hamburg, Gojenbergsweg 112, 21029 Hamburg, Germany
\item Department of Physics, IIT Hyderabad, Kandi, Telangana 502285, India
\item Department of Physics \& Astronomy, University College London, Gower Street, London, WC1E 6BT, UK
\item Institute of Theoretical Astrophysics, University of Oslo. P.O. Box 1029 Blindern, NO-0315 Oslo, Norway
\item Instituto de Fisica Teorica UAM/CSIC, Universidad Autonoma de Madrid, 28049 Madrid, Spain
\item Department of Physics, The Ohio State University, Columbus, OH 43210, USA
\item Center for Astrophysics $\vert$ Harvard \& Smithsonian, 60 Garden Street, Cambridge, MA 02138, USA
\item George P. and Cynthia Woods Mitchell Institute for Fundamental Physics and Astronomy, and Department of Physics and Astronomy, Texas A\&M University, College Station, TX 77843,  USA
\item Observat\'orio Nacional, Rua Gal. Jos\'e Cristino 77, Rio de Janeiro, RJ - 20921-400, Brazil
\item School of Physics and Astronomy, University of Southampton,  Southampton, SO17 1BJ, UK

\end{enumerate}

\institute{ Universit\"at Innsbruck, Institut für Astro- und Teilchenphysik, Technikerstr. 25/8, 6020 Innsbruck, Austria\\ 
              \email{sebastian.grandis@uibk.ac.at}
         \and 
             Max Planck Institute for Extraterrestrial Physics, Giessenbachstrasse 1, 85748 Garching, Germany
         \and 
             Faculty of Physics, Ludwig-Maximilians-Universit\"at M\"unchen, Scheinerstr.\ 1, 81679 Munich, Germany
        \and 
            Argelander-Institut für Astronomie (AIfA), Universität Bonn, Auf dem H\"ugel 71, 53121 Bonn, Germany
        \and 
        Link\"opings Universitet, Institutionen f\"or Systemteknik,
Link\"opings Universitet, Link\"oping 581 83, Sweden
\and 
Kavli Institute for Cosmology, University of Cambridge, Mading-
ley Road, Cambridge CB3 0HA, United Kingdom 
\and 
Institute of Astronomy, University of Cambridge, Madingley
Road, Cambridge, CB3 0HA, United Kingdom
\and 
Physics Department, University of Wisconsin-Madison, Madison, WI 53706, USA
\and 
Argonne National Laboratory, 9700 South Cass Avenue, Lemont,
IL 60439, USA
\and 
Department of Physics and Astronomy, University of Pennsylvania,
Philadelphia, PA 19104, USA
\and 
Department of Physics, Carnegie Mellon University, Pittsburgh,
Pennsylvania 15312, USA
\and 
Instituto de Astrofisica de Canarias, E-38205 La Laguna, Tenerife,
Spain
\and 
Laboratório Interinstitucional de e-Astronomia - LIneA, Rua Gal.
José Cristino 77, Rio de Janeiro, RJ - 20921-400, Brazil
\and 
Universidad de La Laguna, Dpto. Astrofísica, E-38206 La Laguna,
Tenerife, Spain
\and 
Center for Astrophysical Surveys, National Center for Supercom-
puting Applications, 1205 West Clark St., Urbana, IL 61801, USA
\and 
Department of Astronomy, University of Illinois at Urbana-
Champaign, 1002 W. Green Street, Urbana, IL 61801, USA
\and 
Physics Department, William Jewell College, Liberty, MO, 64068, USA
\and 
Department of Astronomy and Astrophysics, University of
Chicago, Chicago, IL 60637, USA
\and 
Kavli Institute for Cosmological Physics, University of Chicago,
Chicago, IL 60637, USA
\and 
Department of Physics, Duke University Durham, NC
27708, USA
\and 
Department of Physics, National Cheng Kung University, 70101 Tainan, Taiwan
\and 
NASA Goddard Space Flight Center, 8800 Greenbelt Rd, Green-
belt, MD 20771, USA
\and 
IRAP, Université de Toulouse, CNRS, UPS, CNES, Toulouse, France
\and 
Jodrell Bank Center for Astrophysics, School of Physics and
Astronomy, University of Manchester, Oxford Road, Manchester,
M13 9PL, United Kingdom
\and 
School of Mathematics and Physics, University of Queensland, Brisbane, QLD 4072, Australia
\and 
Lawrence Berkeley National Laboratory, 1 Cyclotron Road, Berke-
ley, CA 94720, USA
\and 
Fermi National Accelerator Laboratory, P. O. Box 500, Batavia,
IL 60510, USA
\and 
NSF AI Planning Institute for Physics of the Future, Carnegie
Mellon University, Pittsburgh, PA 15213, USA
\and 
Universit\'e Grenoble Alpes, CNRS, LPSC-IN2P3, 38000
Grenoble, France
\and 
Department of Physics and Astronomy, University of Waterloo,
200 University Ave W, Waterloo, ON N2L 3G1, Canada
\and 
Jet Propulsion Laboratory, California Institute of Technology,
4800 Oak Grove Dr., Pasadena, CA 91109, USA
\and 
SLAC National Accelerator Laboratory, Menlo Park, CA
94025, USA
\and 
Institut de F\'isica d’Altes Energies (IFAE), The Barcelona
Institute of Science and Technology, Campus UAB, 08193
Bellaterra (Barcelona), Spain
\and 
Department of Physics and Astronomy, Pevensey Building, University of Sussex, Brighton, BN1 9QH, UK
\and 
School of Physics and Astronomy, Cardiff University, CF24 3AA, United Kingdom
\and 
Department of Astronomy, University of Geneva, ch. d’\'Ecogia 16,
CH-1290 Versoix, Switzerland
\and 
Kavli Institute for Particle Astrophysics \& Cosmology, P. O. Box
2450, Stanford University, Stanford, CA 94305, USA
\and 
Department of Applied Mathematics and Theoretical Physics,
University of Cambridge, Cambridge CB3 0WA, United Kingdom
\and 
Department of Physics, Stanford University, 382 Via Pueblo
Mall, Stanford, CA 94305, USA
\and 
Instituto de F\'isica Gleb Wataghin, Universidade Estadual de
Campinas, 13083-859, Campinas, SP, Brazil
\and 
Physics Program, Graduate School of Advanced Science and Engineering, Hiroshima University, 1-3-1 Kagamiyama, Higashi-Hiroshima, Hiroshima 739-8526, Japan
\and 
Hiroshima Astrophysical Science Center, Hiroshima University, 1-3-1 Kagamiyama, Higashi-Hiroshima, Hiroshima 739-8526, Japan
\and 
Core Research for Energetic Universe, Hiroshima University, 1-3-1, Kagamiyama, Higashi-Hiroshima, Hiroshima 739-8526, Japan
\and 
Department of Physics, University of Genova and INFN, Via
Dodecaneso 33, 16146, Genova, Italy
\and 
Center for Cosmology and Astro-Particle Physics, The Ohio
State University, Columbus, OH 43210, USA
\and 
Centro de Investigaciones Energ\'eticas, Medioambientales y
Tecnol\'ogicas (CIEMAT), Madrid, Spain
\and 
Brookhaven National Laboratory, Bldg 510, Upton, NY 11973,
USA
\and 
Department of Physics and Astronomy, Stony Brook University,
Stony Brook, NY 11794, USA
\and 
D\'epartement de Physique Th\'eorique and Center for Astroparticle Physics (CAP), University of Geneva, 24 quai Ernest Ansermet, 1211 Gen\`eve, Switzerland
\and 
Institut de Recherche en Astrophysique et Plan\'etologie (IRAP), Universit\'e de Toulouse, CNRS, UPS, CNES, 14 Av. Edouard Belin, 31400 Toulouse, France
\and 
Department of Physics, Boise State University, Boise, ID 83725, USA
\and 
McWilliams Center for Cosmology, Department of Physics,
Carnegie Mellon University, Pittsburgh, PA 15213, USA
\and 
Department of Physics, University of Michigan, Ann Arbor, MI
48109, USA
\and 
Astronomy Unit, Department of Physics, University of Trieste, via Tiepolo 11, I-34131 Trieste, Italy
\and 
INAF-Osservatorio Astronomico di Trieste, via G. B. Tiepolo 11, I-34143 Trieste, Italy
\and 
Institute for Fundamental Physics of the Universe, Via Beirut 2, 34014 Trieste, Italy
\and 
Santa Cruz Institute for Particle Physics, Santa Cruz, CA 95064, USA
\and 
Computer Science and Mathematics Division, Oak Ridge National Laboratory, Oak Ridge, TN 37831
\and 
Hamburger Sternwarte, Universit\"{a}t Hamburg, Gojenbergsweg 112, 21029 Hamburg, Germany
\and 
Department of Physics, IIT Hyderabad, Kandi, Telangana 502285, India
\and 
Department of Physics \& Astronomy, University College London, Gower Street, London, WC1E 6BT, UK
\and 
Institute of Theoretical Astrophysics, University of Oslo. P.O. Box 1029 Blindern, NO-0315 Oslo, Norway
\and 
Instituto de Fisica Teorica UAM/CSIC, Universidad Autonoma de Madrid, 28049 Madrid, Spain
\and 
Department of Physics, The Ohio State University, Columbus, OH 43210, USA
\and 
Center for Astrophysics $\vert$ Harvard \& Smithsonian, 60 Garden Street, Cambridge, MA 02138, USA
\and 
George P. and Cynthia Woods Mitchell Institute for Fundamental Physics and Astronomy, and Department of Physics and Astronomy, Texas A\&M University, College Station, TX 77843,  USA
\and 
Observat\'orio Nacional, Rua Gal. Jos\'e Cristino 77, Rio de Janeiro, RJ - 20921-400, Brazil
\and 
School of Physics and Astronomy, University of Southampton,  Southampton, SO17 1BJ, UK
}

\end{document}